\documentclass[twoside]{article}

\usepackage{amsbsy,amssymb,amsmath,setspace,ulem}
\usepackage{mathrsfs}

\usepackage{datetime}
\usepackage{titling}

\usepackage{graphicx,color}
\usepackage[dvipsnames]{xcolor}
\usepackage{moreverb}
\usepackage{dsfont}
\usepackage{upgreek}
\usepackage{bm}
\usepackage{multirow}
\usepackage{soul}
\usepackage{framed,mdframed}

\usepackage{relsize} 

\usepackage{wrapfig} 

\usepackage[sc]{mathpazo} 
\usepackage[T1]{fontenc} 
\usepackage{microtype} 

\usepackage[hmarginratio=1:1,top=32mm,left=18mm,columnsep=17pt]{geometry} %
\usepackage{multicol} 
\usepackage[hang, small,labelfont=bf,up,textfont=it,up]{caption} 
\usepackage{booktabs} 
\usepackage{float} 
\usepackage[backref=page]{hyperref} 
\renewcommand*{\backref}[1]{} 
\renewcommand*{\backrefalt}[4]{%
    \ifcase #1 (Not cited.)%
    \or        (Cited on page~#2.)%
    \else      (Cited on pages~#2.)%
    \fi}

\usepackage{soul} 

\usepackage{lettrine} 
\usepackage{paralist} 

\usepackage{abstract} 

\usepackage{empheq}
\newlength\mytemplen
\newsavebox\mytempbox
\makeatletter
\definecolor{myblue}{rgb}{.8, .8, 1}
\newcommand\mybluebox{%
    \@ifnextchar[
       {\@mybluebox}%
       {\@mybluebox[0pt]}}
\def\@mybluebox[#1]{%
    \@ifnextchar[
       {\@@mybluebox[#1]}%
       {\@@mybluebox[#1][0pt]}}
\def\@@mybluebox[#1][#2]#3{
    \sbox\mytempbox{#3}%
    \mytemplen\ht\mytempbox
    \advance\mytemplen #1\relax
    \ht\mytempbox\mytemplen
    \mytemplen\dp\mytempbox
    \advance\mytemplen #2\relax
    \dp\mytempbox\mytemplen
    \colorbox{myblue}{\hspace{1em}\usebox{\mytempbox}\hspace{1em}}}
\makeatother

\usepackage{bm} 
\usepackage{tikz}\usetikzlibrary{shapes.misc} 

\usepackage{titlesec} 
\usepackage{bold-extra}
\titleformat{\section}[block]{\Large\scshape\centering\color{Blue}}{\thesection.}{1em}{}
\titleformat{\subsection}[block]{\large\color{MidnightBlue}}{\thesubsection.}{1em}{}

\usepackage{fancyhdr} 
\fancypagestyle{firstpagestyle}
{
	\fancyhead[C]{Peshkov, Pavelka, Romenski, Grmela  / \textit{SHTC\&GENERIC}} 
	\fancyhf{}
	\fancyfoot[R]{ \thepage}
	\fancyfoot[L]{\footnotesize\it Preprint submitted to the Journal of 
	Dead-End 
	Ideas}
}

\definecolor{shadecolor}{rgb}{0.85,0.85,0.85}

\newcommand{\AAA}{{\boldsymbol{A}}}

\newcommand{\scH}{\mathscr{H}}
\newcommand{\scS}{\mathscr{S}}
\newcommand{\GG}{{\boldsymbol{G}}}
\newcommand{\rr}{{\boldsymbol{r}}}
\newcommand{\vv}{{\boldsymbol{v}}}
\newcommand{\aaa}{{\boldsymbol{a}}}
\newcommand{\bb}{{\boldsymbol{b}}}

\newcommand{\ww}{{\boldsymbol{w}}}
\newcommand{\mm}{{\boldsymbol{m}}}
\newcommand{\qq}{{\boldsymbol{q}}}
\newcommand{\pp}{{\boldsymbol{p}}}
\newcommand{\xx}{{\boldsymbol{x}}}
\newcommand{\yy}{{\boldsymbol{y}}}
\newcommand{\uu}{{\boldsymbol{u}}}
\newcommand{\ee}{{\boldsymbol{e}}}
\newcommand{\hh}{{\boldsymbol{h}}}
\newcommand{\scE}{{\mathscr{E}}}

\newcommand{\FF}{{\boldsymbol{F}}}
\newcommand{\EE}{{\boldsymbol{E}}}

\newcommand{\DD}{{\boldsymbol{D}}}
\newcommand{\BB}{{\boldsymbol{B}}}
\newcommand{\BBB}{{\mathcal{B}}}
\newcommand{\BBBB}{{\bm{\mathcal{B}}}}
\newcommand{\BBBdev}{{\mathrm{dev}\mathcal{B}}}
\newcommand{\BBBBdev}{{\text{dev}\bm{\mathcal{B}}}}
\newcommand{\JJ}{{\boldsymbol{J}}}
\newcommand{\II}{{\boldsymbol{I}}}
\newcommand{\QQ}{{\boldsymbol{Q}}}

\newcommand{\scU}{{\mathscr{U}}}

\newcommand{\ppi}{\boldsymbol{\pi}}
\newcommand{\ted}{E} 
\newcommand{\tes}{\mathsf{E}} 
\newcommand{\LAlg}{\ell}
\newcommand{\LDer}{\mathcal{L}}

\newcommand{\AF}{\mathscr{A}}
\newcommand{\BF}{\mathscr{B}}
\newcommand{\CF}{\mathscr{C}}
\newcommand{\SF}{\mathscr{S}}
\newcommand{\EF}{\mathscr{E}}

\newcommand{\pd}{\partial}
\newcommand{\dive}{\mathrm{div\,}}
\newcommand{\rmd}{{\rm d}}
\newcommand{\UM}{\bm{I}}

\newcommand{\transpose}{{\rm {\mathsmaller T}}}

\newcommand{\IP}[1]{ \textcolor{blue}   {\small\texttt{
\includegraphics[scale=0.06]{pin_small.jpeg} Ilya: #1}} }
\newcommand{\MP}[1]{ \textcolor{Green}   {\small\texttt{
\includegraphics[scale=0.06]{pin_small.jpeg} Michal: #1}} }

\newcommand{\eps}{\varepsilon}

\newcommand{\Tr}{\text{tr}}

\hypersetup{colorlinks=true,
	linkcolor= [rgb]{0.8, 0.0, 0},
	citecolor=NavyBlue}

\title{\vspace{-15mm}\fontsize{16pt}{10pt}\selectfont{Continuum Mechanics and 
Thermodynamics \\ in the Hamilton and the Godunov-type 
Formulations}}

\author{
\large
\textsc{Ilya Peshkov\thanks{Institut de Math\'{e}matiques de Toulouse, 
Toulouse, France and Sobolev Institute of 
Mathematics, Novosibirsk, Russia, \href{mailto:peshenator@gmail.com}{e-mail}},
\ \
Michal Pavelka\thanks{Mathematical Institute, Faculty of Mathematics and 
Physics, Charles University, Sokolovsk\'{a} 83, 186 75 Prague, Czech Republic, 
\href{mailto:pavelka@karlin.mff.cuni.cz}{e-mail}}, 
\ \
Evgeniy Romenski\thanks{Sobolev 
Institute of Mathematics and Novosibirsk State University, Novosibirsk, 
Russia, \href{mailto:evrom@math.nsc.ru}{e-mail}},\ \
Miroslav Grmela\thanks{\'{E}cole Polytechnique de Montr\'{e}al, C.P.6079 suc. 
Centre-ville, Montr\'{e}al, H3C 3A7, Qu\'{e}bec, Canada, 
\href{mailto:miroslav.grmela@polymtl.ca}{e-mail}
}
}
}
\date{}
\thanksmarkseries{arabic}
\begin{document}

\maketitle

\renewcommand{\abstractname}{Spoiler}
\vspace{-20mm}
\begin{abstract}
\vspace{-5mm}
\noindent Continuum mechanics with dislocations, with the  Cattaneo type heat 
conduction, with mass transfer,  and with electromagnetic fields is put into  
the Hamiltonian form 
and into the form of the  Godunov type system of the first order, symmetric  
hyperbolic partial differential equations (SHTC equations). The compatibility 
with thermodynamics  of the time reversible part of the governing equations is 
mathematically expressed in the former formulation as degeneracy of the 
Hamiltonian structure and in the latter formulation as  the existence of a 
companion conservation law. In both formulations the time irreversible part 
represents gradient dynamics. The  Godunov type formulation brings the 
mathematical rigor (the well-posedness of the Cauchy initial value problem) and 
the possibility to  discretize while keeping the physical content of the 
governing equations (the Godunov finite volume discretization).
\end{abstract}

\thispagestyle{fancy} 
\thispagestyle{firstpagestyle}

{\hypersetup{linkcolor=black} \tableofcontents}

\section{Introduction}\label{sec.Intro}

Results of experimental observations are seen in mathematical  models as 
properties of solutions of their governing equations. Universality of some 
results translates in the models into the universality of the mathematical 
structure of the equations. For instance one of the universal experimental 
observations is the  approach of externally unforced macroscopic systems to the 
thermodynamic equilibrium states at which the behavior is found to be well 
described by the classical equilibrium thermodynamics. The structure 
responsible for this observation has been identified to be the following. The 
irreversible part of the time evolution is of gradient type generated by a 
potential called entropy. This potential is then preserved in the reversible 
part of the time evolution that is Hamiltonian, since it is the part of the 
time evolution directly inherited from the underlying microscopic mechanics.  
The time-evolution equations involving both the
Hamiltonian and the gradient part have appeared first in 
\cite{Dzyaloshinskii1980}, in \cite{GC} (that was presented at
the AMS-IMS-SIAM Joint Summer Research Conference in the Mathematical Sciences 
on Fluids and Plasmas ``Geometry and Dynamics'', held at the University of 
Colorado, Boulder, CO, USA, 17--23 July 1983) and in  
\cite{kauf,mor,GPhD,Beris1994,GrmelaOttingerI,GrmelaOttingerII,Ottinger-book}.  In \cite{GrmelaOttingerI,GrmelaOttingerII}, the 
abstract equation \eqref{eqn.GENERIC}  has been called GENERIC.

In the  smaller pool of governing equations  that arise  in continuum mechanics 
we may expect to identify even stronger common structure. In particular, what 
is missing in  GENERIC is the guarantee that the initial value problem is well 
posed. This is indeed one of the basic validations of a model. The time 
evolution is seen to exist and thus the time evolution seen  in the model 
should also exist. Another aspect that is not included  in GENERIC is the 
adaptability of the governing equation to searching their solutions 
numerically.

The structure addressing both of these aspects have been identified by 
Godunov~\cite{God1961,God1962} for a system of the first order 
partial  
differential equations having the form of local conservation laws (i.e. 
equations having the form of the  time derivative of a field equals divergence 
of a flux). The compatibility with thermodynamics of these equations 
(representing the time reversible part of the time evolution) is expressed by 
the existence of a companion local conservation law for a new field that is an 
appropriately  regular and convex function of the fields playing the role of 
state variables. It is very interesting that this requirement of the agreement 
with experimental observations implies then directly the mathematical 
well-posedness (the companion local conservation law enables to symmetrize the  
system of the first order hyperbolic equations which then implies their well 
posedness). Godunov has  also introduced a discretization (the modification of 
the governing equations needed when computers are brought to assistance in the 
process of finding details of their solutions) preserving the physical content 
of the original partial differential equations~\cite{God1959}.

In this paper we introduce a large family of partial differential equations 
that:
\begin{enumerate}
\item Govern the time evolution of fluids, solids without and with 
dislocations,  the Cattaneo type heat conduction and the electromagnetic 
fields, all coupled together.
\item Possess the GENERIC structure.
\item Possess the SHTC structure, that extends to Godunov structure to a larger 
class of the first order partial differential equations.
\end{enumerate}

The GENERIC route begins with the well known Hamiltonian structure of the Euler 
hydrodynamic equations \cite{Clebsch,Arnold1}. By adopting the field 
of labels into the set of state variables, we pass to the Lagrangian 
formulation and to the solid mechanics without dislocations 
\cite{Miroslav-PLA} and  (by using  the extension developed in this paper) 
with dislocations. This Hamiltonian  formulation of the reversible part of the 
continuum mechanics is then coupled to the Hamiltonian formulation of the 
Cattaneo-like heat conduction and  to the Hamiltonian formulation of the time 
evolution of electromagnetic fields.

The Godunov route begins with the Lagrangian formulation in which the time 
reversible governing equations have the form of the local conservation laws and 
the original Godunov analysis thus applies to them. By passing to the Eulerian 
formulation (the Lagrange $\rightarrow$ Euler passage is worked out in detail 
in this 
paper) the governing equations do not remain to be, at least in general, local 
conservation laws. Their weakly nonlocal extension is however shown to possess 
a new structure (called SHTC structure) that also allows symmetrization, and 
thus the mathematical well posedness, as well as the applicability of the 
Godunov discretization,  is guaranteed for them. The weakly nonlocal extension 
(having the physical meaning, for example, of including dislocations into the 
consideration) consists of:
\begin{enumerate}
\item  Constructing new local conservation laws for new fields that are 
appropriate spatial derivatives of the original fields included in the set of 
the state variables (e.g. the vorticity) by appropriately combining the 
original system of equations with  equations obtained by applying   appropriate 
spatial derivatives on the original system of equations.
\item Adopting the new fields (that are appropriate spatial derivatives of the 
fields included in the original set of the state variables)  as new independent 
fields. 
\end{enumerate}

Both the GENERIC and the Godunov routes lead to the same equations. Having 
proven that the equations  possess both the GENERIC and the SHTC structures, 
all the results and techniques that are available for these structures (in 
particular the physical, the mathematical, and the numerical well posedness) 
are now available for the governing equations investigated in this paper.

\section{Preliminaries}\label{sec.preliminaries}

\subsection{SHTC framework, Symmetric Hyperbolic Equations, Well-Posedness, 
Causality}\label{sec.prem.shtc}

The origin of the SHTC formulation of continuum mechanics should be attributed 
to the work of 
Godunov~\cite{God1961} where he considered ``an interesting class'' of 
nonlinear conservation laws. The key motivation (from the words 
of S.K. Godunov) for this seminal paper was the will to understand what 
physical principles may guarantee the \textit{ well-posedness} of the initial 
value problem (IVP) for a \textit{nonlinear} system of time dependent PDEs. 
Indeed, when one deals with nonlinear dynamical phenomena, and in particular with 
nonlinear 
time-dependent partial
differential equations (PDEs), perhaps, the first examination a new model has 
to pass is to verify if the IVP is well-posed, at least locally, i.e. whether 
the solution to the system of PDEs with given initial data exists, is unique 
and stable (depends continuously on  the initial data). We emphasize
that the well-posedness of the IVP should not be considered as a purely 
mathematical requirement but as a
\textit{fundamental physical observation} about the time evolution of physical 
systems, 
i.e. exactly as we consider
causality, conservation and thermodynamic principles, Galilean or Lorentz 
invariance to be essential features of macroscopic time evolution. In other 
words, a model describing the 
time evolution of a physical
system and having an ill-posed IVP should be regarded as wrong. Moreover, the 
well-posedness of the IVP
is a fundamental property of time-dependent PDEs in order for them to be numerically 
solvable.

Not all nonlinear systems of PDEs, of course, are well posed. 
This even cannot be guaranteed for a first order 
quasi-linear system~\cite{Serre2007,DupretMarchal1986,MullerRuggeri1998},
or for models which were believed to be consistently derived from microscopic 
theories as, for 
example, the Burnett
equations derived from the gas kinetic 
theory~\cite{Bobylev1982,StruchtrupTorrilhonR13,Torrilhon2016a}. However, there 
is a class of nonlinear PDEs for which the IVP is locally well-posed in time, 
by definition, which is the class of 
hyperbolic PDEs. Unfortunately,
it is difficult to prove that a given non-linear first order system is 
hyperbolic in the domain of interest (domain of state parameters) because this 
would 
require to prove the
existence of the full basis of eigenvectors for a matrix whose entries depends
nonlinearly on the state variables.
So how can one guaranty that the IVP for a new nonlinear 
continuum mechanics model is well-posed, i.e. 
that the model is hyperbolic? 

The only possibility we see is to deliberately 
develop the model within a very important \textit{subclass} of first order 
hyperbolic systems 
\begin{equation}\label{eq.intro.symmetric}
\mathsf{A}(\pp)\frac{\pd\pp}{\pd t} + \mathsf{B}_k(\pp)\frac{\pd\pp}{\pd x_k} 
= 0,
\end{equation}
where $ \mathsf{A}^\transpose=\mathsf{A} >0$ and $ 
\mathsf{B}_k^\transpose=\mathsf{B}_k$, for which (local) well-posedness is 
known to hold true~\cite{Kato1975,Serre2007,MullerRuggeri1998,Ruggeri2005}. 
This subclass is called  \textit{symmetric  hyperbolic  systems of PDEs} which 
is a generalization of Friedrichs-symmetrizable linear 
systems~\cite{Friedrichs1958} and is known to be globally hyperbolic by  
definition. However, one should
naturally question how restrictive it is
for a model to be symmetric hyperbolic and, at the same time, represent diverse 
physical phenomena admitting continuum mechanics description? As it was shown 
by Godunov~\cite{God1961,God1962,God1972MHD} and later by
others~\cite{FriedLax1971,Boillat1974,Ruggeri1981Euler} there is an intimate 
connection between the symmetric hyperbolicity and thermodynamics for a 
specific subclass of \textit{system of conservation laws}. 

In particular, if a 
first order system of conservation laws for conservative state 
variables $ \qq = (q_1,q_2,\ldots,q_n) $
\begin{equation}\label{eqn.intro.conslwas}
\frac{\pd \qq}{\pd t} + \frac{\pd \FF^k(\qq)}{\pd x_k} = 0
\end{equation}
admits an extra conservation law
\begin{equation}\label{eqn.intro.energy}
\frac{\pd E(\qq)}{\pd t} + \frac{\pd G^k(\qq)}{\pd x_k} = 0
\end{equation}
for a \textit{strictly convex potential} $ E(\qq) $, which plays the role of 
the 
total energy for the system, then such a system can be parametrized with new 
state variables $ \pp =(p_1,p_2,\ldots,p_n) $ and 
a potential $ 
L(\pp) $ as follows (which shall be refereed to as the \textit{Godunov form} of 
conservation 
laws)
\begin{equation}\label{eqn.intro.Lp}
\frac{\pd L_\pp}{\pd t} +\frac{\pd L^k_\pp}{\pd x_k}=0,
\end{equation}
where, as through out this paper, $ L_\pp $ and $ L^k_\pp $ are the vectors 
with components $ L_{p_i} = \pd L/\pd p_i$, $ L^k_{p_i} = \pd L^k/\pd 
p_i$, see below for the concrete form of these vectors for the conservative 
system. It is obvious that \eqref{eqn.intro.Lp} can be rewritten in a 
symmetric quasilinear form 
\begin{equation}\label{eqn.intro.symm}
L_{p_i p_j}\frac{\pd p_j}{\pd t} + L^k_{p_i p_j}\frac{\pd p_j}{\pd x_k} = 0.
\end{equation}
Moreover, the conservative state variables $ \qq $ and the potential $ E(\qq) $ 
are related to the new variables $ \pp $ and potential $ L(\pp) $ as
\begin{equation}\label{eqn.intro.qprelations}
\qq = L_\pp, \qquad E(\qq) = \pp\cdot L_\pp - L(\pp) = \pp\cdot\qq - L(\pp)
\end{equation}
and vice versa,
\begin{equation}\label{eqn.intro.pqrelations}
\pp = E_\qq, \qquad L(\pp) = \qq\cdot E_\qq - E(\qq) = \qq\cdot\pp - E(\qq).
\end{equation}
The relations $ \pp = E_\qq $ and $ \qq = L_\pp $ are one-to-one relations 
because we assume that $ E(\qq) $ is convex and hence, the Hessian $ 
\frac{\pd\pp}{\pd \qq} = E_{\qq\qq} = (L_{\pp\pp})^{-1} > 0$. Thus, the 
convexity of $ E(\qq) $, in fact, means that \eqref{eqn.intro.symm} is 
symmetric hyperbolic. This result is referred to as the Godunov-Friedrichs-Lax (GFL) 
theorem \cite{Gavrilyuk}.

The key point of the Godunov observation is that the system 
\eqref{eqn.intro.conslwas} and conservation law \eqref{eqn.intro.energy} 
together forms an \textit{overdetermined} 
system of PDEs, i.e. the number of PDEs is $ n+1 $ while the number of state 
variables is $ n $ ($ E $ is not an unknown but the potential $ E(\qq) $), and 
thus, in order such a system has a solution, one of the 
conservation laws should be a consequence of the others. Therefore, the 
simultaneous 
validity of \eqref{eqn.intro.conslwas} and \eqref{eqn.intro.energy} implies 
that 
\begin{equation}\label{eqn.intro.summ}
E_\qq\cdot\left(\frac{\pd \qq}{\pd t} + \frac{\pd \FF^k(\qq)}{\pd x_k} \right) 
\equiv \frac{\pd E(\qq)}{\pd t} + \frac{\pd G^k(\qq)}{\pd x_k},
\end{equation}
where the dot denotes the scalar product.
After introducing the potentials $ L^k(\pp) = E_\qq\cdot\FF^k(\qq) - G^k(\qq) = 
\pp\cdot\FF^k(\qq) - G^k(\qq) $ and noting that $ L^k_\pp = \FF^k $, the energy 
conservation~\eqref{eqn.intro.energy} can be rewritten as
\begin{equation}
\frac{\pd (p_i L_{p_i}-L)}{\pd t} +\frac{\pd (p_iL^k_{p_i}-L^k)}{\pd x_k} = 0.
\end{equation}
In other words, $L$ is the Legendre-conjugate energy and $L^k$ is Legendre 
conjugation of energy flux.
Therefore, returning back to our last question, such a subclass of symmetric 
hyperbolic PDEs can be associated with the \textit{thermodynamically 
compatible} systems of first order nonlinear conservation laws.

However, it turns out that there are some important continuum mechanics models, 
like ideal magnetohydrodynamics (MHD) equations for example, which {\textit{ 
seems}} to be represented by systems of conservation 
laws~\eqref{eqn.intro.conslwas} with an additional conservation 
law~\eqref{eqn.intro.energy} but still can not be symmetrizable in such a way. 
A more attentive analysis by Godunov of this inconsistency of the MHD equations 
in~\cite{God1972} shows that, in fact, the MHD equations written in a classical 
conservative form are not compatible with the energy conservation, i.e. the 
energy conservation law is not the consequence of the mass, momentum, entropy 
and magnetic field conservation laws, like in~\eqref{eqn.intro.summ}. More 
precisely, the energy flux of the MHD equation is inconsistent with the other 
fluxes unless the \textit{non-conservative} product $ B_i v_i \frac{\pd 
B_k}{\pd x_k} \equiv 0$ 
is added to the left-hand side of~\eqref{eqn.intro.summ}, where $ v_i $ and $ 
B_i $ are the velocity and magnetic fields accordingly. We discuss this in more 
details in Section~\ref{sec.Euler.conservative}. Godunov concludes that the 
ideal MHD equations do not admit the original Godunov 
structure~\eqref{eqn.intro.Lp} but generalize it in the following way
\begin{equation}\label{eqn.intro.mhd}
\frac{\pd L_{\pp}}{\pd t} + \frac{\pd (v_k L)_{\pp}}{\pd x_k} + \mathsf{C}_k 
\frac{\pd \pp}{\pd x_k} = 0,
\end{equation}
where $ \mathsf{C}_k^\transpose = C_k $ are symmetric matrices and $ v_k $ 
being the velocity field.

The latter analysis by Godunov and Romenski in nineties in a series of 
papers~\cite{GodRom1995,GodRom1996,GodRom1996a,GodRom1998,GodRom1998,Rom2001,Romenski2002,GodRom2003} of various equations (ideal MHD equations, nonlinear 
elasticity, 
electrodynamics of slowly moving medium, superfluid flow, superconductivity) 
studied in the 
books of Landau and Lifshitz~\cite{Landau1984electrodynamics,Landau-Lifshitz6} 
reveals that for a proper understanding of the \textit{structure} of the 
continuum 
mechanics models it is important to distinguish the Lagrangian and Eulerian 
frames of reference, and that the original Godunov 
structure~\eqref{eqn.intro.Lp} is, in 
fact, rather inherent to the conservation laws written in the Lagrangian frame, 
while their Eulerian counterparts are \textit{inherently non-conservative} time 
evolutions and have more complicated structure~\eqref{eqn.intro.mhd}.

The departure point of the SHTC framework is the invariant under arbitrary 
rotations conservation 
laws in the Lagrangian frame, which were proposed in~\cite{GodRom1996a} and 
can be derived from Hamilton's principle of stationary action, see 
Section~\ref{sec.variation}. Here, we also demonstrate for the first 
time the the Lagrangian SHTC equations can be generated by \textit{canonical 
Poisson 
brackets}. Then, Eulerian non-conservative SHTC 
equations~\eqref{eqn.intro.mhd} are the direct consequence of the 
Lagrange-to-Euler transformation applied to the Lagrangian conservation laws. 
It is the primary goal of this paper to demonstrate that the Eulerian 
non-conservative SHTC equations~\eqref{eqn.intro.mhd} are fully compatible with 
the GENERIC framework presented in the subsequent section. 
Another goal of the paper is to provide the reader with some non-trivial 
details of the Lagrange-to-Euler transformation between the conservative 
Lagrangian SHTC equations and their non-conservative Eulerian counterparts 
which have never been published before. It is thus one of the goal of our paper 
to 
provide these important, though quite technical, details. Moreover, we review 
all properties of the SHTC formulation which were previously published rather 
separately in a series of paper by Godunov and 
Romenski~\cite{GodRom1995,GodRom1996,GodRom1996a,GodRom1998,GodRom1998,Rom2001,Romenski2002,GodRom2003}.

Eventually, we also recall that apart of the well-posedness of the IVP the 
hyperbolicity also naturally accounts for another fundamental observation about 
time evolution of physical 
systems, namely the finite
velocity for any perturbation propagation, i.e. \textit{causality}.

\subsection{GENERIC}

General Equation for Non-Equilibrium Reversible-Irreversible Coupling (GENERIC) 
was first introduced in \cite{GrmelaOttingerI,GrmelaOttingerII}. The 
fundamental idea behind GENERIC is that evolution equations for chosen state 
variables are generated from four building blocks: a Poisson bracket 
$\{\bullet,\bullet\}$, Hamiltonian $\scE = \int E d\rr$, dissipation potential 
$\Xi$ and entropy functional $\scS = \int s d\rr$, 
where the potentials $ E $ and 
$ 
s $ are 
the total energy and entropy densities.  Once these building blocks are at 
hand, 
the evolution equations follow by direct calculation.

The building blocks are required to fulfill certain degeneracies and 
conditions: 
\begin{enumerate}
\item 
Entropy should be a Casimir of the Poisson bracket, which means 
that $\{\AF,\SF\}=0$ for all functionals $\AF$ of the state variables. 
\item Energy is 
not changed by the dissipation potential
\footnote{$\langle\bullet,\bullet\rangle$ denoting integration ($L^2$ scalar product)}, i.e. $\langle \EF_\xx, \Xi_{S_\xx}\rangle 
= 
0$, and 
\item the dissipation potential is a monotonous operator, i.e. $\langle 
\Xi_{\SF_\xx}, \SF_\xx\rangle = 0$ for all entropies.\footnote{Note that 
subscripts 
denote functional derivatives $\SF_\xx = \frac{\delta\SF}{\delta\xx}$. In the special case when 
the functional is a spatial integral over a density which is a real-valued function of the state variables, 
the functional derivative becomes the usual derivative of the density with 
respect to the state variable. For example $\EF = \int E(\xx(\rr))d\rr$ and 
$\EF_\xx = \frac{\partial E}{\partial \xx}$ if $E$ is a real-valued function of 
real variables.}
\end{enumerate}

Apart from these conditions, the Poisson bracket has the following properties: 
\begin{enumerate}
\item 
antisymmetry,  $\{\AF,\BF\}=-\{\BF,\AF\}$, 
\item
Leibniz rule, $\{\AF\BF, \CF\} = 
\{\AF,\CF\}\BF + \AF\{\BF,\CF\}$, 
\item
Jacobi identity, $\{\{\AF,\BF\},\CF\} + \{\{\BF,\CF\},\AF\}+\{\{\CF,\AF\},\BF\} 
= 0$.
\end{enumerate}
 Moreover, the Poisson bracket is constructed 
from a Poisson bivector $L^{ij}$ (antisymmetric twice contravariant tensor 
field) as 
\begin{equation}
\{\AF,\BF\} = \AF_{x^i} L^{ij} \BF_{x^j}.
\end{equation}
Note that summation over an index can be interpreted as integral over fields. 
Finally, Hamiltonian evolution of functionals is given by
\begin{equation}
\frac{\pd \AF}{\pd t} = \{\AF,\EF\}.
\end{equation}

GENERIC then gives the following evolution equations for functionals of state 
variables $\xx$:
\begin{equation}
\frac{\pd \AF}{\pd t} = \{\AF, \EF\} + \langle \AF_\xx, \frac{\delta \Xi}{\delta \SF_\xx}\rangle.
\end{equation}
From antisymmetry of the Poisson bracket and from the degeneracy of the 
dissipation potential it follows that energy is conserved, $\pd\EF/\pd t = 0$. 
From 
the degeneracy of the Poisson bracket and from monotonicity of the dissipation 
potential it follows that entropy grows, $\pd \SF/\pd t \geq 0$. Moreover, when 
energy, entropy and the dissipation potential are taken as even functions with 
respect to time reversal transformation, the Hamiltonian part generates 
reversible evolution 
while the dissipative part generates irreversible evolution 
\cite{Pavelka2014a}. Finally, Onsager-Casimir reciprocal relations are 
satisfied automatically \cite{Pavelka2014a,Ottinger-book}. These results 
manifest compatibility of GENERIC with thermodynamics. 

Instead of working with evolution of functionals, which can be interpreted as a 
sort of weak formulation, one can recover the evolution equations of the state 
variables easily. Taking $\AF = x^i$, we obtain
\begin{equation}\label{eqn.GENERIC}
\frac{\pd x^i}{\pd t} = L^{ij} \EF_{x^j} + \Xi_{\SF_{x^i}}.
\end{equation}
This last equation shows how to 
write down the evolution equations (usually partial differential equations) 
implied by GENERIC.

Since Poisson brackets are a cornerstone of GENERIC, a question often arises 
how the Poisson brackets can be derived from first principles. For example the 
canonical Poisson bracket, that generates Hamilton canonical equations in 
classical mechanics, is generated by the canonical two-form present on the 
cotangent bundle of classical mechanics (positions and momenta of all 
particles). The canonical Poisson bracket can be thus interpreted as a 
geometric result of construction of the cotangent bundle. 

The hydrodynamic Poisson bracket, that generates compressible Euler equations, 
is different from the canonical Poisson bracket. Of course, it generates 
evolution equations for the hydrodynamic fields: density, momentum density and 
entropy density. It can be derived in several ways. One can start for example 
from dynamics of particles in the Lagrangian frame and perform transformation 
to the Eulerian frame \cite{Abarbanel}. The calculation is, however, rather 
complicated. A shorter way is to construct dual of the Lie algebra of the group 
of diffeomorphisms of a volume \cite{MaWe}. Perhaps the shortest way is to use 
projection from the Boltzmann Poisson bracket to the hydrodynamic fields, see 
e.g. \cite{Pavelka2016}. 

It is, in fact, often more difficult to derive the dissipation potential. In 
the simplest cases dissipation potentials can be quadratic functions of 
derivatives of entropy. That is the setting of classical irreversible 
thermodynamics \cite{dGM}. The matrix of second derivatives of the dissipation 
potential is then referred to as the dissipative matrix $M^{ij}$, and the 
evolution equations then become
\begin{equation}
\frac{\pd x^i}{\pd t} = L^{ij}\EF_{x^j} + M^{ij} \SF_{x^j} \qquad \mbox{for} 
\qquad 
\Xi = \frac{1}{2} 
\SF_{x^i} M^{ij}(\xx) \SF_{x^j}.
\end{equation}
Apart from quadratic dissipation potentials, exponential dissipation potentials 
(like hyperbolic cosine) are used in case of chemical reactions and Boltzmann 
equation \cite{GrmelaOttingerI,Grmela2012-PhysD}, and such dissipation 
potentials have also statistical reasoning \cite{Mielke2014}.

In summary, the GENERIC framework is an alternative approach to non-equilibrium 
thermodynamics, which does not rely on classical balance laws and their 
extensions as usual non-equilibrium thermodynamics. The reversible 
(Hamiltonian) part of the evolution can be constructed by various methods, for 
example by means of differential geometry (Lie groups, Lie algebras, semidirect 
products) or projections. The irreversible part is constructed as gradient 
dynamics satisfying the first and second laws of thermodynamics. Applicability 
of GENERIC is very wide, covering for example particle mechanics, kinetic 
theory, motions of fluids and solids, complex fluids, electromagnetism and 
dynamics of interfaces.

\section{SHTC formulation of continuum mechanics}

In this Section, we present the SHTC formulation of continuum mechanics and 
discuss its properties in both Lagrangian and Eulerian frames. We 
first, present the Lagrangian SHTC equations. Exactly these 
equations are fully compatible with original Godunov's 
structure~\eqref{eqn.intro.Lp}. 
The Lagrangian SHTC equations can be also seen as Hamiltonian dynamics 
generated by a canonical Poisson bracket of the cotangent bundle of two vector 
fields, see Section \ref{sec.LSHTC.GEN}.
We then proceed with 
presenting the Eulerian SHTC 
equations which are the equations of principle interest in this paper. The 
derivation of Poisson brackets for the Eulerian SHTC equations and thus 
demonstrating their fully compatibility with GENERIC in 
Section~\ref{sec.GENERIC} and for which we show that 
they have a structure fully compatible with 
GENERIC in 
Section~\ref{sec.GENERIC}.

\subsection{Lagrangian frame}\label{sec.lagr}

In the series of 
papers~\cite{GodRom1995,GodRom1996,GodRom1996a,GodRom1998,GodRom1998,Rom2001,Romenski2002,GodRom2003}, Godunov and Romenski investigated the question of 
whether time evolution equations of various continuum mechanics models share 
some common elements of a general structure. More specifically, 
in~\cite{GodRom1996a}, they examined the following important  question.
Whether it is possible to conclude something meaningful about the 
structure of a PDE system representing a continuum mechanics model without 
specifying the physical settings but being based merely on \textit{universally 
accepted 
fundamental physical principles} such as invariance principles, conservation 
principles, causality principle, the laws of thermodynamics and well-posedness 
of the IVP. In~\cite{GodRom1996a}, this question was considered in the 
Lagrangian coordinates and the following requirements to PDEs were imposed:
\begin{itemize}
\item PDEs are invariant under any rotations of the space
\item PDEs are compatible with an extra conservation law
\item PDEs are \textit{generated} by only one thermodynamic potential
\item PDEs are nonlinear but have well-posed IVP (i.e. they are first-order 
symmetric hyperbolic)
\item PDEs are conservative and the fluxes are generated by invariant
  differential operators only, such as div, grad and curl.
\end{itemize}
Then, based on the analysis of various continuum mechanics models and 
the group representation theory, a quite 
general system (system (25) in \cite{GodRom1996a}) of Lagrangian conservation 
laws satisfying all the above 
requirements was proposed. Moreover, it has 
appeared that even a smaller system of PDEs is sufficient for many practical 
applications and it was proposed to consider a subsystem of the general one 
which we shall refer to as the \textit{master system} and is discussed in the 
following section.

\subsubsection{Energy formulation}

As will be shown thorough this paper, one of the apparent common things 
between the SHTC and 
GENERIC frameworks is the 
use of the state variables and the generating potentials and their conjugate 
ones with respect to the Legendre transformation, see also a discussion 
in~\cite{PeshGrmRom2015}. In both frameworks the primary state variables are 
the 
density like fields and the primary potential is the total energy conservation. 
The conjugate state variables and potential have the meaning of the fluxes and 
a generalized pressure accordingly. In this section, we present 
the Energy formulation of the SHTC Lagrangian equations, i.e. they are written 
in terms of the primary density fields and the total energy potential. The 
conjugate formulation is discussed in Section~\ref{sec.conj.lagr}.

The master system is formulated with respect to the following state variables
\begin{equation}\label{eqn.lagr.q}
\qq=(\mm,\FF,\hh,\ee,\ww,\sigma)
\end{equation}
and the thermodynamic potential
\begin{equation}\label{eqn.lagr.U}
U = U(\qq),
\end{equation}
where $ \mm=[m_i] $ is a vector field with the meaning of the generalized 
momentum, e.g. see~\cite{DPRZ2017}, $ \FF=[F_{ij}] $ is the deformation 
gradient, $ \ee=[e_i] $ and $ \hh=[h_i] $ are vector fields which can be 
interpreted as electromagnetic fields~\cite{DPRZ2017}, $ \sigma $ is a 
scalar 
field which is transported according to the vector field $ \ww=[w_i] $. The 
later two are used for modeling of the mass and heat transfer where $ \sigma $ 
has 
the meaning of mass concentration or the entropy, accordingly. 

In fact, we do not need to give a precise meaning to the fields 
\eqref{eqn.lagr.q} and the further analysis is independent of specific physical 
interpretations. The meanings of the fields~\eqref{eqn.lagr.q}  depend on the 
specification of the generating potential $ U $ in the SHTC framework. For 
instance, in the example 
Section~\ref{sec.Examples}, we show that the same equations can 
describe the heat and mass transfer and the represented physics is fully 
determined by the way we define the generating potential. 

Allover the paper, we use the summation over the repeated indexes and we denote 
the partial derivative of the potentials with respect to the state variables as 
$ U_{q_i} = \pd U/\pd q_i $, $ U_\qq=[U_{q_i}] $, $ U_{m_{i}} = \pd U/\pd m_{i} 
$, etc.

The master system that fulfills all of the requirements listed in 
Section~\ref{sec.prem.shtc} was proposed 
in~\cite{GodRom1996a}. A more general master system however can be easily 
proposed, 
see~\cite{GodRom1996a}, but the minimal system then was selected after analysis 
of a large number of continuum mechanics models. Such a minimal master 
system reads as
\begin{subequations}\label{eqn.lagr.masterU}
	\begin{align}
	& \frac{{\rm d} m_i}{{\rm d} t} - \frac{\pd  U_{F_{i 
	j}}}{\pd y_j} = 0, \label{eqn.lagr.masterU.Momentum}\\[1mm]
	& \frac{{\rm d} F_{ij}}{{\rm d} t} - \frac{\pd  
	U_{m_{i}}}{\pd y_j} = 0, \label{eqn.lagr.masterU.F}\\[1mm]
	& \frac{{\rm d} h_{i}}{{\rm d} 
	t}+\varepsilon_{ijk}\frac{\pd {U}_{e_{k}} }{\pd 
	y_j}=0\,,\label{eqn.lagr.masterU.Magn}\\[1mm]
	& \frac{{\rm d} e_{i}}{{\rm d} 
	t}-\varepsilon_{ijk}\frac{\pd {U}_{h_{k}} }{\pd 
	y_j}=0,\label{eqn.lagr.masterU.Electr}\\[1mm]
	& \frac{{\rm d} w_j}{{\rm d} t} + \frac{\pd  
	U_\sigma}{\pd y_j} = 0, \label{eqn.lagr.masterU.vector}\\[1mm]
	& \frac{{\rm d} \sigma}{{\rm d} t} + \frac{\pd  
	U_{w_j}}{\pd y_j} = 0. \label{eqn.lagr.masterU.scalar}
	\end{align}
\end{subequations}
In addition, an extra conservation 
law is the consequence of the above PDEs
\begin{equation}\label{eqn.lagr.energy}
\frac{{\rm d} U}{{\rm d}t}-\frac{\pd}{\pd 
y_j}\left(U_{v_i}U_{F_{ij}} + \varepsilon_{ijk}U_{e_i}{U}_{h_k} -U_\sigma 
U_{w_j}
\right)=0.
\end{equation}
It can be obtained as the sum of the governing PDEs multiplied by the 
corresponding fluxes:
\begin{equation}\label{eqn.summation.Lagr}
\eqref{eqn.lagr.energy} \equiv U_{m_i}\cdot\eqref{eqn.lagr.masterU.Momentum} + 
U_{F_{ij}}\cdot\eqref{eqn.lagr.masterU.F} + 
U_{e_i}\cdot\eqref{eqn.lagr.masterU.Electr} +
U_{h_i}\cdot\eqref{eqn.lagr.masterU.Magn} +
U_{\sigma}\cdot\eqref{eqn.lagr.masterU.scalar} + 
U_{w_j}\cdot\eqref{eqn.lagr.masterU.vector}.
\end{equation}
Usually, in such a formulation, the potential $ U $ plays the role of the 
\textit{total 
energy density} (per unit mass), i.e. it relates to the total specific energy $ 
\scU $ (per unit volume) as $ U = \rho_0 \scU $, where $ \rho_0 $ is the 
reference 
mass density.

The first pair of equations usually represents the equations of motion, while 
the second pair, \eqref{eqn.lagr.masterU.Electr}, 
\eqref{eqn.lagr.masterU.Magn}, may represent a non-linear generalization of the 
Maxwell equations. However, 
as we already 
mentioned, they may have a drastically 
different interpretations. For example, in~\cite{PeshGrmRom2015}, the equations 
with 
the 
structure of \eqref{eqn.lagr.masterU.Electr}, \eqref{eqn.lagr.masterU.Magn} 
were used to describe the dynamics of flow defects, while in 
Section~\ref{sec.euler.symmhyp}, the Eulerian counterpart of 
\eqref{eqn.lagr.masterU.Magn} is used to describe a weakly non-local dynamics.

\subsubsection{Conjugate formulation and symmetric form}\label{sec.conj.lagr}

System~\eqref{eqn.lagr.masterU} is a system of conservation laws and it is 
compatible with the accompanying conservation law~\eqref{eqn.lagr.energy} and 
hence, according to the Godunov-Friedrichs-Lax theorem discussed in 
Section~\ref{sec.preliminaries}, it can be written in Godunov 
form~\eqref{eqn.intro.Lp} and hence it is symmetric hyperbolic. In order to see 
this explicitly, it is necessary to introduce fluxes 
of~\eqref{eqn.lagr.masterU} as new state variables 
$ \pp=[p_i] $ and a potential $ M(\pp) $ (it has the meaning close to the 
pressure).

The new variables and the potential $ (\pp,M) $ are \textit{thermodynamically 
conjugate} 
to the pair 
$ (\qq,U) $ in the following sense
\begin{equation}\label{eqn.lagr.p}
\pp=U_{\qq},\qquad M(\pp) = q_i U_{q_i} - U = q_ip_i - U,
\end{equation}
i.e. $ M(\pp) $ is the Legendre transformation of $ U(\qq) $. Of course, the 
vice versa is also true
\begin{equation}\label{eqn.lagr.qp}
\qq=M_{\pp},\qquad U(\qq) = p_i M_{p_i} - M = q_ip_i - M.
\end{equation}
In order the relation \eqref{eqn.lagr.p} and \eqref{eqn.lagr.qp} be a 
one-to-one relation, it is necessary that $ U(\qq) $ would be a convex 
function 
($ M(\pp) $ would be then also convex because of the properties of the 
Legendre 
transformation), i.e.
\begin{equation}
U_{\qq\qq}= (M_{\pp\pp})^{-1} > 0.
\end{equation}

\subsubsection{Complimentary structure and odd-even 
parity}\label{sec.compl.parity}

One may clearly note a \textit{complimentary} structure of the 
equations~\eqref{eqn.lagr.masterU}, i.e. when a field under the time 
derivative, say $ m_i $ in~\eqref{eqn.lagr.masterU.Momentum}, then appears in 
the flux of the complimentary equation~\eqref{eqn.lagr.masterU.F}
in the partial derivative $ U_{m_i} $ of the generating potential $ U 
$
Thus 
equations~\eqref{eqn.lagr.masterU} are split into complimentary pairs for $ 
(m_i,F_{ij}) $, $ (h_i,e_i) $ and $ (w_i,\sigma) $.
Therefore, when one constructs a specific model in the SHTC framework, it is 
impossible to take only one PDE in a pair because its flux will be 
undetermined.
However, an arbitrary number of ordinary differential equations
\begin{equation}\label{eqn.lagr.ode}
\frac{\pd q_i}{\pd t} = f_i(\qq)
\end{equation}
can be added to system~\eqref{eqn.lagr.masterU}. Here, $ q_i $ might be 
scalars 
or entries of a vector or tensor field, $ f_i(\qq) $ are sufficiently smooth 
functions of the entire vector of state variables. For example, equations 
\eqref{eqn.lagr.masterU.Momentum}, 
\eqref{eqn.lagr.masterU.Momentum} and ordinary differential equations of 
type~\eqref{eqn.lagr.ode} constitute the nonlinear elastoplasticity model 
studied in~\cite{GodPesh2010}.

The complimentary structure of the SHTC  equations can be also seen from the 
point of view of the \textit{odd-even parity} with respect to 
\textit{time-reversal 
transformation} (TRT)~\cite{Ottinger-book,Pavelka2014a} which has been playing 
a central role in thermodynamics for very long time. If a field changes its 
sign under TRT, i.e. when the velocities of all microscopic particles (and 
magnetic field intensity) are 
inverted, the quantity is odd. If the quantity is not altered by the inversion 
of velocities, it is even. In applications we dealt with so far, in each 
complimentary pair $ (m_i,F_{ij}) $, $ (h_,e_i) $ and $ (w_i,\sigma) $ one of 
the fields has even parity while the second has odd parity. For example, if $ 
m_i $ is treated as the momentum density and $ F_{ij} $ is treated as the 
deformation gradient then $ m_i $ is odd while $ F_{ij} $ is even; if $ h_i $ 
is treated as the magnetic field and $ e_i $ is treated as the electric field 
then $ h_i $ is odd while $ e_i $ is even; if $ w_i $ is treated as the 
momentum of heat carriers and $ \sigma $ is treated as the entropy density then 
$ w_i $ is odd while $ \sigma $ is even. The odd-even parity of the state 
variables is also evident from the geometric form of the Lagrangian SHTC 
equations presented in Section~\ref{sec.LSHTC.GEN}.



\subsubsection{Involution constraints}
Solutions to the master system satisfy some \textit{stationary laws} that are 
conditioned by the structure of the fluxes:
\begin{equation}\label{eqn.constr.u0}
\dfrac{\pd F_{ij}}{\pd y_k}-\dfrac{\pd F_{ik}}{\pd y_j} = 
const\,, \qquad \dfrac{\pd h_i}{\pd y_i}= 
const\,,\qquad 
\dfrac{\pd e_i}{\pd y_i} = const\,, 
\qquad \dfrac{\pd w_{j}}{\pd y_k}-\dfrac{\pd w_{k}}{\pd 
y_j} = const\,,
\end{equation}

In fact, these stationary laws impose constraints on the initial data and hold 
for every $ t>0 $ if they were satisfied at $ t=0 $. Indeed, applying the 
divergence operator, for instance, to equation~\eqref{eqn.lagr.masterU.Electr} 
we 
obtain 
\begin{equation}
\frac{\rmd}{\rmd t }\left(\frac{\pd e_i}{\pd y_i}\right) = 0.
\end{equation}
Thus, the stationary laws \eqref{eqn.constr.u0} should be considered as 
\textit{involution 
constraints}. Moreover, as we shall see in Section~\ref{sec.variation} where 
the master 
system is derived from the variational principle, these involution constraints 
play the role of \textit{integrability conditions} to the equations of motion 
if, however, the constants on the right-hand sides of \eqref{eqn.constr.u0} 
equal to zero.

\subsubsection{Variational nature of the SHTC equations}\label{sec.variation}

The master system was proposed in~\cite{GodRom1996a} based merely on the 
requirements listed at the beginning of Section~\ref{sec.lagr}. In this 
section, we show 
that the master system can be also obtained from Hamilton's principle of 
stationary action if the Lagrangian is taken as a partial Legendre 
transformation (see details below) of the total energy potential 
and stationary 
laws~\eqref{eqn.constr.u0} have zeros on the right-hand sides. Partly, 
these results were published in Chapter 23 of the Russian 
edition~\cite{GodRom1998} of the book~\cite{GodRom2003} as well as in the 
recent paper~\cite{DPRZ2017}.

We start by introducing two vector potentials and two scalar potentials:
\begin{equation}
x_i(t,\yy), \qquad a_i(t,\yy),\qquad \varphi(t,\yy), \qquad 
\chi(t,\yy)
\end{equation}
and denote their derivatives as follows
\begin{subequations}\label{eqn.potentials}
\begin{align}
\hat{v}_i=\dfrac{\pd x_i}{\pd t},\ \ \  & \hat{F}_{ij}=\dfrac{\pd x_i}{\pd 
y_j},\label{eqn.PotentialDeriv1}\\
\hat{e}_i=-\dfrac{\pd a_i}{\pd t} - \dfrac{\pd \varphi}{\pd y_i},\ \ \ & 
\hat{h}_i=\varepsilon_{ijk}\dfrac{\pd a_k}{\pd 
y_j},\label{eqn.PotentialDeriv2}\\
\hat{\sigma}=\dfrac{\pd \chi}{\pd t},\ \ \ & \hat{w}_i=\dfrac{\pd \chi}{\pd 
y_i}.\label{eqn.PotentialDeriv3}
\end{align}
\end{subequations}
The independent variables $ t $, $ \yy=[y_i] $ and $ \xx=[x_i] $ can be seen as 
the time, the Lagrangian and Eulerian
spatial coordinates respectively, $a_i$ and $\varphi$ can be considered as the
conventional electromagnetic potentials, while at this stage it is difficult to 
give a certain meaning to $ \chi $.

Then, we define the action integral
\begin{equation}\label{eqn.Lagrangian}
\mathcal{L} = \int \Lambda \, \rmd \yy \rmd t,
\end{equation}
where $ 
\Lambda=\Lambda(\hat{v}_i,\hat{F}_{ij},\hat{h}_i,\hat{e}_i,\hat{w}_i,\hat{\sigma})
 $ 
is the Lagrangian.

The first variation of $\mathcal{L}$
\begin{equation}
\delta\mathcal{L}=\int\left[\left(\dfrac{\pd \Lambda_{\hat{v}_i}}{\pd t} + 
\dfrac{\pd 
\Lambda_{\hat{F}_{ij}}}{\pd y_j}\right)\delta x_i + 
\left( \dfrac{\pd \Lambda_{\hat{e}_i}}{\pd t} 
+\varepsilon_{ijk}\dfrac{\pd \Lambda_{\hat{h}_k}}{\pd y_j} \right)\delta 
a_i + 
\dfrac{\pd \Lambda_{\hat{e}_j}}{\pd y_j}\delta \varphi + 
\left( \dfrac{\pd \Lambda_{\hat{\sigma}}}{\pd t} +\dfrac{\pd 
\Lambda_{\hat{w}_j}}{\pd y_j} \right)\delta \chi 
\right]\rmd \yy \rmd t
\end{equation}
gives us the Euler-Lagrange equations
\begin{subequations}\label{eqn.EulLagr}
\begin{align}
\dfrac{\pd \Lambda_{\hat{v}_i}}{\pd t} +\dfrac{\pd 
\Lambda_{\hat{F}_{ij}}}{\pd y_j}= 0 &, \label{eqn.EulLagr.moment}\\[2mm]
\dfrac{\pd \Lambda_{\hat{e}_i}}{\pd t} 
+\varepsilon_{ijk}\dfrac{\pd \Lambda_{\hat{h}_k}}{\pd y_j}=0&,  
\ \ \dfrac{\pd \Lambda_{\hat{e}_j}}{\pd y_j}=0,
\label{eqn.EulLagr.e}  \\[2mm]
\dfrac{\pd \Lambda_{\hat{\sigma}}}{\pd t} +\dfrac{\pd 
\Lambda_{\hat{w}_j}}{\pd y_j}= 0 &\label{eqn.EulLagr.c}
\end{align}
\end{subequations}
To this system, the following integrability conditions should be added (they 
are trivial consequences of the definitions \eqref{eqn.potentials})
\begin{subequations}\label{eqn.constr}
\begin{eqnarray} 
	\frac{\pd \hat{F}_{ij}}{\pd t} - \frac{\pd \hat{v}_i}{\pd y_j} & =0\,, & 
	\dfrac{\pd \hat{F}_{ij}}{\pd y_k}-\dfrac{\pd 
	\hat{F}_{ik}}{\pd y_j}  = 0\,, \label{eqn.constr.F}\\
	\frac{\pd \hat{h}_i}{\pd t} +\varepsilon_{ijk}\dfrac{\pd \hat{e}_k}{\pd 
	y_j}     &       =0\,, & \dfrac{\pd \hat{h}_j}{\pd 
	y_j}                                               =0\,. 	
	\label{eqn.constr.h}\\
	\frac{\pd \hat{w}_j}{\pd t}  - \dfrac{\pd \hat{\sigma}}{\pd 
	y_j}     &       =0\,, & \dfrac{\pd \hat{w}_i}{\pd y_j} - \dfrac{\pd 
	\hat{w}_j}{\pd y_i} = 0\,. 	
	\label{eqn.constr.w}
\end{eqnarray}
\end{subequations}

In order to rewrite equations \eqref{eqn.EulLagr}--\eqref{eqn.constr} 
in the form of system \eqref{eqn.lagr.masterU}, let us introduce a 
potential $U$ as a partial Legendre transformation of the Lagrangian $ \Lambda $
\begin{eqnarray}
{\rm d}U&=&{\rm d}(\hat{v}_i 
\Lambda_{\hat{v}_i}+\hat{e}_i\Lambda_{\hat{e}_i}+\hat{\sigma}\Lambda_{\hat{\sigma}}
 -\Lambda)=
\hat{v}_i{\rm d}\Lambda_{\hat{v}_i}+\hat{e}_i {\rm d}
\Lambda_{\hat{e}_i} +\hat{\sigma}_i {\rm d}
\Lambda_{\hat{\sigma}} -\Lambda_{\hat{F}_{ij}} {\rm d}
\hat{F}_{ij} - \Lambda_{\hat{h}_i} {\rm d} \hat{h}_i - \Lambda_{\hat{w}_i} {\rm 
d} \hat{w}_i=
\nonumber \\
&&
\hat{v}_i{\rm d}\Lambda_{\hat{v}_i}+\hat{e}_i {\rm d} \Lambda_{\hat{e}_i} +  
\Lambda_{\hat{F}_{ij}} {\rm d} (-\hat{F}_{ij})  + \Lambda_{\hat{h}_i} {\rm d} 
(-\hat{h}_i) + \Lambda_{\hat{w}_i} {\rm d} 
(-\hat{w}_i).
\end{eqnarray}
Hence, denoting $m_i=\Lambda_{\hat{v}_i} $, $e_i=\Lambda_{\hat{e}_i}$, 
$\sigma=\Lambda_{\hat{\sigma}}$, $ 
F_{ij}=-\hat{F}_{ij} $, $ h_i=-\hat{h}_i $ and $ w_i=-\hat{w}_i $,
we get the thermodynamic identity 
\begin{equation*}
{\rm d}U=U_{m_i} {\rm d} m_i + U_{F_{ij}} {\rm d} F_{ij}  
+ U_{h_{i}}{\rm d} h_{i} + U_{e_i} {\rm d} e_i + U_{w_{i}}{\rm d} w_{i} + 
U_{\sigma} {\rm d} \sigma.
\end{equation*}
Eventually, in terms of the variables
\begin{equation}\label{eqn.VarConsLagr}
\qq=(m_i,F_{ij},h_i,e_{i},w_i,\sigma)
\end{equation}
and the potential $ U=U(\qq) $, equations \eqref{eqn.EulLagr.moment}, 
\eqref{eqn.EulLagr.e}$_1$, \eqref{eqn.EulLagr.c}, \eqref{eqn.constr.F}$ _1 $, 
\eqref{eqn.constr.h}$ _1 $ 
and \eqref{eqn.constr.w}$ _1 $
transform exactly into the equations~\eqref{eqn.lagr.masterU} which should be 
supplemented by stationary constraints~\eqref{eqn.constr.F}$ _2 $, 
\eqref{eqn.EulLagr.e}$ _2 $ 
\eqref{eqn.constr.h}$ _2 $ and \eqref{eqn.constr.w}$ _2 $ which now read as
\begin{equation}\label{eqn.constr.u}
\dfrac{\pd F_{ij}}{\pd y_k}-\dfrac{\pd F_{ik}}{\pd y_j} = 
0\,,\qquad \dfrac{\pd e_i}{\pd y_i}=0\,, \qquad \dfrac{\pd h_i}{\pd y_i}=0\, 
\qquad \dfrac{\pd w_i}{\pd y_j} - \dfrac{\pd w_j}{\pd y_i} = 0.
\end{equation}


\subsection{Eulerian frame}

We now present the Eulerian SHTC equations which later will be shown to be 
compatible with GENERIC in Section~\ref{sec.GENERIC}. We first 
present the equations which are directly obtained from the Lagrangian master 
system~\eqref{eqn.lagr.masterU} by means of the  Lagrange-to-Euler change of 
the spatial variables which results in the change of the time and 
spatial derivatives
\begin{equation}\label{eqn.L2E.change}
\frac{\rmd }{\rmd t} = \frac{\pd }{\pd t} + v_k\frac{\pd }{\pd x_k}, \qquad 
\frac{\pd}{\pd y_j} = F_{kj}\frac{\pd}{\pd x_k}.
\end{equation}
The resulting Eulerian master system have a more complicated structure of 
PDEs than its Lagrangian counterpart~\eqref{eqn.lagr.masterU}. The main 
difference is that, in contrast to the fully conservative Lagrangian master 
system, the Eulerian 
equations are inherently non-conservative ones and do not admit the original 
Godunov structure~\eqref{eqn.intro.Lp} but generalized it in the following way
\begin{equation}\label{eqn.Lp.euler}
\frac{\pd L_{\pp}}{\pd t} + \frac{\pd (v_k L)_{\pp}}{\pd x_k} + \mathsf{C}_k 
\frac{\pd \pp}{\pd x_k} = 0,
\end{equation}
where one can recognize 
the Godunov structure~\eqref{eqn.intro.Lp} in the first two terms, while $ 
\mathsf{C}_k^\transpose = \mathsf{C}_k $ are some symmetric matrices. 
The reason for the insufficiency of the original conservative Godunov 
structure~\eqref{eqn.intro.Lp} is the fact that the involution constraints in 
the Eulerian frame are not 
just auxiliary equations as in the Lagrangian frame but they are an 
intrinsic part of the structure of the Eulerian SHTC equations and contribute 
into matrices $ \mathsf{C}_k $.  

As in the previous section, we start the discussion with the Eulerian SHTC 
equations in terms of the total energy potential and density-type state 
variables. We then 
discuss the role of the involution constraints and only for pedagogical reasons 
we present a fully conservative reformulation of the SHTC equations in 
Section~\ref{sec.inv.constr}. This reformulation results in the lose of 
Galilean invariance property of the Eulerian equations and it is should be 
avoided to thoughtlessly use this conservative formulation in practice. We 
conclude the section by presenting the conjugate 
formulation and demonstrating how the involution constraints essentially affect 
the symmetrization procedure of the SHTC equations in the Eulerian frame.

\subsubsection{Energy formulation}

Transformation from the Lagrangian SHTC master system 
\eqref{eqn.lagr.masterU} to the Eulerian equations has been done in 
\cite{GodRom1996a} while omitting non-trivial details of the calculation, which 
can be now
found in Appendix \ref{sec.SHTC.LagToEu}. The resulting non-conservative 
Eulerian system of 
equations is
\begin{subequations}\label{eqn.SHTC}
	\begin{align}
	\frac{\pd m_i }{\pd t} & + \frac{\pd }{\pd 
	x_k}\left(m_i v_k + \delta_{ik} \left( \rho \ted_\rho  + \sigma \ted_\sigma 
	+ m_l\ted_{m_l} + e_l 
	\ted_{e_l} + h_{l} \ted_{h_l} - \ted \right) - e_k 
	\ted_{e_i} -  h_{k} \ted_{h_i} + w_i\ted_{w_k} + 
	 A_{jk}\ted_{A_{ji}}\right)=0, \label{eqn.SHTC.momentum}\\[2mm]
	\frac{\pd A_{i k}}{\pd t} & + \frac{\pd (A_{il} 
	v_l)}{\pd x_k} + v_j\left(\frac{\pd A_{ik}}{\pd 
	x_j}-\frac{\pd A_{ij}}{\pd x_k}\right)
	= 0,\label{eqn.SHTC.A}\\[2mm]
	\frac{\pd h_i}{\pd t} & + \frac{\pd \left( h_i 
	v_k - v_i h_k + \varepsilon_{ikl} \ted_{e_l} \right)}{\pd x_k} + 
	v_i\dfrac{\pd h_k}{\pd x_k} = 0, \label{eqn.SHTC.Hfield}\\[2mm]
	\frac{\pd e_i}{\pd t} & + \frac{\pd \left( e_i 
	v_k - v_i e_k - \varepsilon_{ikl} \ted_{h_l} \right)}{\pd x_k} + 
	v_i\dfrac{\pd e_k}{\pd x_k} = 0, 
	\label{eqn.SHTC.Efield}\\[2mm]
	\frac{\pd w_k}{\pd t} & + \frac{\pd \left(v_l w_l 
	+ \ted_{\sigma}\right)}{\pd x_k} + v_j\left(\dfrac{\pd 
	w_k}{\pd x_j}-\dfrac{\pd w_j}{\pd x_k}\right)=0.
	\label{eqn.SHTC.w}\\[2mm]
	\frac{\pd \sigma}{\pd t} & + \frac{\pd 
	\left(\sigma v_k + \ted_{w_k} \right)}{\pd 
	x_k}= 0,	\label{eqn.SHTC.entropy}\\[2mm]
	\frac{\pd \rho}{\pd t} & + \frac{\pd (\rho v_k) }{\pd x_k}=0.
	\label{eqn.SHTC.rho}
	\end{align}
\end{subequations}

We note that it is always implied in the SHTC framework (as well as in 
the GENERIC) that the velocity field $ v_i $ and the total momentum field are 
conjugate fields, i.e. $ v_i = \ted_{m_i} $. Thus, in all the equations of 
system~\eqref{eqn.SHTC}, $ v_i $ can be substituted by $ \ted_{m_i} $ but we 
prefer to keep $ v_i $ explicitly written for historical reasons. Later, in 
Section~\ref{sec.GENERIC} we shall always use $ E_{m_i} $ instead of $ v_i $.

Also, note that we use the same notations for the state variables in system 
\eqref{eqn.lagr.masterU} and \eqref{eqn.SHTC}, however they are different 
fields 
which relate to each other by the formulas
\begin{subequations}
\begin{equation}\label{eqn.m1}
\mm'=w \,\mm, \qquad \FF=\AAA^{-1}, \qquad \rho_0 = w\,\rho, \qquad w = 
\det(\FF),
\end{equation}
\begin{equation}\label{eqn.eh1}
\ee' = w \AAA \ee, \qquad \hh' = w \AAA \hh,
\end{equation}
\begin{equation}\label{eqn.w1}
\sigma' = w \,\sigma, \qquad \ww' = \FF^\transpose \ww,
\end{equation}
while the Lagrangian total energy density $ U $ is related to 
Eulerian 
total energy density $ \ted $ as
\begin{equation}\label{eqn.gray.potential1}
U= w \,\ted.
\end{equation}
\end{subequations}
In these relations, $ \mm' $, $ \FF $, $ \hh' $, $ \ee' $, $ \ww' $ and $ 
\sigma' $  are the Lagrangian fields, i.e. exactly those fields in 
equations~\eqref{eqn.lagr.masterU}, while $ \mm $, $ \AAA $, $ \hh $, $ \ee $, 
$ \ww $ and $ \sigma $ are the fields in the Eulerian 
equations~\eqref{eqn.SHTC}. 
Also, $ \rho_0 $ is the reference (Lagrangian) mass density.


Exactly as in the Lagrangian framework (see summation 
rule~\eqref{eqn.summation.Lagr}), the conservation of the total 
energy density $ \ted $,
\begin{multline}\label{eqn.SHTC.energy}
\frac{\pd \ted}{\pd t}+\frac{\pd}{\pd x_k} \left( v_k \ted 
+ v_i 
\left[ \left ( \rho \ted_\rho  + \sigma\ted_\sigma + m_l\ted_{m_l} + e_l 
\ted_{e_l} + h_{l} 
\ted_{h_l} - \ted\right ) \delta_{ik}  - e_k 
	\ted_{e_i} -  h_{k} \ted_{h_i} + w_i\ted_{w_k} + 
	 A_{jk}\ted_{A_{ji}} \right] 
\phantom{1^1}\right.\\ 
\left.   + \varepsilon_{ijk} \ted_{e_i} \ted_{h_j} + \ted_{\sigma} \ted_{w_k} 
\right)=0, 
\end{multline}
is just a consequence of the governing 
PDEs~(\ref{eqn.SHTC}). Namely, it is the sum of PDEs 
\eqref{eqn.SHTC} multiplied by the corresponding factors, the conjugate 
state variables:
\begin{equation}\label{eqn.summation.Euler}
\eqref{eqn.SHTC.energy} \equiv \ted_{m_i}\cdot\eqref{eqn.SHTC.momentum} + 
\ted_{A_{ik}}\cdot\eqref{eqn.SHTC.A} + 
\ted_{h_i}\cdot\eqref{eqn.SHTC.Hfield} + 
\ted_{e_i}\cdot\eqref{eqn.SHTC.Efield} + 
\ted_{w_k}\cdot\eqref{eqn.SHTC.w} +
\ted_{\sigma}\cdot\eqref{eqn.SHTC.entropy} + 
\ted_{\rho}\cdot\eqref{eqn.SHTC.rho}.
\end{equation}
As discussed in Section~\ref{sec.Dissipation}, this summation rule also plays a 
key 
role in the introduction of the dissipative terms in the SHTC framework. 
The calculation can be found in the Russian edition~\cite{GodRom1998} 
of 
the book~\cite{GodRom2003}. However, for completeness, we shall also 
demonstrate it in this paper after we present the conjugate formulation in 
Section~\ref{sec.conj.euler}.

We note that there is no need to use the mass conservation equation in the 
Lagrangian frame because the mass density is $ \rho = \rho_0 w^{-1} $. 
Moreover, as it has been shown several times, 
e.g.~\cite{GodRom1998,PeshGrmRom2015}, 
the mass conservation \eqref{eqn.SHTC.rho} is the consequence of the time 
evolution~\eqref{eqn.SHTC.A} for the distortion matrix $ \AAA $ and 
hence, can be dropped out. But in fact, in the Eulerian frame, for many 
reasons, it is easier to treat 
$ \rho $ as an independent state variable governing by its own time evolution. 
In this sense, the complimentary pair of Lagrangian 
equations~\eqref{eqn.lagr.masterU.Momentum} and \eqref{eqn.lagr.masterU.F} 
corresponds to three equations \eqref{eqn.SHTC.momentum}, 
\eqref{eqn.SHTC.A} and \eqref{eqn.SHTC.rho} of the Eulerian SHTC 
master system. Also, remark that an arbitrary number of equations 
with the structure of~\eqref{eqn.SHTC.rho} can be added to 
system~\eqref{eqn.SHTC}. In practical cases, the entropy conservation law for 
the ideal fluids or 
time evolution for the volume fraction in multi-phase models have the same 
structure as \eqref{eqn.SHTC.rho}, e.g. see Section~\ref{sec.mass.transfer}. 
Thus, \eqref{eqn.SHTC.rho} symbolizes various equations with such a structure. 

Eventually, we mention a few examples in which equations~\eqref{eqn.SHTC} were 
used, see also more details in Section~\ref{sec.Examples}. Thus, 
equations~\eqref{eqn.SHTC.momentum}, \eqref{eqn.SHTC.A}, 
were used 
in~\cite{God1972,Rom1989,GodRom2003,BartonRom2010,Favrie2011,Barton2013,Boscheri2016} for modeling of elastic and elastoplastic deformations in metals, 
while the same equations constitute the unified formulation of continuum 
mechanics proposed in~\cite{HPR2016,DPRZ2016,HYP2016,DPRZ2017} and can be used 
for modeling of fluid flows. Equations 
\eqref{eqn.SHTC.momentum}--\eqref{eqn.SHTC.Hfield} were used 
in~\cite{Rom1998,DPRZ2017} 
to describe the electrodynamics of moving medium, equations 
\eqref{eqn.SHTC.momentum}, \eqref{eqn.SHTC.w} and \eqref{eqn.SHTC.entropy} can 
be used to model multiphase flows, heat conduction, superfluid helium flows,  
see~\cite{Rom1998,RomToro2007,RomDrikToro2010,Romenski2016}.

\subsubsection{Involution constraints}\label{sec.inv.constr}

As we already mentioned, the involution constraints play an exceptional role in 
the Eulerian SHTC equations. This role will show up in all the subsequent 
sections. 

We start the discussion by demonstrating that the following 
\textit{stationary 
constraints}\footnote{The curl 
operator is applied to matrix $ \AAA $ in the column-wise manner.}
\begin{equation}\label{eqn.constr.vec}
\nabla\times\AAA = \BB\,,\qquad 
\nabla \cdot \hh = Q\,, \qquad 
\nabla\cdot\ee = R\,, \qquad 
\nabla\times\ww = \bm{\Omega}\ \ ,
\end{equation}
or, in the component-wise notations,
\begin{equation}\label{eqn.constr.comp}
\varepsilon_{ajk}\frac{\pd A_{ik}}{\pd x_j} = B_{ia}\,, \qquad \dfrac{\pd 
h_k}{\pd 
x_k} = Q\,,\qquad \dfrac{\pd 
e_k}{\pd x_k} = R\,,  
\qquad \varepsilon_{ajk}\frac{\pd w_k}{\pd x_j} = \Omega_a \, ,
\end{equation}
are compatible with system~\eqref{eqn.SHTC}, where the quantities $ \BB $, $ Q 
$, $ R $ and $ \bm{\Omega} $ satisfy the following conservation laws
\begin{subequations}\label{eqn.invol.constr.quantities}
	\begin{align}
	\dfrac{\pd B_{ij}}{\pd t} & + \dfrac{\pd (B_{ij}\, v_k - v_j\, 
			B_{ik})}{\pd x_k}  + v_j \frac{\pd B_{ik}}{\pd x_k}
			= 
			0, 
			\label{eqn.invol.constr.B}\\[2mm]
	\dfrac{\pd R}{\pd t} & + \dfrac{\pd (R\,v_k)}{\pd x_k} = 0, 
	\label{eqn.invol.constr.R}\\[2mm]
	\dfrac{\pd Q}{\pd t} & + \dfrac{\pd (Q\,v_k)}{\pd x_k} = 0, 
	\label{eqn.invol.constr.Q}\\[2mm]
	\dfrac{\pd \Omega_j}{\pd t} & + \dfrac{\pd (\Omega_j\, v_k - v_j\, 
		\Omega_k)}{\pd x_k} + v_j \frac{\pd \Omega_k}{\pd x_k} = 
		0. 
		\label{eqn.invol.constr.O}
	\end{align}
\end{subequations}

Indeed, we are going to 
demonstrate that the following involution constraints
\begin{equation}\label{eqn.invol.constr.euler}
\frac{\pd }{\pd t}\left (\nabla\times\AAA - \BB\right )\equiv 0\,,\qquad 
\frac{\pd }{\pd t}\left (\nabla\cdot\ee - R\right ) \equiv 0\,, \qquad 
\frac{\pd }{\pd t}\left (\nabla \cdot \hh - Q\right ) \equiv 0\,, \qquad
\frac{\pd }{\pd t}\left ( \nabla\times\ww - \bm{\Omega}\right ) \equiv 0\,,
\end{equation}
are satisfied on the solutions to~\eqref{eqn.SHTC} if they hold true at the 
initial moment of time. Indeed, $ \nabla\times\AAA $, $ \nabla\cdot\ee $, $ 
\nabla\cdot\hh $ and $ \nabla\times\ww $ satisfy the following evolution 
equations\footnote{For the derivation of \eqref{eqn.invol.constr.rotA} and 
\eqref{eqn.invol.constr.rotw}, see Appendix~\ref{app.evol.rotw}, while 
equations 
\eqref{eqn.invol.constr.dive} and 
\eqref{eqn.invol.constr.divh} can be easily obtained by applying $ \pd/\pd x_i 
$ to the SHTC equations~\eqref{eqn.SHTC.Efield} and \eqref{eqn.SHTC.Hfield} 
accordingly.
}
\begin{subequations}\label{eqn.invol.constr.oper}
	\begin{align}
	\dfrac{\pd (\nabla\times\AAA)_{ij}}{\pd t} &+ \dfrac{\pd( 
	(\nabla\times\AAA)_{ij}\, v_k - 
	v_j\, (\nabla\times\AAA)_{ik})}{\pd x_k}  + v_j \frac{\pd 
	(\nabla\times\AAA)_{ik}}{\pd x_k} = 0, 
	\label{eqn.invol.constr.rotA}\\[2mm]
	\dfrac{\pd (\nabla\cdot\hh)}{\pd t} &+ \dfrac{\pd (v_k\, 
	(\nabla\cdot\hh))}{\pd 
	x_k} = 0, 
	\label{eqn.invol.constr.divh}\\[2mm]
	\dfrac{\pd (\nabla\cdot\ee)}{\pd t} &+ \dfrac{\pd (v_k\, 
	(\nabla\cdot\ee))}{\pd 
	x_k} = 0, 
	\label{eqn.invol.constr.dive}\\[2mm]
	\dfrac{\pd (\nabla\times\ww)_j}{\pd t} &+ \dfrac{\pd ((\nabla\times\ww)_j\, 
	v_k - v_j\, (\nabla\times\ww)_k)}{\pd x_k} + v_j \frac{\pd 
		(\nabla\times\ww)_k}{\pd x_k}= 0.
	\label{eqn.invol.constr.rotw}
	\end{align}
\end{subequations}
By comparing \eqref{eqn.invol.constr.quantities} and 
\eqref{eqn.invol.constr.oper}, one may conclude that 
\eqref{eqn.invol.constr.euler} is satisfied if it was true at $ t = 0 $. 
 We emphasize that even though the non-conservative terms 
in~\eqref{eqn.invol.constr.rotA} and \eqref{eqn.invol.constr.rotw} are zeros 
they can not be omitted because this would violate the Galilean invariance 
property of these equations and also change the characteristic structure.

In some cases, at $ t = 0 $, one may assume that $ \BB = 0$, $ \hh 
= 0$, $ \ee = 0$
and $ \bm{\Omega} = 0$ and hence, \eqref{eqn.invol.constr.quantities} has 
trivial solution, and \eqref{eqn.invol.constr.euler} becomes 
\begin{equation}\label{eqn.invol.constr.euler0}
\nabla\times\AAA \equiv 0\,,\qquad 
\nabla\cdot\hh \equiv 0\,, \qquad 
\nabla\cdot\ee \equiv 0\,, \qquad
\nabla\times\ww\equiv 0\,
\end{equation}
for any $ t > 0 $. However, in some applications we have to set $ \BB \neq 0$, 
$ \hh \neq 0$,
$ \ee \neq 0$
or $ \bm{\Omega} \neq 0$ in the initial conditions and hence, system  
\eqref{eqn.invol.constr.quantities} has non-trivial time-dependent solution. 
Thus, it is 
necessary to note that quantities $ \BB $, $ \hh $, $ \ee $ and $ \bm{\Omega} $
may have a certain physical meaning. For example, $ \nabla\times\AAA $ has the 
meaning of the density of microscopic defects (dislocation density 
tensor)~\cite{GodRom1998,GodRom2003,PeshGrmRom2015} in the elatoplasticity 
theory. 
Thus, if the material has suffered from plastic deformations in the past ($ t < 
0 $) then at $ t = 0 $ we have $ \nabla\times\AAA\neq 0 $. For example, the 
impact of the dislocation density on the dispersive properties of the elastic 
waves was studied in~\cite{Romenski2011} where it was assumed that the material 
has non-zero initial concentration of defects.

As we shall see later, the involution constraints 
\eqref{eqn.invol.constr.euler} and \eqref{eqn.invol.constr.euler0} have strong 
impact on the symmetrization of system~\eqref{eqn.SHTC}. Thus, if constraints~ 
\eqref{eqn.invol.constr.euler0} hold then the symmetrization is much simpler 
than in the case of  \eqref{eqn.invol.constr.euler}. If only
\eqref{eqn.invol.constr.euler} holds, then system \eqref{eqn.SHTC} can be 
symmetrized in an extended sense. Namely, we shall demonstrate that 
system \eqref{eqn.SHTC} extended by adding to it constraints
\eqref{eqn.invol.constr.quantities} can be symmetrized. This is, in fact, a 
very 
interesting feature of the SHTC formulation because one may note that the 
extended system~\eqref{eqn.SHTC}, \eqref{eqn.invol.constr.quantities} is 
\textit{weakly nonlocal} because the quantities $ \nabla\times\AAA $, $ 
\nabla\cdot\ee $, etc, represent a coarser scale then the state variables $ 
\AAA $, $ \ee $, etc. This will be even more pronounced when we shall discuss 
the irreversible dynamics within the SHTC framework in 
Section~\ref{sec.Dissipation} because in the presence of dissipative processes 
the quantities $ \nabla\times\AAA $, $ \nabla\cdot\ee $, $ \nabla\times\ww $ 
are not zeros even if they were zero at $ t=0 $.

\subsubsection{Weakly non-local conservative non-symmetrizable formulation} 
\label{sec.Euler.conservative}

In contrast to the Lagrangian master system~\eqref{eqn.lagr.masterU}, the 
Eulerian system~\eqref{eqn.SHTC} is not a system of \textit{conservation laws} 
due to 
the presence of the non-conservative differential terms,
\begin{equation}\label{key}
v_j(\pd_j A_{ik} - \pd_k A_{ij}), \qquad v_i \pd_k h_k, \qquad v_i\pd_k e_k, 
\qquad v_j(\pd_jw_k - \pd_k w_j),
\end{equation}
 in 
equations~\eqref{eqn.SHTC.A}, \eqref{eqn.SHTC.Hfield}, 
\eqref{eqn.SHTC.Efield} and \eqref{eqn.SHTC.w} accordingly. Therefore, 
there is a 
temptation to rewrite Eulerian SHTC equations in a fully conservative form 
using the results of Section~\ref{sec.inv.constr} and to symmetrize it in a way 
discussed in Section~\ref{sec.prem.shtc}. Unfortunately, such an idea is not 
viable, and the 
pedagogical goal of this section is to demonstrate this. 


For the sake of simplicity, we consider the ideal MHD equations in this 
section. However, all the analysis can be applied to the entire 
system~\eqref{eqn.SHTC} without any changes. Thus, let us consider the MHD 
equations which using the results of 
Section~\ref{sec.inv.constr} can be written as
\begin{subequations}\label{eqn.heat.cons}
	\begin{align}
\pd_t m_i  & + \pd _k \left (m_i v_k + \delta_{ik} \left( \rho \ted_\rho  + 
\sigma \ted_\sigma 	+ m_l\ted_{m_l} + B_l\ted_{B_l} - \ted \right) - 
B_k\ted_{B_i} \right ) = 
0,\label{eqn.heat.cons.m}\\[1mm]
\pd_t B_i  & + \pd_k \left( B_i v_k - v_i B_k \right) = - v_i\, 
Q,\label{eqn.heat.cons.B}\\[1mm]
\pd_t Q    & + \pd_k(Q\,v_k) = 0,\label{eqn.heat.cons.Q}\\[1mm]
\pd_t \rho & +\pd_k (\rho v_k ) = 0,\label{eqn.heat.cons.rho}\\[1mm]
\pd_t \sigma & +\pd_k (\sigma v_k ) = 0,\label{eqn.heat.cons.s}
	\end{align}
\end{subequations}
\begin{equation}\label{eqn.heat.energy}
\pd_t E + \pd_k \left( v_k \ted 
+ v_i 
\left[ \left ( \rho \ted_\rho  + \sigma\ted_\sigma + m_l\ted_{m_l} + 
B_l\ted_{B_l} - \ted\right 
) \delta_{ik} - B_k\ted_{w_i} \right] \right)=0, 
\end{equation}
which look like the system of conservation laws (i.e. all the differential 
terms are written in a conservative manner) with the algebraic source term $ 
-v_i\, Q $ in the PDE for $ B_i $ which can be ignored in 
the symmetrization process as a low order term. Here,  
instead of the field $ 
h_i $ used in system \eqref{eqn.SHTC} we use a convectional notation $ B_i $ 
for the 
magnetic field, $ \rho $ is the mass density, $ \sigma $ is treated as the 
entropy density. As discussed in Section~\ref{sec.inv.constr}, on the solution 
of the system, the involution constraint 
\begin{equation}
\frac{\pd}{\pd t}\left( \nabla\cdot\BB - Q \right) = 0
\end{equation}
is fulfilled. This involution constraint can be reduced to $ \pd_t Q = 0 $  
because $ \nabla\cdot\BB = 0$ which has the meaning of the absence of magnetic  
monopoles. Therefore, one may even suggest to consider a 
simpler model
\begin{subequations}\label{eqn.heat.cons0}
	\begin{align}
\pd_t m_i  & + \pd _k \left (m_i v_k + \delta_{ik} \left( \rho \ted_\rho  + 
\sigma \ted_\sigma 	+ m_l\ted_{m_l} + B_l\ted_{B_l} - \ted \right) - 
B_k\ted_{B_i} \right ) = 
0,\label{eqn.heat.cons0.m}\\[1mm]
\pd_t B_i  & + \pd_k \left( B_i v_k - v_i B_k \right) = 
0,\label{eqn.heat.cons0.B}\\[1mm]
\pd_t \rho & +\pd_k (\rho v_k ) = 0,\label{eqn.heat.cons0.rho}\\[1mm]
\pd_t \sigma & +\pd_k (\sigma v_k ) = 0,\label{eqn.heat.cons0.s}
	\end{align}
\end{subequations}
as it is done in many works, e.g. see discussion in~\cite{Powell1999}. However, 
neither \eqref{eqn.heat.cons} nor 
\eqref{eqn.heat.cons0} is compatible with the energy conservation 
law~\eqref{eqn.heat.energy}. Indeed, equations \eqref{eqn.heat.cons} and 
\eqref{eqn.heat.energy} form an overdetermined system of PDEs.  Hence, the 
energy conservation should be a combination of equations~\eqref{eqn.heat.cons} 
or \eqref{eqn.heat.cons0}. 
However, as it was shown by Godunov in~\cite{God1972}, the energy flux can not 
be obtained as a combination of the conservative fluxes~\eqref{eqn.heat.cons} 
or \eqref{eqn.heat.cons0} 
because the energy flux is exactly (see details in 
Appendix~\ref{sec.energycons})
\begin{equation}\label{eqn.heat.energyflux}
\text{flux}\eqref{eqn.heat.energy} \equiv 
E_{m_i}\cdot\text{flux}\eqref{eqn.heat.cons0.m} +
E_{B_i}\cdot\text{flux}\eqref{eqn.heat.cons0.B} +
E_{\rho}\cdot\text{flux}\eqref{eqn.heat.cons0.rho} + E_{B_i} v_i 
\pd_k B_k.
\end{equation}
Therefore, the energy conservation is \textit{not a combination} of either 
\eqref{eqn.heat.cons} or \eqref{eqn.heat.cons0} 
and hence the symmetrization is 
impossible by means of the GFL theorem discussed in 
Section~\ref{sec.prem.shtc}. Other ways of  
symmetrization of the MHD equations are unknown. 
Eventually, both formulations \eqref{eqn.heat.cons} 
and \eqref{eqn.heat.cons0} are not Galilean invariant, e.g. 
see~\cite{Powell1999}, and should be avoided in practical use.

The MHD formulation which is compatible with the energy 
conservation~\eqref{eqn.heat.energy} is the following \textit{non-conservative} 
system
\begin{subequations}\label{eqn.heat.noncons}
	\begin{align}
\pd_t m_i  & + \pd _k \left (m_i v_k + \delta_{ik} \left( \rho \ted_\rho  + 
\sigma \ted_\sigma 	+ m_l\ted_{m_l} + B_l\ted_{B_l} - \ted \right) - 
B_k\ted_{B_i} \right ) = 
0,\label{eqn.heat.noncons.m}\\[1mm]
\pd_t B_i  & + \pd_k \left( B_i v_k - v_i B_k \right) + v_i \pd_k B_k = 
0,\label{eqn.heat.noncons.B}\\[1mm]
\pd_t \rho & +\pd_k (\rho v_k ) = 0,\label{eqn.heat.noncons.rho}\\[1mm]
\pd_t \sigma & +\pd_k (\sigma v_k ) = 0.\label{eqn.heat.noncons.s}
	\end{align}
\end{subequations}
This formulation is non-conservative due to the presence of 
the non-conservative differential term $ v_i \pd_k B_k $ in the PDE for $ B_i 
$. However, 
exactly this formulation is compatible with the energy 
law~\eqref{eqn.heat.energy}, it is symmetrizable as was shown 
in~\cite{God1972}, it 
has the structure of the 
equation \eqref{eqn.SHTC.Hfield} of the SHTC master 
system~\eqref{eqn.SHTC}, it is Galilean invariant, and 
exactly equation \eqref{eqn.heat.noncons.B} can be obtained from the Lagrangian 
equation for the magnetic field
\begin{equation}
\frac{\rmd B_i}{\rmd t} =0,
\end{equation}
see details in Section~\ref{sec.app.elmag}.

In summary, it is important not only to pay attention to casting the 
equations into the form of conservation laws, but also to the companion
conservation law, which has to be compatible with the rest of the evolution equations. 
Otherwise, the GFL theorem can not be applied.

\subsubsection{Conjugate formulation}\label{sec.conj.euler}

In the previous section, the governing equations were formulated in terms of 
the state variables $ \qq $ and the total energy density $ \ted(\qq) $. In 
this section, we also provide another, \textit{thermodynamically conjugate}, 
formulation in terms 
of the conjugate state variables $ \pp 
$ (flux fields)  and a potential $ L(\pp) $ which is the 
Legendre transform of $ \ted(\qq) $ and has the physical 
meaning of a generalized 
pressure. This formulation allows to emphasize an exceptional role of the 
generating potentials $ L $  and $ \ted $ and shall be used to prove the 
symmetric hyperbolicity of the governing PDEs in 
Sections~\ref{sec.euler.symmhyp} and~\ref{sec.euler.symmhyp0} within the SHTC 
formalism.

As in the Lagrangian framework, we introduce the conjugate variables
\begin{equation}\label{eqn.p}
\pp=(r,v_i,\alpha_{ik},b_i,d_i,\eta_k,\theta)
\end{equation} 
as the partial derivatives of the 
total energy density $ \ted(\qq) $ with respect to the conservative state 
variables
\begin{equation*}
\begin{array}{c}
v_i=\ted_{m_i},\qquad  \alpha_{ik}=\ted_{A_{ik}},\qquad  
b_i=\ted_{h_i},\qquad d_i=\ted_{e_i}, \qquad
\eta_k=\ted_{w_k},\qquad \theta =\ted_{\sigma}, \qquad r=\ted_\rho,
\end{array} 
\end{equation*}
and the new potential 
\begin{equation}
L(\pp)=\qq\cdot\pp -\ted = r\rho + v_im_i + 
\alpha_{ij}A_{ij} + d_i e_i + b_i h_i + \eta_i w_i + \theta\sigma - \ted
\end{equation}
as the Legendre 
transformation of $ \ted $. In terms of the new variables $ \pp $ and potential 
$ L(\pp) $, system~\eqref{eqn.SHTC} can be 
rewritten as
\begin{subequations}\label{eqn.MasterEulerL}
	\begin{align}
	& \frac{\pd L_{v_i}}{\pd t} + \frac{\pd}{\pd x_k}  
	\left[(v_k L)_{v_i}  + 
	\alpha_{mk}L_{\alpha_{mi}} - \delta_{ik}\alpha_{mn}L_{\alpha_{mn}} 
	- d_i L_{d_k} - b_i L_{b_k} + 
	\eta_{k}L_{\eta_{i}} - \delta_{ik}\eta_n L_{\eta_n}\right] = 
	0, 
	\label{eqn.MasterEulerL.Momentum}\\[2mm]
	& \frac{\pd L_{\alpha_{ik}}}{\pd t} + \frac{\pd  [(v_m 
	L)_{\alpha_{im}}]}{\pd x_k} + 
	v_j\left(\frac{\pd L_{\alpha_{ik}}}{\pd 
		x_j}-\frac{\pd L_{\alpha_{ij}}}{\pd x_k}\right)= 
	0, \label{eqn.MasterEulerL.F}\\[2mm]
	& \frac{\pd L_{b_{i}}}{\pd t}+\frac{\pd[ {(v_kL)_{b_i}} - 
	v_iL_{b_k}+\varepsilon_{ikl}d_l]}{\pd x_k}+v_i \dfrac{\pd L_{b_k}}{\pd 
	x_k}=0,\label{eqn.MasterEulerL.Magn}\\[2mm]
	& \frac{\pd L_{d_{i}}}{\pd t}+\frac{\pd[ {(v_kL)_{d_i}} - 
	v_iL_{d_k}-\varepsilon_{ikl}b_l]}{\pd x_k}+v_i \dfrac{\pd L_{d_k}}{\pd 
	x_k} = 
	0, \label{eqn.MasterEulerL.Electr}\\[2mm]
	& \frac{\pd L_{\eta_k}}{\pd t}+\frac{\pd[ 
	(v_i L)_{\eta_i} + \theta]}{\pd x_k} + v_j\left(\dfrac{\pd 
		L_{\eta_k}}{\pd x_j}-\dfrac{\pd L_{\eta_j}}{\pd x_k}\right)= 0,  
	\label{eqn.MasterEulerL.Heat}\\[2mm]
	& \frac{\pd L_{\theta}}{\pd t}+\frac{\pd[ 
	(v_k L)_\theta+\eta_k]}{\pd x_k}=0,
	\label{eqn.MasterEulerL.Entropy}\\[2mm]
	& \frac{\pd L_r}{\pd t} + \frac{\pd  (v_k 
	L)_r}{\pd x_k} = 0, \label{eqn.MasterEulerL.Mass}
	\end{align}
\end{subequations}
while the energy conservation~\eqref{eqn.SHTC.energy} now reads 
(the prove that it is the consequence of the equations 
\eqref{eqn.MasterEulerL} is given in the Appendix~\ref{sec.energycons})
\begin{multline}\label{eqn.energyL}
\frac{\pd }{\pd t}(r L_r + v_i L_{v_i} + \alpha_{ij} L_{\alpha _{ij}} + d_i 
L_{d_i} + b_i L_{b_i} + \theta L_\theta + \eta_k L_{\eta_k} - L) + \\
\frac{\pd }{\pd x_k}\left(v_k\left(r L_r + v_i L_{v_i} + \alpha_{ij} 
L_{\alpha _{ij}} + d_i L_{d_i} + b_i L_{b_i} + \theta L_\theta + \eta_k 
L_{\eta_k} - L\right) + \right.\\
\left. v_i\left[\left(L - \alpha_{ab} L_{\alpha_{ab}} - \eta_a L_{\eta_a} 
\right) \delta_{ik} + \alpha_{nk} L_{\alpha_{ni}} - d_i L_{d_k} - b_i 
L_{b_k} + \eta_k L_{\eta_i}\right] + \varepsilon_{kij} 
d_i b_j +  \theta \eta_k \right)=0.
\end{multline}

It is still not clear though whether \eqref{eqn.MasterEulerL} can be written as a 
symmetric 
hyperbolic 
system because of the non-trivial structure of the fluxes and non-conservative 
differential terms. Apparently, the Godunov-Friedrichs-Lax theorem discussed in 
Section~\ref{sec.prem.shtc} can not be applied because \eqref{eqn.MasterEulerL} 
is not a system of \textit{conservation laws}. Moreover, it can not be treated 
as a conservative system in the extended sense as discussed in 
Section~\ref{sec.Euler.conservative} because  energy 
conservation would be \textit{incompatible} with such conservative 
equations\footnote{See details in Appendix~\ref{sec.energycons}, where we prove 
that if the non-conservative terms are ignored then the energy flux is not a 
consequence of the other fluxes.} and 
 the GFL theorem would still be inapplicable.

The source of the problem is in the following. Unlike the 
Lagrangian frame, in the Eulerian frame, it has 
appeared~\cite{God1972,Godunov1996,Rom1998,Rom2001} that the involution 
constraints \eqref{eqn.invol.constr.euler} play an important role in the 
symmetrization of system \eqref{eqn.MasterEulerL}. If the initial conditions 
were chosen in such a way that 
\eqref{eqn.invol.constr.euler0} holds true, then the symmetrization is much 
easier and is discussed in Section~\ref{sec.euler.symmhyp0} while the general 
case for non curl-free and divergence-free involution constraints 
\eqref{eqn.invol.constr.euler} is discussed in Section~\ref{sec.euler.symmhyp}.

\subsubsection{Symmetric form of conjugate formulation for curl-free 
and divergence-free involution constraints}\label{sec.euler.symmhyp0}

As in the Lagrangian framework, the parametrization of the governing equations 
in terms of the conjugate variables $ \pp $ and generating potential $ L(\pp) $ 
allows 
to rewrite the system in a symmetric quasilinear form. However, this is not 
that straightforward as it is in the Lagrangian frame. In order, to show that 
\eqref{eqn.MasterEulerL} is equivalent to a quasilinear symmetric system it is 
necessary to add the following linear combination of the involution 
constraints~\eqref{eqn.invol.constr.euler0}
\begin{equation}\label{eqn.symm.combin0}
\alpha_{jk}\left(\frac{\pd L_{\alpha_{jk}}}{\pd x_i}-\frac{\pd 
L_{\alpha_{ji}}}{\pd x_k}\right) + d_i \frac{\pd L_{d_k}}{\pd x_k} + b_i 
\frac{\pd L_{b_k}}{\pd x_k} + \eta_k\left(\frac{\pd L_{\eta_{k}}}{\pd 
x_i}-\frac{\pd 
L_{\eta_{i}}}{\pd x_k}\right) \equiv 0
\end{equation}
to equation \eqref{eqn.MasterEulerL.Momentum} and \textit{formally} brake its 
conservative form. After that and some trivial term 
rearrangements in the rest of the equations, \eqref{eqn.MasterEulerL} can be 
rewritten as
\begin{subequations}\label{eqn.MasterEulerL.symmetric}
	\begin{align}
	& \displaystyle\frac{\pd L_{v_i}}{\pd t} + \frac{\pd(v_k 
	L)_{v_i}}{\pd x_k} + L_{\alpha_{ij}}\frac{\pd \alpha_{kj}}{\pd x_k} - 
		L_{\alpha_{jk}}\frac{\pd \alpha_{jk}}{\pd x_i}- L_{d_k}\frac{\pd 
		d_i}{\pd 
		x_k} - L_{b_k}\frac{\pd 
	b_i}{\pd x_k} + L_{\eta_i}\frac{\pd \eta_k}{\pd x_k} - 
	L_{\eta_j}\frac{\pd \eta_j}{\pd x_i}= 0, 
	\label{eqn.MasterEulerL.symmetric.a}\\[2mm]
	& \displaystyle\frac{\pd L_{\alpha_{ij}}}{\pd t} + \frac{\pd  (v_k 
	L)_{\alpha_{ij}}}{\pd x_k} + L_{\alpha_{mj}}\frac{\pd v_m}{\pd x_i} - 
	L_{\alpha_{ij}}\frac{\pd v_k}{\pd x_k} = 0, \\[2mm]
	& \displaystyle\frac{\pd L_{b_{i}}}{\pd t}+\frac{\pd 
	(v_kL)_{b_i}}{\pd x_k} - L_{b_k}\frac{\pd v_i}{\pd x_k} + 
		\varepsilon_{ikj}\frac{\pd d_j}{\pd x_k}= 0,\\[2mm]
	& \displaystyle\frac{\pd L_{d_{i}}}{\pd t}+\frac{\pd 
	(v_kL)_{d_i}}{\pd x_k} - L_{d_k}\frac{\pd v_i}{\pd x_k} - 
	\varepsilon_{ikj}\frac{\pd b_j}{\pd x_k}= 0,\\[2mm]
	& \displaystyle\frac{\pd L_{\eta_i}}{\pd t} + \frac{\pd 
	(v_k L)_{\eta_i}}{\pd x_k} + L_{\eta_j}\frac{\pd v_j}{\pd x_i} - 
	L_{\eta_i}\frac{\pd v_k}{\pd x_k} + \frac{\pd \theta}{\pd x_i}= 0, 
	\label{eqn.MasterEulerL.symmetric.f}\\[2mm]
	& \displaystyle\frac{\pd L_{\theta}}{\pd t}+\frac{\pd 
	(v_k L)_\theta}{\pd x_k} + \frac{\pd \eta_k}{\pd x_k}=0, 
	\label{eqn.MasterEulerL.symmetric.e}\\[2mm]
	& \displaystyle\frac{\pd L_r}{\pd t} + \frac{\pd  (v_k 
	L)_r}{\pd x_k} = 0.
	\end{align}
\end{subequations}
Now, these modified equations can be rewritten as a quasilinear system
\begin{equation}\label{eqn.symm.quasilinear}
\mathsf{A}\frac{\pd \pp}{\pd t} + \mathsf{B}_k \frac{\pd \pp}{\pd x_k} + 
\mathsf{C}_k\frac{\pd \pp}{\pd x_k} = 0 
\end{equation}
which is apparently symmetric because the first two terms in each 
equation in~(\ref{eqn.MasterEulerL.symmetric}) form symmetric matrices $ 
\mathsf{A}(\pp)=L_{\pp\pp} $ and $ \mathsf{B}_k(\pp) = 
(v_k 
L)_{\pp\pp} $ as 
second order derivatives of the potentials $ L $ and 
$ v_k L $ while the rest terms contribute to these matrices only in a symmetric 
way via symmetric matrices $ \mathsf{C}_k(\pp) $. Moreover, if $ L $ is a 
convex 
potential, then system 
\eqref{eqn.MasterEulerL} is \textit{symmetric hyperbolic} because $ \mathsf{A} 
= L_{\pp\pp} > 0 $. Recall that due the 
properties of the Legendre transformation, the convexity of $ L(\pp) $ in $ \pp 
$ is equivalent to convexity of $ \ted(\qq) $ in $ \qq $. 

To conclude this section, we demonstrate the symmetric structure of the matrix 
$ \mathsf{C}_k $. For example, for $ k=1 $ the submatrix of $ \mathsf{C}_1 $ 
corresponding to the variables $ v_i $, $ \eta_i $ and $ \theta $  is 
\begin{equation}
\left(
\begin{array}{ccccccc}
0      &      0      & 0 &      0      & -L_{\eta _2} &   -L_{\eta_3} & 0\\
0      &      0      & 0 & L_{\eta _2} &      0  &  0      & 0 \\
0      &      0      & 0 & L_{\eta _3} &      0    &       0  & 0     \\
	     0       & L_{\eta _2} & L_{\eta _3} & 0 &      0      &      0       
	     &       1       \\
	-L_{\eta _2} &      0      &      0      & 0 &      0      &      0       
	&       0       \\
	-L_{\eta _3} &      0      &      0      & 0 &      0      &      0       
	&       0\\
0      &      0      & 0 &      1      &      0    	     &       0  & 0
\end{array}
\right).
\end{equation}

\subsubsection{Symmetric form of conjugate weakly non-local formulation in 
general case}
\label{sec.euler.symmhyp}

In the previous section, we have demonstrated that conjugate 
form~\eqref{eqn.MasterEulerL} of master equations~\eqref{eqn.SHTC} can be 
written as a symmetric quasilinear system. The key 
step was a modification of the momentum 
equation~\eqref{eqn.MasterEulerL.Momentum} by \textit{formally} breaking its 
conservative 
form via adding to it the linear combination of involution 
constraints~\eqref{eqn.symm.combin0}. In the case of non curl-free and 
non divergence-free involution constraints~\eqref{eqn.invol.constr.euler} such 
a 
step is not valid. However, the symmetrization is still possible but in an 
extended sense when the PDEs~\eqref{eqn.invol.constr.quantities} for 
the new state variables $ \BB $, $ Q $, $ R $ and $ \bm{\Omega} $ are added to 
the 
master system. Because these quantities coincide with $ \nabla\times\AAA $, 
$ \nabla\cdot\hh $, $ \nabla\cdot\ee $ and $ \nabla\times\ww $ accordingly 
at each time instant, the extended system should be treated as a \textit{weakly 
non-local} extension system of the original Eulerian SHTC 
equations~\eqref{eqn.SHTC}. Here, we follow the idea presented 
in~\cite{Romenski2002} and for the sake of simplicity, we demonstrate the 
symmetrization 
algorithm 
for the 
complimentary couple $ (w_i,\sigma) $, or in fact, for their conjugate couple $ 
(\eta_i,\theta) $. As will be shown in 
the example section~\ref{sec.Examples}, these equations constitute an important 
class of equations based on which heat and mass transfer hyperbolic models are 
developed within the both frameworks, the SHTC and GENERIC formalism. All the 
steps of the algorithm then can be applied for 
symmetrizing the PDEs for the couples $ (m_i,A_{ij}) $ and $ (e_i,h_i) $ and 
eventually the entire system~\eqref{eqn.SHTC}.  Thus, we consider the 
following system for the $ 
(w_i,\sigma) $-pair which is coupled with the evolution PDE for the vector 
field $ \Omega_i $ from which we expect to be equal to $ \Omega_i = 
\varepsilon_{ijk}\pd_j w_k $
\begin{subequations}\label{eqn.symm.heat.E}
	\begin{align}
\pd_t m_i & + \pd _k \left (m_i v_k + \delta_{ik} \left( \rho \ted_\rho  + 
\sigma \ted_\sigma 	+ m_l\ted_{m_l} + \Omega_{l} \ted_{\Omega_l} - \ted 
\right) -  \Omega_{k} \ted_{\Omega_i} + w_i\ted_{w_k} \right ) = 0,\\[1mm]
\pd_t w_k & + \pd_k (w_a v_a + \ted_\sigma ) + v_j (\pd_j w_k - \pd_k w_j ) = 
0, \\[1mm]
\pd_t \sigma & +\pd _k(\sigma v_k + \ted_{w_k})=0,\\[1mm]
\pd_t \Omega_i & + \pd_k(\Omega_i v_k - \Omega_k v_i) + v_i \pd_k \Omega_k = 
0,\label{eqn.symm.heat.O}\\[1mm]
\pd_t \rho & +\pd_k (\rho v_k )=0.
	\end{align}
\end{subequations}
Here, it is implied that we have extended the set of state variables by adding 
to it the vector field $ \bm{\Omega} $. However, the addition of a new state 
variable and its evolution PDE can not be done 
arbitrary in the SHTC framework because this might be incompatible with the 
energy conservation law. In order to make it compatible, we have to add the 
extra stress $-\Omega _k E_{\Omega_i} $ in the momentum flux, otherwise the 
summation rule~\eqref{eqn.summation.Euler} does not give the energy 
conservation. Thus, the energy density and the 
conjugate potential
\begin{equation}
\ted = \ted(m_i,w_i,\sigma,\Omega_i,\rho), \qquad L= L 
(v_i,\eta_i,\theta,\omega_i,r)
\end{equation}
now also depend on $ \Omega_i $ and its conjugate $ \omega_i = 
\ted_{\Omega_i} $ accordingly. 
The entire system~\eqref{eqn.symm.heat.E} including new equation for $ \Omega_i 
$ forms a 
sub-system of the master 
system~\eqref{eqn.SHTC}.  For example, one may clearly notice that the new PDE 
for $ 
\Omega_i 
$ has the same structure as the PDE~\eqref{eqn.SHTC.Hfield} for the field $ h_i 
$ which, in turn, also has the corresponding extra stress $ -h_k \ted_{h_i} $. 
The following involution constraints are implied as well 
\begin{equation}\label{eqn.symm.inv}
\frac{\pd}{\pd t}\left(\varepsilon_{ijk}\pd_j w_k - \Omega_i\right) = 0, \qquad
\frac{\pd}{\pd t} \left(\pd_k \Omega_k \right) = 0.
\end{equation}
Therefore, the extended system 
\eqref{eqn.symm.heat} is still formulated within the SHTC framework.

The conjugate formulation of \eqref{eqn.symm.heat.E} is
\begin{subequations}\label{eqn.symm.heat}
	\begin{align}
\pd_t L_{v_i} & + \pd _k \left ((v_k L)_{v_i} + \eta_{k}L_{\eta_i} - 
\delta_{ik}\eta_n 
L_{\eta_n} - \omega _iL_{\omega _k}\right ) = 0,\\[1mm]
\pd_t L_{\eta _k} & +\pd _k (L_{\eta _a}v_a + \theta ) + v_j (\pd_j	L_{\eta_k} 
- \pd_k L_{\eta_j} )= 0,\\[1mm]
\pd_t L_{\theta } & +\pd _k((v_k L)_{\theta } + \eta_k)=0,\\[1mm]
\pd_t L_{\omega _i} & +\pd _j(L_{\omega _i}v_j-L_{\omega 
_j}v_i) + v_i \pd_k L_{\omega_k} = 0, 
	\end{align}
	\begin{align}
\pd_t L_r & +\pd _k \left(v_kL\right)_r=0,
	\end{align}
\end{subequations}
while the involution constraints \eqref{eqn.symm.inv} are
\begin{equation}\label{eqn.symm.heat.constr}
\frac{\pd }{\pd t}\left ( \nabla\times L_{\bm{\eta}} - L_{\bm{\omega}} \right ) 
\equiv 0\,, 
\qquad \frac{\pd}{\pd t} \left(\nabla\cdot L_{\bm{\omega}} \right) \equiv 0.
\end{equation}

The energy conservation law for such an extended formulation now reads as
\begin{multline}\label{eqn.symm.heat.energy}
\frac{\pd }{\pd t}(r L_r + v_i L_{v_i} + \omega_i L_{\omega_i} + \theta 
L_\theta + \eta_k 
L_{\eta_k} - L) + \frac{\pd }{\pd x_k}\left(v_k\left(r L_r + v_i L_{v_i} + 
\omega_i L_{\omega_i} + \theta L_\theta + \eta_k L_{\eta_k} - L\right) + 
\right.\\
\left. v_i\left[\left(L - \eta_a L_{\eta_a} 
\right) \delta_{ik} - \omega_i L_{\omega_k}\right] + \theta \eta_k \right)=0,
\end{multline}
which, we emphasize, has the fluxes fully compatible with the spatial 
differential terms of the governing equations~\eqref{eqn.symm.heat}, i.e. it 
can be obtained by 
the summation 
rule~\eqref{eqn.summation.Euler}.

Now, we are in position to demonstrate that the conjugate 
formulation~\eqref{eqn.symm.heat} is indeed symmetrizable.
At the first step of the symmetrization algorithm we open the brackets 
in the PDE for $ L_{\eta_k} $, which, after that, becomes
\begin{equation}\label{eqn.symm.heat.step1}
\pd _tL_{\eta _k} + L_{\eta _a}\pd _kv_a+v_j\pd _jL_{\eta 
_k}+\pd _k\theta =0,
\end{equation}

At the second step, we modify this PDE by adding to it $ 
L_{\eta_k}\pd_j v_j - L_{\eta_k} \pd_j v_j \equiv 0 $. We have
\begin{equation}
\pd _tL_{\eta _i}+\pd _k\left(v_kL\right)_{\eta _i}-L_{\eta _i}\pd 
_kv_k+L_{\eta _a}\pd _iv_a+\pd _i\theta =0.
\end{equation}

At the last step of the algorithm, we formally ``destroy'' the conservative 
form of the momentum equation by adding to it the linear combination of the 
constraints~\eqref{eqn.symm.heat.constr}
\begin{equation}
0\equiv \eta_k (\pd_i L_{\eta_k} - \pd_k L_{\eta_i}) - \varepsilon_{ijk} 
\eta_j L_{\omega_k} + \omega_i\pd_k L_{\omega_k}.
\end{equation}
Eventually, we have
\begin{subequations}\label{eqn.symm.heat.step2}
	\begin{align}
\pd _tL_{v_i} & + \pd _k\left(v_kL\right)_{v_i}+L_{\eta _i}\pd _k\eta 
_k-L_{\eta 
_a}\pd _i\eta _a-L_{\omega _k}\pd _k\omega
_i=\varepsilon _{i j k}\eta _jL_{\omega _k},\label{eqn.symm.heat.step2.v}\\[1mm]
\pd _tL_{\eta_i} & + \pd _k\left(v_kL\right)_{\eta _i}-L_{\eta _i}\pd 
_kv_k+L_{\eta _a}\pd _iv_a+\pd _i\theta =0,\\[1mm]
\pd _tL_{\theta } & + \pd _k\left(v_kL\right)_{\theta }+\pd_k\eta_k=0,\\[1mm]
\pd _tL_{\omega _i} & + \pd_k\left(v_k L\right)_{\omega _i} - L_{\omega_k}\pd_k 
v_i=0,\\[1mm]
\pd _tL_r & + \pd _k \left(v_kL\right)_r=0,
	\end{align}
\end{subequations}
which can be written as a symmetric quasilinear 
system~\eqref{eqn.symm.quasilinear} but with an algebraic source term $ 
\mathsf{S} = (\varepsilon _{i j k}\eta _jL_{\omega _k},0,0,0,0)$:
\begin{equation*}
\mathsf{A}\frac{\pd \pp}{\pd t} + \mathsf{B}_k \frac{\pd \pp}{\pd x_k} + 
\mathsf{C}_k\frac{\pd \pp}{\pd x_k} = \mathsf{S}.
\end{equation*}
Indeed, because the non-conservative differential terms have the 
same structure as in~\eqref{eqn.MasterEulerL.symmetric}, they
contribute in a symmetric way into the matrices $ \mathsf{C}_k $. The key idea 
of such a symmetrization is that the source 
term $ \varepsilon _{i j k}\eta _jL_{\omega _k} $ in the momentum 
equation~\eqref{eqn.symm.heat.step2.v} is an algebraic 
term due to the fact that we 
treat $ \nabla\times\ww $ as the new independent state variable $ 
\bm{\Omega} = L_{\bm{\omega}} $. This 
source term is therefore treated as a low order term and does not affect the 
type of the PDEs (i.e. it does not contribute into the structure of the 
coefficient matrices $ \mathsf{A} $, $ \mathsf{B}_k $ and $ \mathsf{C}_k $).

Let us demonstrate the symmetric structure of the matrix 
$ \mathsf{C}_k $. For example, for $ k=1 $ the submatrix of $ \mathsf{C}_1 $ 
corresponding to the variables $ v_i $, $ \eta_i $, $ \theta $ and $ \omega_i $ 
is 
\begin{equation}
\left(
\begin{array}{cccccccccc}
0 & 0 & 0 & 0 & -L_{\eta _2} & -L_{\eta _3} & -L_{\omega _1} & 0 & 0 & 0 \\
0 & 0 & 0 & L_{\eta _2} & 0 & 0 & 0 & -L_{\omega _1} & 0 & 0\\
0 & 0 & 0 & L_{\eta _3} & 0 & 0 & 0 & 0 & -L_{\omega _1} & 0\\
 0 & L_{\eta _2} & L_{\eta _3} & 0 & 0 & 0 & 0 & 0 & 0 & 1 \\
 -L_{\eta _2} & 0 & 0 & 0 & 0 & 0 & 0 & 0 & 0 & 0 \\
 -L_{\eta _3} & 0 & 0 & 0 & 0 & 0 & 0 & 0 & 0 & 0 \\
 -L_{\omega _1} & 0 & 0 & 0 & 0 & 0 & 0 & 0 & 0 & 0 \\
 0 & -L_{\omega _1} & 0 & 0 & 0 & 0 & 0 & 0 & 0 & 0 \\
 0 & 0 & -L_{\omega _1} & 0 & 0 & 0 & 0 & 0 & 0 & 0 \\
0 & 0 & 0 & 1 & 0 & 0 & 0 & 0 & 0 & 0
\end{array}
\right).
\end{equation}

The equations for the pairs $ (\mm,\AAA) $ and $ (\hh,\ee) $ are symmetrizable 
exactly in the same way. For example, in order to symmetrize the equations for 
the $ (\mm,\AAA) $-pair, one has to add the time 
evolution~\eqref{eqn.invol.constr.B} for $ \BB=\nabla\times\AAA $ to the 
system, extend the set of state variables by considering the total energy as $ 
E = 
E(\mm,\AAA,\BB,\rho) $ and repeat the above steps.

In summary, extending the idea from~\cite{Romenski2002}, we have demonstrated 
that in case the constraints $ \nabla\times\AAA = 0 $, $ \nabla\cdot\hh = 0 $, 
$ \nabla\cdot\ee = 0 $ and $ \nabla\times\ww = 0 $ are not met, the Eulerian 
SHTC equations~\eqref{eqn.SHTC} can be symmetrized only after a weakly 
non-local 
extension when the quantities $ \nabla\times\AAA $, $ \nabla\cdot\hh $, $ 
\nabla\cdot\ee $ and $ \nabla\times\ww $ starts to play the role of the state 
variables governed by their own time 
evolutions~\eqref{eqn.invol.constr.quantities}.

\section{GENERIC form of SHTC equations}\label{sec.GENERIC}
After having recalled the SHTC equations, let us turn to the question whether 
equations \eqref{eqn.SHTC} can be seen as a Hamiltonian evolution. The answer 
to that question is affirmative, and the equation are thus fully compatible 
with GENERIC framework. Moreover, focusing on the Poisson bracket provides an 
alternative way of derivation and generalization of the SHTC equations.

We, first, demonstrate in sub-section~\ref{sec.LSHTC.GEN} that the Lagrangian 
SHTC equations can be generated by \textit{canonical} Poisson brackets, while 
the rest of the section is dedicated to the central and more challenging 
question of whether the Eulerian SHTC system \eqref{eqn.SHTC} can be seen as a 
Hamiltonian evolution.

\subsection{Lagrangian SHTC system as Hamiltonian 
evolution}\label{sec.LSHTC.GEN}
Consider the cotangent bundle of two vector fields $\qq(\rr)$ and $\pp(\rr)$, 
where the Poisson bracket is canonical, i.e.
\begin{equation}\label{eq.PB.can.VF}
\{\AF,\BF\}=\int (\AF_\qq \cdot \BF_{\pp}-\BF_\qq \cdot \AF_{\pp}) d\rr.
\end{equation}

Let us first construct a Poisson bracket for the pair 
\eqref{eqn.lagr.masterU.Momentum}, \eqref{eqn.lagr.masterU.F}. Let us denote 
$\qq$ as $\mm$ and assume that the functionals depend only on gradients of 
$\pp$, namely on $F_{ij} = -\pd_i p_j$. Poisson bracket \eqref{eq.PB.can.VF} 
then becomes
\begin{equation}\label{eq.PB.VF.1}
\{\AF,\BF\} = \int (\AF_{m_i} \pd_j \BF_{F_{ij}}-\BF_{m_i} \pd_j \AF_{F_{ij}}) 
d\rr,
\end{equation}
which generates the first couple of evolution equations \eqref{eqn.lagr.masterU}. 

Let us now denote $\qq$ as $\ee$ and assume that the functionals depend only on 
curl of $\pp$, namely on $h_i= -\eps_{ijk}\pd_j p_k$. Poisson bracket 
\eqref{eq.PB.can.VF} then becomes
\begin{equation}\label{eq.PB.VF.2}
\{\AF,\BF\} = \int (\AF_{e_i} \eps_{ijk}\pd_j \BF_{h_k}-\BF_{e_i} 
\eps_{ijk}\pd_j \AF_{h_k}) d\rr,
\end{equation}
which generates the second couple of evolution equations \eqref{eqn.lagr.masterU}. 

Finally, let us denote $\qq$ as $\ww$ and assume that the functionals depend only on divergence of $\pp$, namely on $s = \pd_j p_j$. Poisson bracket \eqref{eq.PB.can.VF} then becomes
\begin{equation}\label{eq.PB.VF.3}
\{\AF,\BF\} = \int \left (\AF_{w_i} (-\pd_i \BF_s)-\BF_{w_i} (-\pd_i 
\AF_s)\right ) d\rr,
\end{equation}
which generates the last couple of evolution equations \eqref{eqn.lagr.masterU}. 

The Lagrangian STHC evolution equations \eqref{eqn.lagr.masterU} can be thus 
seen as canonical Hamiltonian evolution of a triplet of cotangent bundles, 
Poisson bracket of which is the sum of brackets \eqref{eq.PB.VF.1}, 
\eqref{eq.PB.VF.2} and \eqref{eq.PB.VF.3}. The triplet of operations, namely 
the rotationally invariant differential operators
gradient, curl and divergence, forms a sort of complete set of operations in 
three-dimensional space, see p. 181 of book \cite{Fecko} and \cite{GodRom1996a}.

\subsection{Hydrodynamics of ideal fluids}
Let us start with the Hamiltonian formulation of the hydrodynamics of ideal 
fluids. The 
Poisson bracket for Euler equations of ideal fluids, i.e. for fields 
$(\mm,\rho,s)$, 
is (e.g. see \cite{Clebsch,Arnold1,MaWe,GPhD,Morrison1998,Pavelka2016})
\begin{equation}\label{eq.PB.CH}
\{\AF,\BF\}^{(Euler)} = \int \left[ m_i(\BF_{m_j} \pd_j \AF_{m_i} - 
\AF_{m_j} 
\pd_j \BF_{m_i}) 
+ 
\rho(\BF_{m_i} \pd_i \AF_\rho - \AF_{m_i} \pd_i \BF_\rho) + s(\BF_{m_i} 
\pd_i \AF_s - 
\AF_{m_i} \pd_i \BF_s) \right] d\rr,
\end{equation}
where $ \AF(\mm,\rho,s) $ and $ \BF(\mm,\rho,s) $ are two arbitrary 
\textit{functionals} and 
$ \AF_\rho $, $ \BF_\rho $, $ \AF_{m_i} $, $ \BF_{m_i} $, etc. are the 
functional 
derivatives, i.e. $ \AF_\rho = \frac{\delta \AF}{\delta \rho}$, $ \AF_{m_i} 
= 
\frac{\delta \AF}{\delta m_i}$, etc.

The Poisson bracket expresses kinematics of the state variables, which are the 
fields $\xx=(\mm,\rho,s)$. Taking an arbitrary functional $ \AF(\mm,\rho,s) $, 
evolution of that functional is given by the Poisson bracket 
\begin{equation}
\frac{\pd \AF}{\pd t} = \{\AF,\scH\}^{(Euler)},
\end{equation}
together with the Hamiltonian
\begin{equation}
\scH(\mm,\rho,s) = \int E(\mm,\rho,s)d\rr,
\end{equation}
where the integrand $ E(\mm,\rho,s) $ is the total energy potential.
Poisson bracket and the Hamiltonian thus generate reversible evolution of 
functionals of state variables, and the evolution equations for the fields $ 
(\mm,\rho,s) $  are
\begin{equation}\label{eqn.PDE.bracket}
\frac{\pd m_i}{\pd t} = \{m_i,\scH\}^{(Euler)}, \qquad \frac{\pd \rho}{\pd t} = 
\{\rho,\scH\}^{(Euler)}, 
\qquad \frac{\pd s}{\pd t} = \{s,\scH\}^{(Euler)}.
\end{equation}

Alternatively, evolution of a functional $ \AF(\xx) =\int A(\xx)d \rr$ can be 
given by 
\begin{equation}
\frac{\pd \AF}{\pd t} = \int \frac{\delta \AF}{\delta x^i} \frac{\pd 
x^i}{\pd t} d\rr ,
\end{equation}
and hence, by collecting terms multiplying derivatives of the functional $\AF$ 
in the Poisson bracket, i.e. by rewriting the bracket as
\begin{equation}\label{eqn.bracket.funcPDE}
\{\AF,\scH\}^{(Euler)} = \int \left [  \AF_{m_i}\cdot 
(\dots) + \AF_{\rho} \cdot (\dots) + \AF_s \cdot (\dots)\right] d\rr, 
\end{equation}
one can directly read the evolution equations of the state variables denoted as
``$\dots$'' in \eqref{eqn.bracket.funcPDE}.


Eventually, by or using 
\eqref{eqn.bracket.funcPDE}, one can obtain the Euler equations of 
compressible hydrodynamics
\begin{subequations}\label{eq.evo.CH}
\begin{align}
\frac{\pd m_i}{\pd t} &= -\pd_j(m_i 
\ted_{m_j}) - \rho\pd_i \ted_\rho -m_j \pd_i \ted_{m_j} - s\pd_i 
\ted_s = -\pd_j(m_i 
\ted_{m_j}) - \partial_i p\\[1mm]
\frac{\pd \rho}{\pd t} &= -\pd_i (\rho \ted_{m_i})\\[1mm]
\frac{\pd s}{\pd t} &= -\pd_i \left(s \ted_{m_i}\right),
\end{align}
where the pressure is identified as
\begin{equation}
p = \rho \ted_\rho + s \ted_s  + m_i \ted_{m_i} - \ted.
\end{equation}
This construction of pressure is general 
within the context of Hamiltonian dynamics and STHC formalism.
\end{subequations}
Compressible Euler equations can be thus seen as Hamiltonian evolution.

\subsection{Distortion matrix}

The goal of this section is to formulate the Poisson 
brackets 
generating the time evolutions~\eqref{eqn.SHTC.momentum}, 
\eqref{eqn.SHTC.A} and \eqref{eqn.SHTC.rho} for the momentum, 
distortion and the mass density fields of the SHTC formulation, and hence to 
demonstrate that these SHTC equations are compatible with GENERIC. The 
distortion field is a very important concept, which is used to 
describe elastic and inelastic deformation of continuum (viscous flows, 
elastoplastic deformations) in the SHTC theory.
Before discussing the dynamics of distortion matrix, let us first consider 
dynamics of labels advected by the fluid particles. 

\subsubsection{Dynamics of labels}
When describing dynamics of solids, the usual approach is to formulate the 
evolution equations in the Lagrangian coordinates, since dynamics of the 
material points is then simply related to classical Newtonian mechanics. Since 
the balance equations (of energy, momentum, mass and angular momentum) are 
formulated in the Eulerian coordinates, one needs to use the mapping from the 
Lagrangian coordinates to the Eulerian coordinates. The classical treatment of 
elasticity, e.g. \cite{Gurtin}, is formulated this way. After the 
transformation to the Eulerian frame, 
the standard equations consist of balance laws (conservation of mass, momentum 
and entropy (or energy)) equipped with an equation expressing advection of the 
Lagrangian coordinates of the particles, i.e. \textit{labels} of the particles.

However, when the material undergoes plastic deformations or even 
chemical reactions or phase transitions, where the material points disappear 
and are created, the mapping from the Lagrangian coordinates to the Eulerian 
coordinates loses its physical significance. The reference configuration itself 
loses its physical significance and it plays merely the role of an initial 
condition. 

Because the balance equations still keep their physical meaning even when the 
reference configuration does not, it seems to be advantageous to formulate the 
evolution equations of solids purely within the Eulerian coordinates. GENERIC 
provides a way for such a construction \cite{Miroslav-PLA}, and the result 
turns out to be compatible with the Godunov-Romenski approach 
\cite{GodRom1972,God1978,GodPesh2010,BartonRom2010,Favrie2011,HPR2016,DPRZ2016,HYP2016}.
Let us now develop the GENERIC way. 

Note that dynamics of solids has already been formulated within GENERIC. For example in 
\cite{hutter-plastic} evolution equation for the deformation tensor is generated by a 
Poisson bracket which is constructed by using the requirements of antisymmetry and momentum conservation.

The aim here is to find evolution equations for solids capturing balance of mass, 
momentum and entropy (or energy) as well as dynamics of individual points of 
the continuum (particle labels). Therefore, the state variables will be density $\rho$, momentum 
density $\uu$, entropy density $s$ and volume density of the labels $\bb$. The 
volume density of labels is a vector field that expresses where the material at 
given Eulerian position belongs. 

The reversible evolution of the state variables is required to be Hamiltonian, 
i.e. generated by a Poisson bracket. Poisson bracket governing evolution of 
variables $\rho$, $\mm$ and $s$ is the standard hydrodynamic Poisson bracket 
\eqref{eq.PB.CH}. The bracket expresses that mass density, momentum density 
and entropy density are passively advected by the momentum density. 
Therefore, it is clear how to add an another passively advected field -- the 
volume density of labels $\bb$. The overall bracket becomes
\begin{align}\label{eq.PB.CHL}
\{\AF,\BF\}^{(Euler + Labels)} = \int &\left[ m_i(\pd_j \AF_{m_i} 
\BF_{m_j}-\pd_j 
\BF_{m_i}\AF_{m_j}) + \rho(\pd_i \AF_\rho 
\BF_{m_i}-\pd_i 
\BF_\rho \AF_{m_i})
+
\right.\nonumber\\
& \left. s(\pd_i \AF_s \BF_{m_i}-\pd_i \BF_s \AF_{m_i})
+b_i(\pd_j \AF_{b_i}\BF_{m_j} - \pd_j \BF_{b_i}\AF_{m_j}) \right] d\rr 
\end{align}
for any two functionals $\AF$ and $\BF$ of the state variables. This is the 
Poisson bracket expressing kinematics of hydrodynamic fields and volume density 
of labels.

However, each label should correspond to a concrete particle of the material 
rather than to a volume of the material, and thus we perform transformation from $\bb$ 
to
\begin{equation}
\aaa \stackrel{def}{=} \frac{\bb}{\rho}\ ,
\end{equation}
which is the mass-density of labels. Bracket \eqref{eq.PB.CHL} then transforms 
(by plugging in functionals of type $\AF(\mm,\bb(\rho,\aaa),\rho,s)$) to
\begin{equation}\label{eq.PB.Lin}
\{\AF,\BF\}^{(Lin)} = \{\AF,\BF\}^{(Euler)}+\int \pd_j a_i 
\left(\AF_{m_j}\BF_{a_i}-\BF_{m_j}\AF_{m_i}\right)d\rr.
\end{equation}
This Poisson bracket was first introduced in \cite{Miroslav-PLA}, where it was 
referred to as the Lin Poisson bracket. The evolution equations implied by 
bracket \eqref{eq.PB.Lin} are
\begin{subequations}\label{eq.evo.CHL}
\begin{align}
\label{eq.evo.u.CHL}\frac{\pd m_i}{\pd t} &= 
-\pd_j(m_i \ted_{m_j}) - \rho\pd_i \ted_\rho -m_j 
\pd_i \ted_{m_j} - s\pd_i \ted_s + \ted_{a_j}\pd_i a_j\\[1mm]
\frac{\pd \rho}{\pd t} &= -\pd_i (\rho \ted_{m_i})\\[1mm]
\frac{\pd s}{\pd t} &= -\pd_i \left(s \ted_{m_i}\right)\\[1mm]
\frac{\pd a_i}{\pd t} &= -\ted_{m_j}\pd_j a_i.
\end{align}
When the field $\aaa$ is interpreted as the Lagrangian coordinate of a particle, the evolution equations are the same as the standard equations of continuum mechanics transformed into the Eulerian frame.
\end{subequations}

In order to write down the equations explicitly, we have to choose an energy 
potential $ \ted (\mm,\aaa,\rho,s)$. Besides the 
kinetic energy, which is of course $ \mm^2/(2\rho) $, and internal energy, 
which is a function of $ \rho $ and $ s $, the deformation energy should be 
taken into account. This energy can not depend on the field $\aaa$ itself 
but only on gradient of the field due to the invariance with respect to 
translations of the whole body. Thus, the energy can be taken as
\begin{equation}
\ted = \ted(\mm,\nabla \aaa,\rho,s).
\end{equation}
Derivative of energy with respect to the field of labels $\aaa$ is then
\begin{equation}
E_{a_i} = -\pd_j \frac{\pd E}{\pd \pd_j a_i} = -\pd_j E_{A_{ij}}, 
\qquad A_{ij} 
= \pd_j a_i,
\end{equation}
and the evolution equation for the field of labels becomes 
\begin{equation}
\frac{\pd \aaa}{\pd t} + \vv \cdot \nabla \aaa =0,
\end{equation}
where velocity $\vv$ is identified with $\ted_{\mm}$. The evolution 
equation for momentum density now reads 
\begin{equation}
\frac{\pd m_i}{\pd t} = 
-\pd_j(m_i \ted_{m_j}) - \rho\pd_i \ted_\rho -m_j 
\pd \ted_{m_j} - s\pd_i \ted_s +\pd_j \sigma_{ij},\\[1mm]
\end{equation}
where the stress tensor 
generated by the labels is
\begin{equation}
\sigma_{ij} = E \delta_{ij} -\pd_i a_l \frac{\pd E}{\pd 
\pd_j a_l} = E \delta_{ij} -A_{li} \ted_{A_{lj}}.
\end{equation}
The rest of 
evolution equations 
\eqref{eq.evo.CHL} is as in classical hydrodynamics. We have thus explicitly written down the evolution equations for the hydrodynamic variables with an advected field of particle labels. These equations can be regarded as standard equations for continuum mechanics in the Eulerian frame.

\subsubsection{Differential geometric derivation}\label{sec.labels.geo}
In this section we provide an alternative derivation of Poisson bracket 
\eqref{eq.PB.Lin} by means of differential geometry.
More specifically, we employ the formulas for matched pairs and semidirect 
products derived in \cite{elmag}, in particular formula (43) from that paper, 
\begin{align}\label{eq.PB.matched}
\{\AF,\BF\}^{(\LAlg^* \times T^*V)} &= \{\AF,\BF\}^{(\LAlg^*)} + 
\{\AF,\BF\}^{(T^*V)}\nonumber\\
&+\langle \AF_v, \BF_\xi \rhd v\rangle-\langle \BF_v, \AF_\xi \rhd 
v\rangle\nonumber\\
&+\langle \alpha, \AF_\xi \rhd \BF_\alpha \rangle-\langle \alpha, \BF_\xi \rhd 
\AF_\alpha \rangle.
\end{align}
The bracket expresses kinematics of a Lie algebra dual $\LAlg^*$, $\xi\in 
\LAlg^*$, coupled with a cotangent bundle $T^*V$, $(v, \alpha)\in T^*V$. The 
coupling is just one-sided, the Lie algebra dual acts on the cotangent bundle. 
The $\langle\bullet,\bullet\rangle$ brackets denote a scalar product, here 
standard $L^2$ scalar product, i.e. simple integration. Finally, for example 
action $\BF_\xi \rhd v$ stands for $-\LDer_{\BF_\xi} v$, $\LDer$ being the Lie 
derivative, and the other actions are interpreted analogically. 
This is the meaning of bracket \eqref{eq.PB.matched}.

Let us suppose that the functionals do not depend on $\alpha$, i.e. elements of 
the dual to the vector space $V$. Bracket \eqref{eq.PB.matched} then reduces to 
\begin{equation}\label{eq.PB.matched.V}
\{\AF,\BF\}^{(\LAlg^* \times V)} = \{\AF,\BF\}^{(\LAlg^*)} +\langle \AF_v, 
\BF_\xi 
\rhd 
v\rangle-\langle \BF_v, \AF_\xi \rhd v\rangle.
\end{equation}
The hydrodynamic fields $(\rho,\mm)$ can be seen as elements of a Lie algebra 
dual $\LAlg^*$, see e.g. \cite{MaWe} or \cite{Pavelka2016}. Consider now the 
vector field of labels interpreted as an element of a vector space $V$. Bracket 
\eqref{eq.PB.matched.V} then becomes
\begin{align}\label{eq.PB.labels.geo}
\{\AF,\BF\}^{(\LAlg^* \times V)} = \int \rho\left(\pd_i \AF_\rho 
\BF_{m_i}-\pd_i 
\BF_\rho \AF_{m_i}\right)d\rr &+ \int m_i\left(\pd_j \AF_{m_i} \BF_{m_j}-\pd_j 
\BF_{m_i} 
\AF_{m_j}\right)d\rr \nonumber\\
&+\int  (\BF_{a_i} \AF_{m_j} \pd_j a_i -\AF_{a_i} \BF_{m_j} \pd_j a_i) d\rr,
\end{align}
which is the same Poisson bracket as bracket \eqref{eq.PB.Lin}. We have thus 
interpreted the Lin bracket in terms of differential geometry.

\subsubsection{Dynamics of general distortion matrix}\label{sec.general.A}

Evolution equations \eqref{eq.evo.CHL} for hydrodynamics with labels are required
to be invariant with respect to shift of the field of labels by a constant. To 
fulfill such a requirement, it was necessary to let the energy depend only on 
the spatial gradient of the field of labels. Let us now reformulate the 
dynamics solely in terms 
of the gradient $ \nabla\aaa $, which is referred to as the distortion matrix 
(following 
\cite{God1978,GodRom1998,GodRom2003,HPR2016,DPRZ2016}), 
\begin{equation}\label{eq.def.A}
A_{ij} = \pd_j a_i,
\end{equation}
and expresses local variations of the field of labels. This definition of the 
distortion matrix implies the integrability conditions
\begin{equation}\label{eq.A.gradient}
\pd_k A_{ij} = \pd_j A_{ik} \ \forall\, i, j, k,\qquad \text{ that is}\qquad 
\varepsilon_{ljk}\pd_j A_{ik} = 0\ \forall\, l \quad \iff \quad \nabla \times 
\AAA = 0.
\end{equation}
We first propose a Poisson bracket for the time evolution of the 
compatible distortion, i.e. for which the integrability conditions are 
fulfilled and which is thus the gradient  of the field of labels. Our 
main goal, however, is to provide a Poisson bracket for generally incompatible 
distortion fields, for which $ \nabla\times\AAA\neq 0 $. The incompatible 
distortion field is the most important case because such distortions are used 
to describe irreversible deformations of viscous fluids and elastoplastic 
solids~\cite{GodRom2003,HPR2016,DPRZ2016,HYP2016}, see also examples in 
Section~\ref{sec.Examples}, and non-Newtonian fluids, see Sec. \ref{sec.B}.

Transforming Poisson bracket \eqref{eq.PB.CHL} into variables $(\mm, \AAA, 
\rho,  s)$, which also means that the two functionals depend only on the new 
variables, leads to the Poisson bracket of hydrodynamics with the distortion 
matrix,
\begin{equation}\label{eq.PB.CHA}
\{\AF,\BF\}^{(Euler+A)} = \{\AF,\BF\}^{(Euler)} + \int 
A_{ij}\left(\BF_{m_j}\pd_k 
\AF_{A_{ik}} 
-\AF_{m_j}\pd_k \BF_{A_{ik}}\right)d\rr.
\end{equation}
This bracket is a Poisson bracket, in particular it fulfills Jacobi identity 
(checked with program \cite{kroeger2010}), 
if and only if the integrability conditions \eqref{eq.def.A} hold. However, it 
can be expected that irreversible evolution violates 
these conditions and hence the distortion matrix loses the connection 
\eqref{eq.A.gradient} with the field of labels. Indeed, having $\nabla\times 
\AAA\neq 
0$ also means that 
integral of $\AAA$ over a closed loop does not necessarily give zero, which 
means that there might be a discontinuity in the field of labels -- the 
material 
undergoes irreversible deformation and identity of the particles (the labels) 
are altered. We need to extend the Poisson bracket so that it fulfills
the Jacobi identity also for the incompatible distortion fields.

The extension should be done in such a way that for curl-free distortion 
matrices bracket \eqref{eq.PB.CHA} is recovered, which means that the extra terms 
(to be added to the bracket) should thus be 
multiplied by $\nabla \times \AAA$. Moreover, in order to generate reversible 
evolution, the extra terms should provide 
additional coupling between $\AAA$ (even with respect to time reversal) and momentum $\mm$ (odd) because $\mm$ is the only 
variable 
with opposite parity than $\AAA$, see Section~\ref{sec.compl.parity} and 
\cite{PRE15}. The new distortion 
matrix bracket then reads
\begin{equation}\label{eq.PB.DM}
\{\AF,\BF\}^{(DM)} = \{\AF,\BF\}^{(Euler + A)} + \int \left(\pd_k A_{ij}-\pd_j 
A_{ik}\right)\left(\AF_{A_{ik}} 
\BF_{m_j}-\BF_{A_{ik}}\AF_{m_j}\right) d\rr,
\end{equation}
which fulfills the Jacobi identity unconditionally (checked with program 
\cite{kroeger2010}). Note that the extra terms can not be multiplied by any 
number or function without violating the Jacobi identity, which makes the 
choice of the extension unique. This Poisson bracket leads to evolution equations
\begin{subequations}\label{eq.evo.DM}
\begin{align}
\label{eq.evo.u.DM}\frac{\pd m_i}{\pd t} &= 
-\pd_j(m_i \ted_{m_j}) - \rho\pd_i \ted_\rho -m_j 
\pd_i \ted_{m_j} - s\pd_i \ted_s - A_{ji} \pd_k \ted_{A_{jk}} 
-(\pd_k A_{ji}-\pd_i A_{jk})\ted_{A_{jk}}\\[1mm]
 &= -\pd_i p-\pd_j(m_i \ted_{m_j}) +\pd_j 
 \sigma_{ij}\nonumber\\
\label{eq.evo.A}\frac{\pd A_{ik}}{\pd t} &= -\pd_k (A_{ij} 
\ted_{m_j})+(\pd_k A_{ij}-\pd_j A_{ik})\ted_{m_j},\\[1mm]
\frac{\pd \rho}{\pd t} &= -\pd_i (\rho \ted_{m_i})\label{eq.evo.DM.rho}\\[1mm]
\frac{\pd s}{\pd t} &= -\pd_i \left(s \ted_{m_i}\right).\label{eq.evo.DM.s}
\end{align}
\end{subequations}
where the generalized pressure $ p $ and extra stress tensor $ \sigma_{ij} $ 
were identified as 
(taken $ \scH = \int E(\mm,\AAA,\rho,s) d\rr$)
\begin{equation}
p = \rho \ted_\rho + s\ted_s + m_i \ted_{m_i} + A_{ij}\ted_{A_{ij}} - 
\ted,\qquad 
 \quad \sigma_{ij}= -A_{ki}\ted_{A_{kj}} + A_{kl}\ted_{A_{kl}} \delta_{ij}.
\end{equation}
Equation \eqref{eq.evo.DM} are the same as the Eulerian SHTC 
equation~\eqref{eqn.SHTC.momentum}, \eqref{eqn.SHTC.A} (recall that 
equations \eqref{eq.evo.DM.rho} and \eqref{eq.evo.DM.s} have the 
same structure and thus are represented by one equation~\eqref{eqn.SHTC.rho} in 
system \eqref{eqn.SHTC}). It should be 
borne in mind that the distortion matrix is no longer required to have zero 
curl. Also, evolution equations \eqref{eq.evo.DM} are Galilean invariant, as 
they can be 
rewritten in terms of material derivatives $\pd_t + \vv\cdot\nabla$ and 
gradients of velocity.

Kinematics of the field of labels is the same as kinematics of 
Lagrangian coordinates of the fluid particles in the Eulerian frame. Therefore, 
kinematics of the distortion matrix expressed by Poisson bracket 
\eqref{eq.PB.CHA}, where only matrices with zero curl are allowed, is the same 
as kinematics of inverse of the deformation tensor, which is standard in 
continuum mechanics. The extended Poisson bracket \eqref{eq.PB.DM}, however, 
allows for distortion matrices with non-zero curl, and the resulting kinematics 
thus lies outside of the scope of classical continuum mechanics because the 
Lagrangian coordinates can not be reconstructed uniquely.

In summary, the hydrodynamic Poisson bracket was first enriched with the field 
of labels, kinematics of which is pure advection. That is the setting of 
classical continuum mechanics in the Eulerian frame. Due to the required 
invariance with respect to spatial shifts, only spatial gradients of the field of 
labels (the distortion matrix) were kept among the state variables, and a new 
Poisson bracket followed. The bracket fulfilled Jacobi identity only for 
curl-free distortion matrices, and to satisfy the identity also for distortion 
matrices with non-zero curl, it was necessary to add (in a unique way) extra 
terms to the Poisson bracket. The Poisson bracket expressing kinematics of non 
curl-free distortion matrices has then been found, which generates the same 
evolution equations~\eqref{eqn.SHTC.momentum}, 
\eqref{eqn.SHTC.A} and \eqref{eqn.SHTC.rho} of the SHTC framework.
To our best knowledge, these results are new.

\subsubsection{Dynamics of the left Cauchy-Green (or Finger) tensor}\label{sec.B}


One of the main motivations for developing GENERIC framework was 
modeling of rheological properties of 
complex fluids~\cite{GrmelaOttingerI,GrmelaOttingerII}, 
where it is popular to include a symmetric positive definite tensor into the
set of state variables, e.g. the conformation tensor. Here we choose the 
left Cauchy-Green tensor (or Finger tensor), which has been shown useful when describing
complex fluids ~\cite{Rajagopal-B,Malek-B,Hron-B}. 

The goal of this section is to discuss the 
relation of a tensor $\BBBB$,
which becomes the left Cauchy-Green tensor (or Finger tensor)
in the case of curl-free distortion field, e.g. in the absence of dissipation. 
Indeed, the distortion matrix is 
identical to the inverse deformation gradient if dissipation is 
absent~\cite{GodRom2003}. In particular, we demonstrate that a Poisson bracket 
for the Left Cauchy-Green tensor can be derived from the Poisson bracket 
\eqref{eq.PB.DM}, which generates reversible evolution of the hydrodynamic 
fields and the general distortion matrix.

The left Cauchy-Green tensor is defined as
\begin{equation}\label{eq.B.def}
\BBB_{ij} = A^{-1}_{ik} A^{-1}_{jk}.
\end{equation}
Derivative of 
a functional $\AF(\mm,\BBBB, \rho, s)$ 
with respect to the distortion matrix then becomes
\begin{equation}
\AF_{A_{mn}} = \frac{\delta \AF}{\delta A_{mn}}=\frac{\delta \AF}{\delta 
\BBB_{ij}}\frac{\pd 
\BBB_{ij}}{\pd  A_{mn}}.
\end{equation}
By differentiating definition \eqref{eq.B.def}, we can obtain that 
\begin{equation}
\frac{\pd \BBB_{ij}}{\pd A_{mn}} = -A^{-1}_{im}\BBB_{nj} - 
A^{-1}_{jm}\BBB_{in},
\end{equation}
from which it follows that
\begin{equation}
\AF_{A_{ik}} = -\AF_{\BBB_{mn}}A^{-1}_{mi}\BBB_{kn} - \AF_{\BBB_{mn}}A^{-1}_{ni}\BBB_{mk}.
\end{equation}
Substituting this last relation into equation \eqref{eq.PB.DM} leads to the 
following Poisson 
bracket
\begin{eqnarray}\label{eq.PB.B}
\{\AF,\BF\}^{(Finger)} &=& \{\AF,\BF\}^{(Euler)} + \int \BBB_{ik} \left(\pd_j 
\AF_{\BBB_{ik}} \BF_{m_j}-\pd_j \BF_{\BBB_{ik}} \AF_{m_j}\right) d\rr \nonumber\\
&&+\int \BBB_{km}\left(\left(\AF_{\BBB_{mj}}+\AF_{\BBB_{jm}}\right)\pd_k 
\BF_{m_j}-\left(\BF_{\BBB_{mj}}+\BF_{\BBB_{jm}}\right)\pd_k 
\AF_{m_j}\right) d\rr \nonumber\\
&&+\int \BBB_{ik}\left(\AF_{\BBB_{ik}} \pd_j \BF_{m_j}-\BF_{\BBB_{ik}} \pd_j 
\AF_{m_j}\right) d\rr,
\end{eqnarray}
which generates the following reversible evolution equations:
\begin{subequations}\label{eq.evo.B}
\begin{align}
\frac{\pd m_i}{\pd t} &= 
-\pd_j(m_i \ted_{m_j}) - \rho\pd_i \ted_\rho -m_j 
\pd_i \ted_{m_j} - s\pd_i \ted_s 
-\BBB_{jk}\pd_i \ted_{\BBB_{jk}} 
+\pd_k\left(\BBB_{km}(\ted_{\BBB_{mi}}+\ted_{\BBB_{im}})\right)
+\pd_i \left(\BBB_{jk}\ted_{\BBB_{jk}}\right)\\[1mm]
&=-\pd_j(m_i \ted_{m_j}) - \rho\pd_i \ted_\rho -m_j 
\pd_i \ted_{m_j} - s\pd_i \ted_s 
+ \ted_{\BBB_{jk}}\pd_i \BBB_{jk}
+\pd_k\left(\BBB_{km}(\ted_{\BBB_{mi}}+\ted_{\BBB_{im}})\right)
\\[1mm]
\frac{\pd \BBB_{ij}}{\pd t} &= -\ted_{u_k}\pd_k \BBB_{ij} 
+\BBB_{ki}\pd_k \ted_{u_j} + \BBB_{kj} \pd_k \ted_{u_i},\\[1mm]
\frac{\pd \rho}{\pd t} &= -\pd_i(\rho \ted_{m_i})\\
\frac{\pd s}{\pd t} &= -\pd_i(s \ted_{m_i}).
\end{align}
\end{subequations}
This means that the left Cauchy-Green tensor has zero upper-convected derivative
unless dissipation is introduced.

Note also that kinematics of the tensor $\BBBB$ can be thus seen as a consequence 
of kinematics of the distortion matrix \eqref{eq.PB.DM}, which in turn lies 
outside of the scope of classical continuum solid mechanics due to the 
possibility of 
non-zero curl of the matrix.

\subsection{Electrodynamics of slowly moving medium}


The goal of this section is to demonstrate that SHTC equations 
\eqref{eqn.SHTC.momentum}, \eqref{eqn.SHTC.Hfield}, \eqref{eqn.SHTC.Efield} and 
\eqref{eqn.SHTC.rho} governing the dynamics of the fields $ 
(\mm,\hh,\ee,\rho,s) 
$ has the structure which is compatible with GENERIC, i.e. these equations 
can be generated by a Poisson bracket. However, as usually in GENERIC, while 
constructing the Poisson brackets, we have to be more specific and thus the 
fields $ (\mm,\hh,\ee,\rho,s) $ have to be endowed with certain physical 
meanings.  
For example, the electrodynamics of slowly moving continuous media is 
formulated in~\cite{DPRZ2017} within the SHTC framework. In this paper, the 
fields $ (\ee,\hh) $ were recognized as
\begin{equation}\label{eqn.elmag.eh}
\ee = \epsilon (\EE + \vv\times(\vv\times \EE)), \qquad
\hh = \BB + \vv\times(\vv\times \BB),
\end{equation}
while the conjugate fields are
\begin{equation}\label{eqn.elmag.db}
\ted_\ee = \EE + \vv\times\BB, \qquad \ted_\hh = \BB - \epsilon\, 
\vv\times\EE,
\end{equation}
where $ \epsilon $ is the electric permittivity, $ \EE $ and $ \BB $ are the 
electric and magnetic fields in the 
laboratory frame and the speed of light is assumed to be one.

Thus, if the motion is absent ($ \vv = 0 $) then the standard electric 
displacement field $ \DD $ and 
magnetic field are recovered
\begin{equation}\label{eqn.DB}
\ee =  \epsilon \EE = \DD, \qquad  \hh = \BB .
\end{equation}
It has appeared that a Poisson bracket governing evolution of exactly the 
fields $ (\DD,\BB) $ and fields $ \mm 
$, $ \rho $ and $ s $ has been referred to as the 
\textit{electromagnetohydrodynamic} Poisson 
bracket (EMHD) in \cite{Holm-EMHD} or in \cite{elmag} (see equation (66) there), and it reads
\footnote{
Bracket \eqref{eq.PB.EMHD.modif} can be derived from Eq. (65) in paper 
\cite{elmag} by rewriting action $\rhd$ as minus Lie derivative when 
interpreting fields $\DD$ and $\BB$ as vector fields, i.e. $H_\DD$ and $H_\BB$ 
as covector fields (in notations of \cite{elmag}). By applying constraints 
\eqref{eq.EM.constraints}, bracket 
\eqref{eq.PB.EMHD.modif} becomes bracket (66) from paper \cite{elmag}. 
Constraints \eqref{eq.EM.constraints}, which are two of four Maxwell equations, 
can be seen either as results of gauge invariance of the theory of 
electromagnetism coupled with matter \cite{MaWe} or as conditions under which 
Poisson bracket (66) from paper \cite{elmag} becomes a projection from kinetic 
theory coupled with matter, equation (58) in \cite{elmag}.
}
\begin{align}\label{eq.PB.EMHD.modif}
&\left\{\AF,\BF\right\}^{(EMHD)}=\{\AF,\BF\}^{(Euler)} + 
\{\AF,\BF\}^{(EM)}  \nonumber\\
&+\int D_i(\BF_{m_j}\pd_j \AF_{D_i} -\AF_{m_j}\pd_j 
\BF_{D_i})d\rr
+\int \pd_jD_j (\AF_{m_i} \BF_{D_i} -\BF_{m_i} 
\AF_{D_i}) d\rr
+\int (\AF_{m_i} D_j \pd_j \BF_{D_i} - \BF_{m_i} D_j \pd_j 
\AF_{D_i}) d\rr \nonumber\\
&+\int B_i( \BF_{m_j}\pd_j \AF_{B_i} -\AF_{m_j}\pd_j 
\BF_{B_i}) d\rr 
+\int \pd_jB_j (\AF_{m_i} \BF_{B_i} -\BF_{m_i} 
\AF_{B_i}) d\rr
+\int (\AF_{m_i} B_j \pd_j \BF_{B_i} - \BF_{m_i} B_j\pd_j \AF_{B_i}) d\rr.
\end{align}
Note that the momentum density $ \mm $ is the coupled matter-field momentum 
in~\cite{elmag} and \cite{DPRZ2017}. This Poisson bracket contains also the 
electromagnetic 
Poisson bracket,
\begin{equation}
\{\AF,\BF\}^{(EM)} = \int (\AF_{D_i}\eps_{ijk}\pd_j 
\BF_{B_k}-\BF_{D_i}\eps_{ijk}\pd_j \AF_{B_k}) d\rr,
\end{equation}
which generates the Maxwell equations. 



By taking a Hamiltonian $ \scH = \int E(\mm,\BB,\DD,\rho,s) d\rr$ and 
rewriting  
bracket \eqref{eq.PB.EMHD.modif} in a form similar 
to~\eqref{eqn.bracket.funcPDE}, one directly obtains the following evolution 
equations
\begin{subequations}
\begin{align}\label{eqn.GEN.elemag}
\frac{\pd m_i}{\pd t} &= - \pd_j (m_i  \ted_{m_j}) 
-\rho \pd_i \ted_{\rho} - m_j 
\pd_i \ted_{m_j} - s \pd_i \ted_s -D_j \pd_i \ted_{D_j} - B_j \pd_i 
\ted_{B_j} 
+ \pd_j\left(D_j \ted_{D_i} + B_j 
\ted_{B_i}\right),\\
\frac{\pd B_i}{\pd t} &= -\pd_j (B_i E_{m_j} - E_{m_i} B_j + \eps_{ijk}\pd_j 
\ted_{D_k}) - E_{m_i}\pd_j B_j,\label{eqn.GEN.elemag.B}\\
\frac{\pd D_i}{\pd t} & = -\pd_j (D_i E_{m_j} - E_{m_i} D_j - \eps_{ijk}\pd_j 
\ted_{B_k}) - E_{m_i}\pd_j D_j,\label{eqn.GEN.elemag.D}\\
\frac{\pd \rho}{\pd t} &= -\pd_i \left( \rho \ted_{m_i} \right),\\
\frac{\pd s}{\pd t} &= -\pd_i \left(s \ted_{m_i}\right),
\end{align}
where pressure $p$ is defined as
\begin{equation}
p =  \rho \ted_\rho + s \ted_s + m_i \ted_{m_i} + D_i \ted_{D_i} + B_i
\ted_{B_i} - \ted,
\end{equation}
and $E(\mm,\BB,\DD,\rho,s)$ being the total energy density.
These evolution equations are also equipped with two constraints imposed on the 
Poisson bracket 
\begin{equation}\label{eq.EM.constraints}
 \dive (\DD) = \frac{z e}{\epsilon}\rho, \qquad 
 \dive (\BB) =  0,
\end{equation}
\end{subequations}
where $ z $ is number of elementary charges per particle and $ e $ is the 
elementary charge.

By subtracting and adding $ \pd_i E $  to the momentum equation, it can be 
transformed exactly into the SHTC momentum conservation law 
\eqref{eqn.SHTC.momentum}, while the equations for $ \BB $ and $ \DD $ already 
has exactly the structure of equations \eqref{eqn.SHTC.Hfield} and 
\eqref{eqn.SHTC.Efield} accordingly.
Thus, the entire GENERIC formulation 
and the SHTC formulation for electrodynamics of moving medium are fully 
compatible.

%
%

\subsection{Ballistic transport}


The goal of this section is to present a Poisson bracket which 
generates equations~\eqref{eqn.SHTC.momentum}, \eqref{eqn.SHTC.entropy} and 
\eqref{eqn.SHTC.w}, and thus to prove their compatibility with GENERIC. In 
applications, these equations can be used for modeling of heat and mass 
transfer if equipped with appropriate dissipative algebraic source terms (see 
Section~\ref{sec.Dissipation} and \ref{sec.Examples} for details). However, in 
this section we shall use the terminology for heat conduction and we shall call 
the scalar field $ s $ the entropy density while the vector field $ \ww $ as 
the 
conjugate entropy flux. Absence 
of the dissipative terms, i.e. absence of the interaction between the mass or 
heat carriers and the medium, corresponds to the most non-equilibrium state, 
\textit{ballistic transport}, which explains the title of this Section.

\subsubsection{The SHTC-compatible bracket}

Consider first a rigid heat-conducting body. We wish to prescribe a field of 
entropy density $s$ expressing local thermodynamic state of the body. Let us 
now address dynamics of the field.
Taking a Lagrangian $\Lambda(s, \dot{s})$, the Legendre transformation 
introduces the 
corresponding Hamiltonian, 
\begin{equation}\label{eqn.Lagrangian.w}
\frac{\delta}{\delta \dot{s}}\left(-\Lambda(s,\dot{s}) + \int \psi(\rr) 
\dot{s}(\rr)\right) d\rr = 0 \quad \Leftrightarrow \quad \Lambda_{\dot{s}} = 
\psi,
\end{equation}
and the conjugate scalar field $\psi(\rr,t)$. The couple $(\psi,s)$ is equipped 
with the canonical Poisson bracket
\begin{equation}
\{\AF,\BF\}^{(\psi,s)} = \int (  \AF_\psi \BF_s- \BF_\psi \AF_s) d\rr.
\end{equation}
However, the aim is to work with a vector (covariant) field that plays the role 
of conjugate entropy flux. Therefore, taking functionals dependent only on $\ww 
= -\nabla \psi$, the canonical bracket becomes
\begin{equation}\label{eq.PB.sw}
\{\AF,\BF\}^{(\ww,s)} = \int (\pd_k \AF_s \BF_{w_k} - \pd_k \BF_s \AF_{w_k}) 
d\rr,
\end{equation}
evolution equations generated by bracket \eqref{eq.PB.sw} are
\begin{subequations}\label{eq.Cat.sw.evo}
\begin{align}
\label{eq.Cat.s.can}\frac{\pd s}{\pd t} &= -\pd_k E_{w_k},\\[1mm]
\label{eq.Cat.s.can.w}\frac{\pd w_k}{\pd t} &= -\pd_k E_s,
\end{align}
equipped with the condition that $\ww$ is a potential vector field, or that 
\begin{equation}\label{eq.w.deg}
\pd_i w_j = \pd_j w_i.
\end{equation}
\end{subequations}
From \eqref{eq.Cat.s.can} it is clear why $\ww$ is referred to as the 
conjugate entropy flux. Equations \eqref{eq.Cat.sw.evo} express
evolution of entropy within rigid heat conductors.

How can one couple the dynamics of heat with hydrodynamics, i.e. with transport of matter? As in 
section \ref{sec.labels.geo}, the hydrodynamic bracket for $(\mm,\rho)$ can be 
extended to describe also action of hydrodynamics on elements of a vector space 
$V$, $\psi \in V$, and the resulting Poisson bracket would be analogical to 
bracket \eqref{eq.PB.labels.geo} with $a_i$ replaced by $\psi$. Here, however, 
we will consider coupling between the hydrodynamic Lie algebra and a 
cotangent bundle $(\psi, s)\in T^*V$, kinematics of which is expressed by 
bracket \eqref{eq.PB.matched},
\begin{equation}
\{\AF,\BF\}^{(\mm,\rho) \times (\psi,s)} = \{\AF,\BF\}^{(Euler)} + \int 
(\AF_\psi \BF_s - \AF_s \BF_\psi ) d\rr \nonumber\\
+\int (\BF_{\psi} \AF_{m_j} \pd_j \psi - \AF_{\psi} \BF_{m_j} \pd_j \psi) d\rr.
\end{equation}
Taking functionals dependent only on gradients of $\psi$, denoted by $w_i = 
-\pd_i \psi$, this last bracket becomes
\begin{equation}\label{eq.PB.CatHyd.geo}
\{\AF,\BF\}^{(\mm,\rho) \times (\ww,s)} = \{\AF,\BF\}^{(Euler)} + \int 
(\BF_{w_i} \pd_i 
\AF_s - \AF_{w_i} \pd_i \BF_s) d\rr + \int w_j\left(\pd_i \AF_{w_i} \BF_{m_j} - 
\pd_i \BF_{w_i} 
\AF_{m_j}\right) d\rr.
\end{equation}
This Poisson bracket expresses kinematics of variables $(\mm,\rho,\ww,s)$ 
provided $\ww$ is a potential vector field, i.e. $\nabla\times\ww = 0$. Indeed, 
Jacobi identity is fulfilled only when this involution constraint is met (as in 
the case of the transition from kinematics of labels to kinematics of 
distortion matrix). 

Validity of Jacobi identity for bracket \eqref{eq.PB.CatHyd.geo} even for 
fields $\ww$ with non-zero curl can be achieved by extending the Poisson 
bracket as in the case of distortion matrix in Section~\ref{sec.general.A}. A 
new term has to be added to the 
bracket proportional to $\pd_i w_j -\pd_j w_i$ (the term can not be multiplied 
by any constant). Moreover, the term has to 
provide coupling between $\ww$, which is an odd variable, and an another odd 
field, for example with the field $\mm$. The only way to extend the bracket 
\eqref{eq.PB.CatHyd.geo} is then as follows (the Jacobi identity for this 
bracket 
has been checked by program 
\cite{kroeger2010})
\begin{equation}\label{eq.PB.CatHyd}
\{\AF,\BF\}^{(Heat)} = \{\AF,\BF\}^{(\mm,\rho) \times 
(\ww,s)} + \int \left(\pd_i w_j -\pd_j 
w_i\right)\left(\AF_{w_i}\BF_{m_j}-\BF_{w_i}\AF_{m_j}\right) d\rr,
\end{equation}
which\footnote{This bracket was first proposed in \cite{Ottinger1998}.} after rearranging the terms as in~\eqref{eqn.bracket.funcPDE}, leads to 
the time evolutions equations
\begin{subequations}\label{eq.evo.CatHyd}
\begin{align}
\frac{\pd m_i}{\pd t} &= -\pd_k(m_i \ted_{m_k}) - \rho\pd_i 
\ted_\rho - s\pd_i \ted_s - m_k \pd_i \ted_{m_k} - w_i \pd_k \ted_{w_k} - 
\ted_{w_k}(\pd_a w_i 
- \pd_i w_k),\\[1mm]
\frac{\pd w_k}{\pd t} &= -\pd_k(w_j  \ted_{m_j} + \ted_s) - (\pd_j w_k - \pd_k 
w_j) \ted_{m_j},\label{eq.evo.CatHyd.w}\\[1mm]
\frac{\pd s}{\pd t} &= -\pd_k \left(s \ted_{m_k} + \ted_{w_k}\right),\\[1mm]
\frac{\pd \rho}{\pd t} &= -\pd_k (\rho \ted_{m_k})
\end{align}
\end{subequations}
which coincide with the Eulerian SHTC equations~\eqref{eqn.SHTC.momentum}, 
\eqref{eqn.SHTC.w}, \eqref{eqn.SHTC.entropy} and \eqref{eqn.SHTC.rho}.

In summary, Poisson bracket \eqref{eq.PB.CatHyd.geo} has been derived as a 
result of coupling between the Lie algebra dual of classical hydrodynamics and 
the cotangent bundle of dynamics of entropy. The Jacobi identity is valid only 
for potential vector fields $\ww$. Validity of Jacobi identity can be extended 
to fields $\ww$ with non-zero curl by extending the bracket by a term coupling 
fields $\ww$ and $\mm$, and such an extension can be carried out in only one 
way. The final extended Poisson bracket is bracket \eqref{eq.PB.CatHyd}, which 
generates the SHTC evolution equations \eqref{eq.evo.CatHyd} for ballistic 
transport of the scalar 
field $ s $.

\subsubsection{Alternative bracket, Cattaneo hydrodynamics}
A Poisson bracket expressing the Cattaneo-like dynamics of heat was proposed in 
\cite{Grmela2011a}, 
\begin{equation}\label{eq.PB.Cat.MG}
\{\AF,\BF\}^{(Cat)} = \int \pi_i \left(\pd_j \AF_{\pi_i}\BF_{\pi_j} - 
\pd_j \BF_{\pi_i}\AF_{\pi_j}\right) d\rr  + \int s \left(\pd_i \AF_s 
\BF_{\pi_i} - \pd_i \AF_s \BF_{\pi_i}\right)d\rr ,
\end{equation}
where $\ppi$ is a momentum flux associated with transport of entropy. 
There are at least two reasons why such a bracket plays a role. Firstly, it can 
be obtained by projection from the Boltzmann Poisson bracket expressing 
kinematics of phonons. Secondly, it can be seen as the Poisson bracket of the 
Lie algebra dual of the semidirect product where entropy is advected by its 
momentum.

It is a 
matter 
of straightforward calculation that by transformation 
\begin{equation}
\ppi = \ww s
\end{equation}
and by using condition \eqref{eq.w.deg}, bracket \eqref{eq.PB.Cat.MG} becomes 
bracket \eqref{eq.PB.sw}. The two brackets are thus equivalent and related by 
change of variables provided the vector field $\ww$ is curl-free.

However, bracket \eqref{eq.PB.Cat.MG} is not constrained to the case of 
curl-free $\ww$ fields. In fact, it can be derived by projection to fields $s$ 
(a scalar function of the phonon distribution function) and $\ppi$ (first 
moment of the distribution function) from the Boltzmann Poisson bracket 
governing dynamics of phonons. Bracket \eqref{eq.PB.Cat.MG} is the hydrodynamic 
bracket expressing kinematics of phonons.

When bracket \eqref{eq.PB.Cat.MG} has to be coupled with hydrodynamics, 
one can proceed simply by summing the hydrodynamic bracket for density and 
momentum of matter, $(\rho, \uu)$, given by \eqref{eq.PB.CH}, and bracket 
\eqref{eq.PB.Cat.MG}, 
\begin{equation}\label{eq.PB.CatHydAlt.semidirect}
\{\AF,\BF\}^{(\uu,\rho,\ppi,s)} = \{\AF,\BF\}^{(\uu,\rho)} + 
\{\AF,\BF\}^{(Cat)}.
\end{equation}
Coupling between those brackets is introduced by transformation from matter and 
phonon momentum onto total momentum and phonon momentum, $(\uu,\rho,\ppi,s) 
\rightarrow (\mm = \uu+\ppi, \rho,\ppi, s)$. After this transformation,
Poisson bracket \eqref{eq.PB.CatHydAlt.semidirect} becomes
\begin{align}
\{\AF,\BF\}^{(\mm,\rho,\ppi,s)} &= \{\AF,\BF\}^{(Euler)} + \int s 
\left(\pd_i A_s B_{\pi_i}-\pd_i B_s A_{\pi_i}\right) d\rr \nonumber\\
&+\int \pi_i \left[\left(\pd_j A_{m_i} B_{\pi_j}-\pd_j B_{m_i} A_{\pi_j}\right) 
+\left(\pd_j A_{\pi_i}B_{m_j}-\pd_j B_{\pi_i}A_{m_j}\right) 
+\left(\pd_j A_{\pi_i}B_{\pi_j}-\pd_j B_{\pi_i}A_{\pi_j}\right)\right]d\rr.
\end{align}
Further transformation from $(\rho,\mm,s,\ppi)$ to $(\rho,\mm,s,\ww=\ppi/s)$ 
leads to bracket
\begin{equation}\label{eq.PB.CatHydAlt}
\{\AF,\BF\}^{(CatHydAlt)} = \{\AF,\BF\}^{(Heat)} + \int d\rr 
\frac{1}{s}(\partial_i w_j-\partial_j w_i) 
A_{w_i}B_{w_j},
\end{equation}
where the first term on the right hand side is given by Eq. 
\eqref{eq.PB.CatHyd.geo} and the second term disappears if condition 
\eqref{eq.w.deg} is satisfied. The condition, however, does not need to be 
satisfied in general, especially in the case when dissipation (usually 
proportional to $\ted_{\ww}$) is taken into account. Poisson bracket \eqref{eq.PB.CatHydAlt} is new.

In summary, Poisson bracket \eqref{eq.PB.CatHydAlt} has been derived in a way 
alternative to the way bracket \eqref{eq.PB.CatHyd} was derived. These 
brackets, 
however, are nearly the same. The difference lies in how the Jacobi identity is 
ensured for fields $\ww$ with non-zero curl. In the case of bracket 
\eqref{eq.PB.CatHyd} Jacobi identity was satisfied by adding a term coupling 
field $\ww$ with momentum $ \mm $ whereas in the case of bracket 
\eqref{eq.PB.CatHydAlt} a yet another term coupling field $\ww$ with itself 
appeared. We have thus arrived at a 
SHTC-compatible ballistic heat conduction expressed in Poisson bracket 
\eqref{eq.PB.CatHyd} and an alternative bracket, \eqref{eq.PB.CatHydAlt}, that 
contains advection of field $\ww$ by itself. It will be a matter of future 
research to compare the alternatives with results on non-Fourier heat 
transport, 
e.g.~\cite{Van2017a} or \cite{Van-book}.


Poisson bracket \eqref{eq.PB.CatHydAlt} generates evolution equations
\begin{subequations}\label{eq.evo.CatHydAlt}
\begin{align}
\frac{\pd m_i}{\pd t} &= -\pd_j(m_i 
\ted_{m_j}) -\pd_j(w_i \ted_{w_j}) - \rho\pd_i 
\ted_\rho -m_j \pd_i \ted_{m_j} - s\pd_i \ted_s 
-w_k \pd_i \ted_{w_k}+\pd_i(\ted_{w_k} w_k)\\
\frac{\pd w_k}{\pd t} &= -\pd_k \ted_s -\pd_k(w_j  \ted_{m_j}) +(\pd_k w_j - 
\pd_j 
w_k)\left(\ted_{m_j}+\frac{1}{s}\ted_{w_j}\right),\label{eq.evo.CatHydAlt.w}\\
\frac{\pd s}{\pd t} &= -\pd_k \left(s \ted_{m_k}+\ted_{w_k}\right),\\
\frac{\pd \rho}{\pd t} &= -\pd_k (\rho \ted_{m_k}).
\end{align}
The difference between the SHTC Cattaneo hydrodynamics and this alternative 
Cattaneo hydrodynamics lies in the last term of \eqref{eq.evo.CatHydAlt.w} with 
prefactor $\frac{1}{s}$.
\end{subequations}


\subsection{Complete Poisson bracket}
The complete Poisson bracket generating SHTC Eulerian 
equations~\eqref{eqn.SHTC} is then the 
combination of brackets \eqref{eq.PB.CH}, \eqref{eq.PB.DM}, 
\eqref{eq.PB.EMHD.modif} 
and \eqref{eq.PB.CatHyd},
\begin{equation}\label{eq.PB.SHTC}
\{\AF,\BF\}^{(SHTC)} = -2\{\AF,\BF\}^{(Euler)} + \{\AF,\BF\}^{(DM)} + 
\{\AF,\BF\}^{(EMHD)} + \{\AF,\BF\}^{(Heat)}.
\end{equation}
Bracket $ \{\AF,\BF\}^{(Euler)} $ is contained in the last three terms and therefore we have to subtract it twice so that it is contained only once in the final Poisson bracket.
Validity of the Jacobi identity for this bracket was checked by the program 
developed in \cite{kroeger2010}.

\section{Dissipation}\label{sec.Dissipation}


Dissipative evolution is usually meant as the evolution that 
raises entropy. In the sense of the hyperbolic conservation laws (SHTC) entropy 
density is among state variables and entropy is thus conserved unless 
dissipative terms are added to equations \eqref{eqn.SHTC}. In the sense of 
GENERIC, 
entropy is required to be a Casimir of the Poisson bracket, i.e. $\{\SF,\EF\}=0$ 
for all possible energies $\EF$, and it is thus also conserved unless some 
dissipative terms are added. It is the goal of this section to compare the way 
dissipation is added in the SHTC and GENERIC formalisms. 

Before introducing concrete forms of dissipative terms, let us make a comment 
on reversibility and irreversibility. Reversible evolution equations is such 
that it is intact by the \textit{time-reversal transformation} (TRT), where 
velocities 
of all particles and magnetic field are inverted as well as the sign of the 
time increment in time derivatives. Evolution 
equations in non-equilibrium thermodynamics usually contain both reversible and 
irreversible parts. Assuming that the left hand sides of the evolution 
equations are constituted by only partial time derivatives of the state 
variables, the reversible parts of the right hand sides transform under TRT 
exactly as the left hand 
sides of the corresponding equations (partial time derivatives of the state 
variable) while the 
irreversible parts of the right hand sides gain opposite signs than the 
corresponding left 
hand sides. Reversible evolution is invariant with respect to TRT while the 
irreversible changes its sign.

In order to be able to apply TRT to an equation, however, one has to first 
introduce parities with respect to TRT. A quantity is called \textit{even with 
respect 
to TRT} (parity equal to $1$) if TRT does not alter the quantity, and it is 
called \textit{odd} (parity equal to $-1$) if TRT changes 
sign of the quantity. How can we determine whether a quantity is even or odd? 
The odd 
quantities are those that change its sign when velocities of particles and 
magnetic field are inverted. For example, density is even, momentum is odd, 
entropy and energy are even. Electric displacement field is even and magnetic 
field is odd. There are, however, quantities with no definite parity, which 
means that TRT does neither leave them intact nor changes their sign, e.g. the 
one-particle distribution function in kinetic theory. In that 
case one has to consider a more geometric definition of TRT (a push-forward on 
vector fields), see \cite{PRE15}. Finally, we require (as is usual in 
non-equilibrium thermodynamics) that the reversible parts of evolution equation 
do not change entropy (i.e. are non-dissipative) while the irreversible terms 
raise entropy (are dissipative), and the adjectives irreversibility and 
dissipativity will be then considered equivalent.

In general, dissipative (or irreversible) evolution within the SHTC framework 
is constructed by means of a quadratic dissipation potential, i.e. the 
irreversible terms are linear in derivatives of energy (conjugate state 
variables within SHTC). Within GENERIC we also 
prefer using a dissipation potential although some authors prefer using a 
possibly non-symmetric dissipative bracket \cite{hutter2013,Ottinger-book}. The geometric 
character and 
statistical interpretation of dissipation potentials \cite{Mielke2014}, discussed also in 
\cite{JNET-EntProdMax}, convince us to prefer dissipation potentials. The 
dissipation potential is, however, used in a 
slightly different way in SHTC than within GENERIC, which differ by the way 
conjugate variables are constructed from the state variables. In the SHTC 
formalism conjugate variables are constructed as derivatives of energy while in 
GENERIC they are derivatives of entropy. Let us now discuss all the mentioned 
possibilities.

\subsection{Irreversible dynamics}

\subsubsection{Energy representation}

In the SHTC framework, irreversible dynamics is modeled via only algebraic 
(independent of spatial gradients) 
dissipative terms which do not harm the hyperbolic type of the PDEs. In the 
light of our recent 
results~\cite{HPR2016,DPRZ2016,DPRZ2017} on modeling of dissipative 
processes such as viscous momentum, heat and charge transfer, the modeling of 
dissipative phenomena with algebraic terms of relaxation type does not seem to 
be something restrictive. 

In order to simplify the notations, in this sections, we also denote the 
conjugate variables as
\begin{equation}
A_{ij}^* = \ted_{A_{ij}},\quad  e_i^* = \ted_{e_i} ,\quad  h_i^* = 
\ted_{h_i},\quad  s^* = \ted_s,\quad  w_i^* = \ted_{w_i}.
\end{equation}

In what follows we provide a recipe on how to introduce dissipative terms into 
system~\eqref{eqn.SHTC}. It is strongly motivated by the key fact around the 
SHTC equations, the summation rule~\eqref{eqn.summation.Euler}. In this 
consideration we do not consider the momentum and mass conservation as they are 
pure reversible time evolutions. Thus, we add, so far, arbitrary functions 
of the state variables into the right hand side of each of the  remaining 
equations:
\begin{subequations}\label{eqn.SHTC.diss}
	\begin{align}
	&\frac{\pd A_{i k}}{\pd t}+\pd_k (A_{il} 
	v_l) + v_j(\pd_j A_{ik} - \pd_k A_{ij})
	= S^A_{ik},\label{eqn.SHTC.diss.deformation}\\[1mm]
	&\frac{\pd h_i}{\pd t} + \pd ( h_i v_k - v_i h_k + \eps_{ikl} \ted_{e_l} ) 
	+ v_i\pd_k h_k = S^h_i, \label{eqn.SHTC.diss.Hfield}\\[1mm]
	&\frac{\pd e_i}{\pd t} +  \pd_k ( e_i v_k - v_i e_k - \eps_{ikl} 
	\ted_{h_l}) + v_i\pd_k e_k = S^e_i, 
	\label{eqn.SHTC.diss.Efield}\\[1mm]
	&\frac{\pd w_k}{\pd t} + \pd_k (v_l w_l + \ted_{s}) + v_j(\pd_j	w_k - \pd_k 
	w_j) = S^w_k,
	\label{eqn.SHTC.diss.heatflux}\\[1mm]
	&\frac{\pd s}{\pd t} + \pd_k (s v_k + \ted_{w_k} ) = S^s.
	\label{eqn.SHTC.diss.entropy}
	\end{align}
\end{subequations}
Because the functions $ S^A_{ij} $, $ S^h_i $, $ S^e_i $, $ S^w_i $ and $ S^s $ 
are quite arbitrary, we can not guaranty that
\begin{equation}\label{eqn.summation.source}
A_{ij}^* S^A_{ij} + h_i^*S^h_i + e_i^* S^e_i + w_i^*S^w_i  + s^*S^s \equiv 0,
\end{equation}
which is required by the summation rule~\eqref{eqn.summation.Euler} and energy 
conservation~\eqref{eqn.SHTC.energy}. Therefore, 
the energy conservation, the first law of thermodynamics, might be, in general, 
violated. Thus, the central question is  how can we introduce the dissipative 
sources in~\eqref{eqn.SHTC.diss} which fulfills the first 
law~\eqref{eqn.SHTC.energy}? It seems that 
the 
only degree of freedom 
we have is to assign to one 
of the variables the role of the entropy. Bearing in mind the 
Hamiltonian nature of the reversible part of the SHTC equations, only a Casimir 
of the corresponding Poisson bracket \eqref{eq.PB.SHTC} can be chosen as the 
entropy. The field $s$ indeed fulfills this criterion.
We emphasize that the introduction of the entropy in the SHTC framework 
is not a spontaneous 
step but it is fully motivated by the necessity to respect the  first law of 
thermodynamics.


We now show how the entropy field can be used to fix the energy conservation, 
i.e. to guarantee the zero on the right-hand side of~ 
\eqref{eqn.summation.source}. It is evident that if we 
define the entropy source term as
\begin{equation}\label{eqn.summation.entropy}
S^s =- \frac{1}{s^* }( A_{ij}^* S^A_{ij} + h_i^*S^h_i + e_i^* S^e_i + 
w_i^*S^w_i)
\end{equation}
then \eqref{eqn.summation.source} holds automatically (recall that $ s^* = 
\ted_{s} $ is usually interpreted as the temperature). However, there is 
also 
the second law of thermodynamics which postulates the positivity of the 
entropy production, $ S^s \geq 
0$. How can we guarantee the second law is fulfilled? Perhaps, the simplest (but 
not 
exhaustive)
solution is as follows. Let us introduce the vectors of state variables and 
conjugate variables
\begin{equation}
\QQ = (\qq,s), \qquad \QQ^* = (\qq^*,s^*),
\end{equation}
where $ \qq = (A_{ij} ,  h_i , e_i ,  w_i )$ and $ \qq^* = (A_{ij}^* ,  
h_i^* , e_i^* ,  w_i^* )$, i.e. they do not 
include $ s  $ and $ s^* $ accordingly. Then, we define the 
dissipative sources as the partial derivatives
\begin{equation}\label{eqn.sources.form}
S^A_{ij} =-\frac{\pd \Psi}{\pd A_{ij}^*}, \quad S^h_i =-\frac{\pd \Psi}{\pd 
h_i^*},\quad  S^e_i 
=-\frac{\pd \Psi}{\pd e_i^*}, \quad S^w_i =-\frac{\pd \Psi}{\pd w_i^*},
\end{equation}
of a potential $ \Psi(\qq^*)$ which does not \textit{explicitly} depend on $ 
s^* $ and 
is 
taken as
\begin{equation}\label{eqn.diss.potetnial.SHTC}
\Psi(\qq^*) = \frac{1}{2} {\qq^*} \Lambda \qq^*,
\end{equation}
where $ \Lambda = \Lambda(\QQ) $ being a symmetric positive semidefinite 
matrix, $ 
\Lambda^\transpose 
= \Lambda \geq 0$, whose entries may depend on the primary state variables $ 
\QQ $ (as well as on other state  variables whose time evolutions are 
reversible and are omitted in this section, e.g. $ \rho $) and will 
be 
associated with the inverse of the characteristic 
dissipation times in Section~\ref{sec.Examples}. Therefore, with the 
choice~\eqref{eqn.sources.form}, \eqref{eqn.diss.potetnial.SHTC}, the second 
law, i.e. the inequality
\begin{equation}
S^s \geq 0,
\end{equation}
is automatically fulfilled. 

We note that the main nonlinearity of the dissipative 
processes is not defined by the nonlinearity of the dissipation 
potential~\eqref{eqn.diss.potetnial.SHTC} or nonlinearity of the conjugate 
variables $ \qq^* $ but it is, in fact, hidden in the entries of $ \Lambda(\qq) 
$. 
For 
example, when dealing with 
irreversible deformations in 
solids, the entries of $ \Lambda $ are associated with the inverse strain 
relaxation 
times~\cite{God1978,Rom1989,GodRom1998,GodRom2003,GodPesh2010,BartonRom2010,Favrie2011,Boscheri2016} which may vary over several orders of magnitude.

Therefore, we have shown that the SHTC irreversible time evolution can be 
expressed in terms of the quadratic dissipation 
potential $ \Psi(\qq^*) $~\eqref{eqn.diss.potetnial.SHTC} 
as
\begin{subequations}\label{eq.SHTC.irr}
\begin{equation}
\left (\frac{\pd q_i}{\pd t}\right )_{\text{irr}} = -\Psi_{q_i^*},
\end{equation}
except for the right hand side of the evolution equation for entropy density, 
which is constructed as 
\begin{equation}\label{eqn.irr.SHTC.s}
\left (\frac{\pd s}{\pd t} \right )_{\text{irr}} = \frac{1}{s^*} 
q^*_i \, \Psi_{q^*_i} \geq 0.
\end{equation}
\end{subequations}
This particular choice of the dissipative source terms respects the both laws 
of 
thermodynamics. 




\subsubsection{Entropy representation and gradient dynamics}
In the previous Section, we have shown that the irreversible 
part of the SHTC time evolution of all the state variables 
can 
be 
generated by the potential $ \Psi(\qq^*) $ which does not depend explicitly on 
$ 
s^* $, and $ \qq^* $ were defined as the conjugate variables with respect 
to the energy, $ \qq^* = E_\qq $. Whereas, within GENERIC, conjugate variables 
in the irreversible 
part are usually identified as derivatives of 
entropy with respect to the state variables and irreversible time evolution can 
be 
formulated as 
\textit{nonlinear gradient dynamics} with a prefactor playing the role of 
temperature
\begin{equation}\label{eq.Generic.irr}
\left (\frac{\pd \qq}{\pd t}\right )_{irr} = \frac{1}{s_\ted}\frac{\delta 
\Xi}{\delta 
s_{\qq}} 
\qquad \text{ 
and } \qquad
\left (\frac{\pd \ted}{\pd t}\right )_{irr} = \frac{1}{s_\ted}\frac{\delta 
\Xi}{\delta 
s_\ted},
\end{equation}
where $\delta/\delta$ stands for a functional derivative and $\ted$ is the 
total 
energy density. The prefactor was introduced for example in \cite{PRE15}. Is 
the SHTC irreversible evolution compatible with this GENERIC irreversible 
evolution?

The answer is affirmative, but in order to 
demonstrate it we have to first recall transformations between the energetic 
representation, where state variables are $(\qq,s)$, and entropic 
representation, where state variables are $(\qq,\ted)$.
In the energetic representation conjugate variables are identified with 
derivatives of energy with respect to the state variables, $(\qq^*,s^*) = 
(E_\qq, E_s)$, while in the entropic representation the conjugate 
variables are derivatives of entropy, $(\pp^*, \ted^*) = (s_\qq, s_\ted)$. The 
two representations thus have different conjugate variables.

In order to show the  compatibility of both formalisms, we need to recall the 
following relations between conjugate variables in the two representations. 
\begin{equation}\label{eq.Ex.Sx}
\left(\frac{\pd E}{\pd q_i}\right)_s = - \left(\frac{\pd 
E}{\pd s}\right)_{\qq} \cdot\left(\frac{\pd s}{\pd q_i}\right)_\ted
\qquad \text{ and } \qquad \frac{\pd E}{\pd s} = \frac{1}{\frac{\pd 
s}{\pd \ted}},
\end{equation}
or 
\begin{equation}\label{eq.Ex.Sx.2}
q^*_i = - s^* p^*_i
\qquad \text{ and } \qquad s^* = \frac{1}{\ted^*}.
\end{equation}
From \eqref{eq.Ex.Sx} it follows that
\begin{subequations}
\begin{eqnarray}
\left(\frac{\pd q^*_i}{\pd p^*_j}\right)_{\ted^*} = -s^* \delta_{ij}, 
\qquad
\left(\frac{\pd q^*_i}{\pd \ted^*}\right)_{\pp^*}= 
\frac{p^*_i}{(\ted^*)^2}\\[2mm]
\left(\frac{\pd s^*}{\pd p^*_i}\right)_{\ted^*} = 0, 
\qquad
\left(\frac{\pd s^*}{\pd \ted^*}\right)_{\pp^*}
=-\frac{1}{(\ted^*)^2}.
\end{eqnarray}
\end{subequations}

Derivatives of the dissipation potential then transform as 
\begin{subequations}\label{eq.Xi.Sx.Ex}
\begin{eqnarray}
\frac{\pd \Xi}{\pd p^*_i} & = & \frac{\pd \Xi}{\pd q^*_j} 
\frac{\pd q^*_j}{\pd p^*_i} + \frac{\pd \Xi}{\pd s^*} 
\frac{\pd s^*}{\pd p^*_i}
=-s^* \frac{\pd \Xi}{\pd q^*_i}\label{eq.Xi.Sx.Ex.1}\\[2mm]
\frac{\pd \Xi}{\pd \ted^*} & = & \frac{\pd \Xi}{\pd 
q^*_j} \frac{\pd q^*_j}{\pd \ted^*} + \frac{\pd \Xi}{\pd s^*} 
\frac{\pd s^*}{\pd \ted^*}
= -s^* \left(q^*_j\frac{\pd \Xi}{\pd q^*_j} + s^* \frac{\pd 
\Xi}{\pd s^*}\right).\label{eq.Xi.Sx.Ex.2}
\end{eqnarray}
\end{subequations}
Irreversible GENERIC evolution equations \eqref{eq.Generic.irr} can be then 
reformulated as
\begin{subequations}\label{eq.Generic.irr.ene}
\begin{align}
\left(\frac{\pd q^i}{\pd t}\right)_{irr} &= - (s^*)^2 \Xi_{q^*_i}\\
\left(\frac{\pd \ted}{\pd t}\right)_{irr} &= - (s^*)^2 
\left(q^*_j\frac{\pd \Xi}{\pd q^*_j} + s^* \frac{\pd 
\Xi}{\pd s^*}\right).
\end{align}
Since the STHC dissipation potential $\Psi$ is always assumed algebraic, i.e. 
does not contain spatial gradients, the sought corresponding GENERIC 
dissipation potential $\Xi$ should be also algebraic. Energy conservation then 
requires that the right hand side of the evolution equation for total energy 
density disappears, i.e.
\begin{equation}
\Xi_{\ted^*} = 0 = q^*_j\frac{\pd \Xi}{\pd q^*_j} + s^* \frac{\pd \Xi}{\pd 
s^*},
\end{equation}
which means that the dissipation potential must be a zero-homogeneous function 
in the energetic representation.
These evolution equations imply the irreversible evolution of entropy density, 
\begin{equation}
\left(\frac{\pd s}{\pd t}\right)_{irr} = \frac{p^*_j}{\ted^*} \Xi_{p^*_j}  
=q^*_j s^* \Xi_{q^*_j}.
\end{equation}
\end{subequations}
We have arrived at the algebraic energy-conserving irreversible GENERIC 
evolution equations in the energetic representation.

Let us now go back to the irreversible evolution formulated within the SHTC 
framework, Eqs. \eqref{eq.SHTC.irr}, which are generated by the dissipation 
potential $\Psi$, which is independent of $s^*$, and by the requirement of 
energy conservation. In order to find the gradient dynamics in the entropic 
representation that is compatible with the SHTC irreversible evolution, we need 
to find the dissipation potential $\Xi$ such that Eqs. 
\eqref{eq.Generic.irr.ene} are the same as Eqs. \eqref{eq.SHTC.irr}. This is 
achieved by choosing 
\begin{equation}
\Xi(\qq^*, s^*) = \Psi(\qq^*)|_{\qq^* := \qq^*/s^*}.
\end{equation}
Such a dissipation potential clearly conserves energy because it can be written 
as a function of $\pp^*$ only (not $\ted^*$). Derivatives of this dissipation 
potential are
\begin{equation}
\Xi_{\qq^*} = \frac{1}{s^*}\Psi_{\qq^*}|_{\qq^* := \qq^*/s^*} 
\qquad\mbox{ and } \qquad \Xi_{s^*} = 
-\frac{q^*_i}{(s^*)^2}\Psi_{q^*_i}|_{\qq^* := \qq^*/s^*},
\end{equation}
and evolution equations \eqref{eq.Generic.irr.ene} become
\begin{subequations}\label{eq.Generic.SHTC}
\begin{align}
\left(\frac{\pd q^i}{\pd t}\right)_{irr} &= - s^* \Psi_{q^*_i}|_{\qq^* := 
\qq^*/s^*}\\[1mm]
\left(\frac{\pd s}{\pd t}\right)_{irr} &= q^*_i \Psi_{q^*_i}|_{\qq^* := 
\qq^*/s^*},
\end{align}
which are exactly the same as Eqs. \eqref{eq.SHTC.irr} for quadratic potentials 
$\Psi$.
\end{subequations}


In summary, energy-conserving algebraic irreversible GENERIC gradient dynamics 
with the temperature prefactor thus completely compatible with the 
irreversible STHC equations provided the SHTC dissipation potential is 
quadratic. If the potential is not quadratic, extra dependence on temperature 
pops out, which does not play any crucial role because of the possible 
dependence of the phenomenological equations on temperature. One can even start 
with purely gradient dynamics without any prefactor, as is usual within 
GENERIC, and the resulting evolution equations in the energetic representation 
would be compatible with the SHTC evolution up to a prefactor given by a power 
of temperature.

\subsection{Involution constraints in case of dissipation}

It is important to remark that there is an intimate connection between 
the 
dissipation and involution constraints discussed in 
Section~\ref{sec.inv.constr}. The presence of the dissipative algebraic source 
terms $ S^A_{ik} $, $ S^h_i $, $ S^e_i $ and $ S^w_i $ in 
\eqref{eqn.SHTC.diss} also affects the time evolutions of $ 
\BB=\nabla\times\AAA 
$, $ Q=\nabla\cdot\hh $, $ R=\nabla\cdot\ee $ and $ \bm{\Omega}=\nabla\times\ww 
$ in the following way
\begin{subequations}\label{eqn.diss.invol.constr}
	\begin{align}
	\dfrac{\pd B_{ij}}{\pd t} & + \pd_k (B_{ij}\, v_k - v_j\, B_{ik} + 
	\eps_{jkl}S^A_{il})  + v_j \pd_k B_{ik} = 0,\label{eqn.diss.constr.B}\\[2mm]
	\dfrac{\pd Q}{\pd t} & + \pd_k (Q\,v_k + S^h_k)= 0, 
	\label{eqn.diss.constr.Q}\\[2mm]
	\dfrac{\pd R}{\pd t} & + \pd_k (R\,v_k + S^e_k) = 0, 
	\label{eqn.diss.constr.R}\\[2mm]
	\dfrac{\pd \Omega_j}{\pd t} & + \pd_k (\Omega_j\, v_k - v_j\, 
		\Omega_k + \eps_{jkl}S^w_l) + v_j \pd_k \Omega_k = 0, 
		\label{eqn.diss.constr.O}
	\end{align}
\end{subequations}
i.e. the dissipative sources emerge as constitutive fluxes in the time 
evolutions for the fields $ \BB $, $ Q $, $ R $ and $ \bm{\Omega}$. Therefore, 
in contrast to the reversible dynamics, these fields deviate from zero values 
even if 
initially ($ t = 0 $) $ \BB = 0  $, $ Q = 0 $, $ R = 0 $ and $ \bm{\Omega}=0$ 
and 
hence, may serve as indicators of irreversible dynamics. See also 
Section~\ref{sec.Examples} with some examples for more discussions.

\subsection{Dispersive dynamics}

It is not obligatory in the SHTC theory that addition of the algebraic source 
terms to the master equations results in that the dynamics becomes 
irreversible, i.e. rise the entropy. In fact, the source terms can be added in 
such a way that both, the entropy and the total energy, are \textit{conserved}. 
This can be achieved by introducing new state variables representing the 
dynamics at a microscale and designing the source terms which models the energy 
conversion between the macro- and micro- scales without any lose. For example, 
the impact of the dislocation density on the dispersive properties of the 
elastic waves was studied in~\cite{Romenski2011} where it was assumed that the 
material has non-zero initial concentration of defects and the energy  converts 
reversibly between the kinetic energy carried by the elastic waves and the 
energy stored in the microscopic defects. 

We recall that in the classical continuum mechanics, the dispersive dynamics, 
as well as the dissipative dynamics, is modeled by means of high-order PDEs. 
However, as we have demonstrated in~\cite{HPR2016,DPRZ2016,DPRZ2017} that the 
classical dissipative models can be successfully modeled with first-order PDEs, 
the dispersive properties of the continuous media can be also modeled with 
merely first-order PDEs, e.g. see~\cite{Romenski2011,Favrie2017,Mazaheri2016}.

\section{Examples}\label{sec.Examples}

\subsection{MaxEnt reduction}\label{sec.MaxEnt}
In non-equilibrium thermodynamics one often deals with the problem of how to 
reduce some detailed evolution to a less detailed evolution. In particular, one 
often has a fast state variable which is to be eliminated from the evolution 
equations. The standard procedure for elimination of the fast variable is the 
Chapman-Enskog reduction, see e.g. \cite{dGM}. However, the reduction often 
relies on demanding calculations and rigorous origin of the method can also be 
questioned.

An alternative method to the Chapman-Enskog reduction was proposed in 
\cite{PRE15}. The method relies on identification of state and conjugate 
variables in the evolution equations. Note that the Chapman-Enskog reduction 
does not use this extra information. It is becoming fashionable to formulate 
non-equilibrium thermodynamics within contact geometry, where the conjugate 
variables have their \textit{independent} meaning and own evolution equations, 
see e.g. 
\cite{Grmela2014a}. In this sense, it is reasonable to let conjugate and state 
variables relax with different speeds. For example, in the reduction proposed 
in \cite{PRE15}, let us refer to it as to the MaxEnt reduction, the fast 
variable is set to the corresponding value given by maximization of entropy 
while the conjugate fast variable is set to a value solving the stationary fast 
evolution equation. This way the conjugate fast variable becomes a function of 
the other state variables, and due to the coupling between the fast and slow 
variables, evolution equations for the slow variables obtain irreversible terms 
caused by relaxation of the fast variable. Let us demonstrate this reduction on 
the following examples.

\subsection{Navier-Stokes dynamics}\label{sec.NSE}
Compressible Navier-Stokes (NS) equations consist of a reversible and 
irreversible part. The reversible part, the compressible Euler equations, is of 
course implied by the SHTC and GENERIC formalisms. Taking for example Poisson 
bracket \eqref{eq.PB.CH}, the implied evolution equations are equations 
\eqref{eq.evo.CH}, which represent the Euler equations.

The irreversible part of NS equations is more difficult to obtain. The NS 
equations can not be incorporated into the SHTC framework because of the 
parabolic character of the irreversible terms (Laplacian of velocity). 
Within GENERIC, the NS equations can be recovered by taking a quadratic 
dissipation potential, or a dissipative bracket, which is composed of second 
derivatives of the potential, as in \cite{Ottinger-book},
\begin{equation}
\Xi(\xx^*) = \frac{1}{2} S_{x^i} M^{ij} S_{x^j},
\end{equation}
where state variables are $\xx = (\mm, \rho, \ted)$ and $\xx^*$ are the 
corresponding conjugate (with respect to entropy) variables.

Our goal, however, is to see the Navier-Stokes equation as an approximation of 
the SHTC equations. This, in particular, was demonstrated  
in~\cite{HPR2016,DPRZ2016}, but this time we would like also to demonstrate 
the work of the MaxEnt reduction. For this purpose, let us consider 
dynamics of the Finger (or left Cauchy-Green) tensor from Sec. \ref{sec.B}, 
which is a consequence of kinematics of the generalized distortion matrix.
Let the energy potential depend on the fields through
\begin{equation}
E = \eps(\rho,s) + \frac{1}{2}\alpha^2 \left(\BBBBdev\right)^2 + 
\frac{1}{2}\beta^2 
\left(\Tr \BBBB - 3 \right)^2 + 
\frac{\mm^2}{2\rho}, 
\end{equation}
derivative of which with respect to the Finger tensor is 
\begin{equation}
E_{\BBBB} = \alpha^2 \BBBBdev + \beta^2 \left(\Tr \BBBB-3\right) \UM.
\end{equation}
Note that $\BBBBdev$ is the deviatoric (traceless) part of tensor $\BBBB$ and 
$\alpha$ and $\beta$ are assumed to be material-dependent constants (typically 
sound 
speeds, e.g. see~\cite{DPRZ2016}).
The derivative $E_\BBBB$ is zero for $\BBBB=\UM$. Similarly, one can express 
entropy as function of $(\mm,\BBBB,\rho,E)$ and derivative of entropy with 
respect to $\BBBB$ will be zero only for Finger tensor equal to the unit 
matrix. 
The 
MaxEnt value of $\BBBB$ is thus the unit matrix $\UM$. 



Putting the MaxEnt value of $\BBBB$ into the evolution equations 
\eqref{eq.evo.B}, 
and 
adding the shear and volume dissipative terms\footnote{The dissipative terms 
can be seen as derivatives of a quadratic dissipation potential with respect to 
$\BBBB^*$, see Section~\ref{sec.Dissipation}. It should be also borne in mind that 
$\BBBB^*$ is interpreted as an element of the dual space to the space where $\BBBB$ lives, 
but in the precise variant of the evolution, Eqs. \eqref{eq.evo.B}, it can be identified with
derivative of energy with respect to $\BBBB$.} $-\tau_S^{-1} \BBBBdev^*$ and 
$-\tau^{-1}_V\Tr(\BBBB^*)\UM$ to the 
right hand side of the equation for $\BBBB$, leads to 
\begin{subequations}\label{eq.evo.B.MaxEnt}
\begin{align}
\frac{\pd m_i}{\pd t} &= 
-\pd_j(m_i m^*_j) - \rho\pd_i \rho^* -m_j 
\pd_i m^*_j - s\pd_i s^* 
+\pd_k\left(\BBB^*_{ki}+\BBB^*_{ik}\right)\\
\label{eq.B.relaxed}0 &= 
\pd_i m^*_j + \pd_j m^*_i - 
\frac{1}{\tau_S}\BBBdev^*_{ij}-\frac{1}{\tau_V}\Tr(\BBBB^*)\delta_{ij},\\
\frac{\pd \rho}{\pd t} &= -\pd_i(\rho m^*_i)\\
\frac{\pd s}{\pd t} &= -\pd_i(s m^*_i).
\end{align}
\end{subequations}
The last equation means that
\begin{equation}
\BBBdev^*_{ij} = \tau_S\left(\pd_i m^*_j + \pd_j m^*_i - \frac{2}{3}\pd_k m^*_k \delta_{ij}\right) 
\qquad\mbox{ and }\qquad 
\Tr(\BBBB^*)=\frac{2}{3}\tau_V \pd_k m^*_k,
\end{equation}
which, after substitution into the equation for $\mm$ and evaluating the conjugate variables as the corresponding derivatives of energy, leads to the Navier-Stokes equations 
\begin{align}
\frac{\pd m_i}{\pd t} &= 
-\pd_j(m_i \ted_{m_j}) - \rho\pd_i \ted_\rho -m_j 
\pd_i \ted_{m_j} - s\pd_i \ted_s \nonumber\\
&+\pd_k\left(2\tau_S \left(\pd_i \ted_{m_k} + \pd_k \ted_{m_i}-\frac{2}{3}\pd_l \ted_{m_l}\delta_{ik}\right)\right)
+\pd_i\left(\frac{4}{3}\tau_V \pd_k \ted_{m_k}\right),
\end{align}
where $\tau_S$ and $\tau_V$ are  related to shear and volume viscosities. 

Note that if the volumetric relaxation time, $\tau_V$, goes to infinity, it follows from Eq. \eqref{eq.B.relaxed} that the spherical part of the $\BBBB$ tensor 
remains undetermined and the condition of incompressibility, $\nabla\cdot \ted_{\mm}=0$,
is enforced after the MaxEnt reduction.  This way the incompressible Navier-Stokes 
equation can be recovered, and the undetermined spherical part of $\BBBB$ 
plays the role of pressure in incompressible fluids.

In summary, compressible (and also incompressible) Navier-Stokes equation is an approximation of the evolution of the Finger tensor, 
which is implied by evolution of the generalized distortion matrix, compatible with the SHTC equations.

%

\subsection{Elastic and elastoplastic solids}

Classical evolution of elastic solids can be formulated by means of the 
distortion matrix easily~\cite{God1978,GodRom1998,GodRom2003}. The distortion 
matrix is 
then interpreted as inverse 
of the deformation gradient, and evolution equations \eqref{eq.evo.DM} or 
\eqref{eqn.SHTC.momentum}, \eqref{eqn.SHTC.A}, \eqref{eqn.SHTC.rho} 
become 
equivalent with the balance of mass, balance of momentum, and 
evolution equation for the inverse deformation gradient in the Eulerian frame. 
Recall that the fact that the distortion matrix is the inverse deformation 
gradient is expressed by the integrability condition $ \nabla\times\AAA = 0 $.

Equations \eqref{eq.evo.DM} or \eqref{eqn.SHTC.A}  are however much 
more general than the mentioned 
Eulerian equations of classical continuum theory because matrix $ \AAA $ is 
allowed to have 
non-zero curl, in which case the interpretation as a \textit{global} 
deformation field becomes 
invalid. The case of distortion matrix with non-zero curl corresponds to 
irreversible deformations, where indeed one can not reconstruct the original 
Lagrangian frame precisely, e.g. 
see~\cite{God1978,Rom1989,GodRom1998,GodRom2003}. 
Non-zero curl of $ \AAA $ is caused by 
dissipative term $ \frac{1}{\tau'}\ted_{\AAA} $ (see \eqref{eqn.diss.constr.B}) 
in 
the evolution equation for the 
distortion matrix because the time evolution for the Burgers tensor $ \BB = 
\nabla\times\AAA $ (dislocation density tensor) is, e.g. 
see Appendix~\ref{app.evol.rotw} or 
~\cite{God1978,GodRom1998,GodRom2003,PeshGrmRom2015},
\begin{equation}\label{eqn.Burg.PDE}
\frac{\pd B_{ij}}{\pd t} + \pd_k \left (B_{ij} v_k - v_j B_{ik} + 
\eps_{jkm}\frac{1}{\tau'}\ted_{A_{im}}\right ) + v_j \pd_k B_{ik} = 0,
\end{equation}
from which it follows that even if $ B_{ij} = 0 $ initially it will, in 
general, distinct from zero at later times. One can clearly see that this 
equation has the SHTC structure of equation~\eqref{eqn.SHTC.Hfield}. However, 
to have the full SHTC and GENERIC structure, this equation should be 
accompanied by a time evolution for a complimentary tensor field $ \DD $, 
exactly as $ 
\hh $ is the complimentary field to $ \ee $, see 
Section~\ref{sec.compl.parity}. In the theory of dislocations, the 
complimentary field $ \DD $ has the meaning of the flux dislocation density, 
e.g. see a discussion in~\cite{PeshGrmRom2015}.

It is interesting to note that, in case $ \BB \neq 0 $, i.e. in case of 
irreversible deformations, the SHTC Eulerian 
equations \eqref{eqn.SHTC.momentum}, \eqref{eqn.SHTC.A} can be symmetrized only 
in the extended weakly non-local sense, see Section~\ref{sec.euler.symmhyp}, 
i.e. when the Burgers tensor $ \BB $ and the flux dislocation density tensor $ 
\DD $ are considered as new independent state variables and the total energy 
potential $ E = E(\mm,\AAA,\BB,\DD,\rho,s) $ depends on the extended set of 
state variables. The later leads to the appearing of extra stress in the 
momentum flux~\eqref{eqn.SHTC.momentum} due to dislocation motion:
\begin{equation}\label{eqn.disl.momentum}
\frac{\pd m_i }{\pd t} + \frac{\pd }{\pd 
	x_k}\left(m_i \ted_{m_k} + \delta_{ik} \left( \rho \ted_\rho  + \sigma 
	\ted_\sigma 
	+ m_l\ted_{m_l} + B_{jl} \ted_{B_{jl}} + D_{jl} \ted_{D_{jl}} - \ted 
	\right) - B_{jk} \ted_{B_{ji}} -  D_{jk} \ted_{D_{ji}} + 
	A_{jk}\ted_{A_{ji}}\right)=0.
\end{equation}

Thus, evolution of the distortion matrix $ \AAA $ gives, in fact, exact 
evolution of the Burgers tensor, defining dynamics of dislocations~ 
\cite{God1978,GodRom1998,GodRom2003}. The evolution 
equation of the distortion matrix coupled with mass, momentum and energy 
conservation, thus represents an Eulerian framework covering elastic and 
elasto-plastic solids as well as fluids, see also 
\cite{HPR2016,DPRZ2016,HYP2016,Rom1989,BartonRom2010,Boscheri2016,Favrie2011}. 
In practice, the transition between the elastic and plastic response of solids 
is governed by the dependence of the strain relaxation time on the state 
parameters, $ \tau = \tau (\AAA,\BB,\DD,\rho,s) 
$, see details in~\cite{HYP2016,Rom1989,BartonRom2010,Boscheri2016,Favrie2011}.

\subsection{Non-Newtonian fluid dynamics}

Dynamics of viscoelastic fluids (or solids) can be also 
formulated within the 
SHTC and GENERIC formalisms by the proper choice of the dissipative source term 
in the time evolution for the distortion matrix:
\begin{equation}
	\frac{\pd A_{ik}}{\pd t} + \pd_k (A_{im} v_m) + v_j(\pd_j A_{ik} - \pd_k 
	A_{ij}) =-\frac{1}{\tau'}\ted_{A_{ik}}\,
\end{equation}
or by using the Finger tensor as in Sec. \ref{sec.B}.
Here, the relaxation function $ \tau' = \tau'(\tau) $ depends on the real 
relaxation time $ \tau $ and typically is defined as 
$\tau' \sim \tau\,b^2,$
while its further specification depends on the specific choice of the total 
energy, see~\cite{HPR2016,DPRZ2016}. Here, $ b $ is the shear sound speed (the 
speed of propagation of small transversal perturbations). For the further 
discussion, 
let us recall the physical interpretation of the strain dissipation time $ \tau 
$. In the unified flow theory~\cite{HPR2016,DPRZ2016,HYP2016}, $ \tau $ is 
understood as a continuum interpretation of the seminal idea 
of the so-called \textit{particle settled life time} of 
Frenkel~\cite{Frenkel1955}, who applied it to describe the ability of liquids 
to flow, see also recent promising experimental and theoretical 
advances~\cite{brazhkin2012two,bolmatov2013thermodynamic,bolmatov2015revealing,Bolmatov2015a,Bolmatov2016} confirming and further developing Frenkel's idea. 
Thus, in our continuum approach, the time $ \tau $ is the time taken by a given 
continuum particle (finite volume) to ``escape'' from the cage composed of its 
neighbor particles, 
i.e. the time taken to rearrange with one of its neighbors. The more viscous a 
fluid is, the larger the time $ \tau $, i.e. the longer the continuum  
particles stay in contact with each other. 

Such a concept of the strain 
dissipation time, in contrast to the phenomenological viscosity concept, allows 
to 
cover the entire spectrum of material responses, 
from ideal fluids ($ \tau = 0 $) to elastic solids ($ \tau = \infty $) through 
the viscous fluids, either Newtonian or non-Newtonian ($ 0 < \tau < \infty $). 
Note that the viscosity concept does not allow to model deformable solids. 
Moreover, in the proposed framework, the division on the Newtonian and 
non-Newtonian fluids becomes artificial because it dose not rely on any 
constitutive laws designed specifically for either Newtonian or non-Newtonian 
flows. For example, to apply our model to modeling Newtonian flows one may 
even do not know that Newton's viscous law exists. All flows are treated as 
being non-equilibrium and no assumptions 
on the closeness to the global equilibrium was assumed in both formalisms.
Thus, according to our approach, simple fluids as water for example exhibit 
elastic response at time scales $ T < \tau $, viscoelastic at time scales $ T 
\sim \tau $ and Newtonian response at time scales $ T \gg \tau $, where $ T $ 
is the characteristic flow time scale. We emphasize that the viscoelastic and 
even solid like properties of simple liquids at small time scales is an 
experimental 
fact, see e.g. 
\cite{brazhkin2012two,bolmatov2013thermodynamic,bolmatov2015revealing,Bolmatov2015a,Bolmatov2016} and was predicted by Frenkel~\cite{Frenkel1955}.
 
In practice, as in the modeling of elastoplastic solids, the user have to 
provide a suitable function $ \tau=\tau(\AAA,\rho,s) $ to cover diverse 
non-Newtonian responses. We emphasize that, in contrast to the viscosity 
approach, there is no need to use the dependence of $ \tau $ on the strain rate 
because our relaxation model is rate dependent by the construction, e.g. see 
Fig.~1 in~\cite{HPR2016}. Moreover, in contrast to the nonlinear viscosity, the 
switching from $ \tau = \text{const} $ to nonlinear $ \tau(\AAA,\rho,s) $ dose 
not increase the complexity of the model. Thus, in classical phenomenological 
approach the substitution of the constant viscosity onto a nonlinear viscosity 
transforms a linear parabolic PDE system into a nonlinear parabolic system 
which is 
a 
way more difficult problem than the linear theory from both, the theoretical as 
well as numerical viewpoint.  In our case, $ 
\tau(\AAA,\rho,s) $ does not affect the hyperbolic differential part 
(reversible part of the time evolution) of the 
model at all.

\subsection{Heat and mass transfer}\label{sec.CatFou}

In this section, we demonstrate the viewpoint discussed in the Introduction 
section on 
the \textit{universality of the mathematical structure of the equations}. We 
shall demonstrate that equations~\eqref{eqn.SHTC.w}, 
\eqref{eqn.SHTC.entropy} equipped with proper dissipative source terms and 
coupled with momentum and mass conservation, i.e.
\begin{subequations}\label{eqn.heat}
	\begin{align}
	\frac{\pd m_i }{\pd t} & + \pd_k (m_i v_k + [ \rho \ted_\rho  + 
	\sigma \ted_\sigma 	+ m_l \ted_{m_l} - \ted ]\delta_{ik} + w_i\ted_{w_k} 
	)=0, 
	\label{eqn.heat.momentum}\\[1mm]
	\frac{\pd w_k}{\pd t} & + \pd_k (v_l w_l + \ted_{\sigma}) + v_j(\pd_j 
	w_k - \pd_k w_j) = -\frac{1}{\tau}\ted_{w_k},	\label{eqn.heat.w}\\[1mm]
	\frac{\pd \sigma}{\pd t} & + \pd_k(\sigma v_k + \ted_{w_k}) = 
	\frac{1}{\ted_\sigma \tau}\ted_{w_i}\ted_{w_i},	
	\label{eqn.heat.entropy}\\[1mm]
	\frac{\pd \rho}{\pd t} & + \pd_k (\rho v_k) = 0,
	\label{eqn.heat.rho}
	\end{align}
\end{subequations}
possess a sort of universal structure because they 
can be applied to quite different physical settings. Namely, to the transport 
of heat 
and transport of mass. In particular, we shall demonstrate that the solution to 
this system satisfies
the equilibrium laws of heat and mass transfer of Fourier and Fick accordingly, 
if the energy potential $ E $ is properly designed. The non-Fourier heat 
conduction and non-Fickian diffusion corresponds to $ \tau \sim T $ and are 
beyond the scope of this paper. Here, $ \tau $ is the relaxation parameter 
in~\eqref{eqn.heat.w}, and $ T $ is the characteristic macroscopic time scale. 
Let us start with the discussion of the heat transfer.


\subsubsection{Heat conduction}

Energy conservation law~\eqref{eqn.SHTC.energy} for system \eqref{eqn.heat} 
reduces to
\begin{equation}\label{eqn.heat.energy.examp}
\frac{\pd \ted}{\pd t} + \pd_k ( v_k \ted 
+ v_i [ ( \rho \ted_\rho  + \sigma\ted_\sigma + m_l\ted_{m_l} - \ted) 
\delta_{ik} + w_i\ted_{w_k}] + \ted_{\sigma} \ted_{w_k} 
)=0, 
\end{equation}
from which one can recognize the heat flux
\begin{equation}\label{eqn.heat.flux}
q_i = \ted_\sigma \ted_{w_i}.
\end{equation}
By taken the energy potential in the most simple form
\begin{equation}\label{eqn.heat.E}
E = E^{hydro}(\rho,s) + \frac{\alpha^2}{2}\ww^2 + \frac{1}{2\rho} \mm^2
\end{equation}
and expanding the solution in a series in $ \tau $, e.g. $ \ww = \ww_0 + \tau 
\ww_1 
+ \tau^2\ww + \ldots $, one can show (exactly as in~\cite{DPRZ2016}) that in 
the leading terms, the Fourier law of heat conduction is recovered as
\begin{equation}\label{eqn.Fourier.law}
q_i = -\rho \theta \tau \alpha^2 \frac{\pd \theta}{\pd x_i}, \qquad \theta = 
\ted_\sigma,
\end{equation}
where $ \theta $ is the temperature. From this relation, it is easy to 
recognize the heat conductivity $ \kappa $ as $ \kappa = \rho\, \theta\, \tau \,
\alpha^2 $.

Let us now demonstrate that similar results can be obtained with the help of 
MaxEnt approach (see Section~\ref{sec.MaxEnt}) which we have already applied to 
the Navier-Stokes equations in Section~\ref{sec.NSE}. Again, for this purpose,
it is more convenient to use equivalent form~\eqref{eq.evo.CatHyd} of 
\eqref{eqn.heat}. Thus, the goal is to reduce the Cattaneo-type hydrodynamics, 
represented 
by the equations
\begin{subequations}
\begin{align}
\frac{\pd m_i}{\pd t} &= -\pd_k(m_i \ted_{m_k}) - \rho\pd_i 
\ted_\rho - s\pd_i \ted_s - m_k \pd_i \ted_{m_k} - w_i \pd_k w^*_k - 
w^*_k(\pd_a w_i 
- \pd_i w_k),\\
\frac{\pd w_k}{\pd t} &= -\pd_k \ted_s -w_j \pd_k \ted_{m_j} - 
\ted_{m_j}\pd_j w_k +\frac{1}{s}(\pd_k w_j - \pd_j w_k)w^*_j - 
\frac{1}{\tau} w^*_k,\\
\frac{\pd s}{\pd t} &= -\pd_k (s \ted_{m_k} + w^*_k)
\end{align}
\end{subequations}
into hydrodynamics with Fourier heat conduction. Here, we use the notation $ 
w^*_i = \ted_{w_i} $ for the $ \ww $ conjugate field.

As always, the energy is assumed to be a convex function of $\ww$ and thus 
the entropy is a concave function of 
$\ww$. Derivative of entropy with respect to $\ww$ is zero for $\ww =0$, which 
is the value to which field $\ww$ is set in the MaxEnt reduction. This leads to 
equations
\begin{subequations}
\begin{align}
\frac{\pd m_i}{\pd t} &= -\pd_j(m_i 
\ted_{m_j})  - \rho\pd_i \ted_\rho -m_j \pd_i \ted_{m_j} - s\pd_i \ted_s \\
0 &= -\pd_k \ted_s - \frac{1}{\tau} w^*_k,\\
\frac{\pd s}{\pd t} &= -\pd_k \left(s \ted_{m_k}+w^*_k\right).
\end{align}
The second equation gives 
\begin{equation}
w^*_k = -\tau \pd_k \theta, \qquad \theta = \ted_s,
\end{equation}
where $\theta$ being the temperature.
\end{subequations}

Plugging the relaxed value of $\ww^*$ into the equation for entropy yields
\begin{subequations}
\begin{align}
\frac{\pd m_i}{\pd t} &= -\pd_j(m_i 
\ted_{m_j})  - \rho\pd_i \ted_\rho -m_j \pd_i \ted_{m_j} - s\pd_i \ted_s \\
\frac{\pd s}{\pd t} &= -\pd_k \left(s \ted_{m_k} - \tau \ted_s\right),
\end{align}
which represent Euler equations with Fourier heat conduction. 
\end{subequations}

Eventually we note that, by combining the Cattaneo-type 
hydrodynamics with the dynamics of the Finger tensor, Sec. \ref{sec.B}, one 
can obtain the full Navier-Stokes-Fourier (NSF) system of equations by the 
MaxEnt reduction from the SHTC framework although the NSF system is not part of 
the framework itself.

\subsubsection{Mass transfer}\label{sec.mass.transfer}

In this section we continue exploiting equations~\eqref{eqn.heat} and, this 
time, show that 
the classical Fickian diffusion usually described by parabolic PDE can be 
successfully modeled by equations~\eqref{eqn.heat}. However, for this purpose 
we shall prescribe different meanings to the variables $ (\ww,\sigma) $.

Let us consider a mixture of two inviscid fluids. We characterize the mixture 
morphology by 
three independent scalars, the mass density of the mixture $ \rho $, and the 
mass and volume fractions of one of the phases (denoted by the index 
``1'') $ c $ and $ \alpha  $ 
correspondingly. These three scalars relate to each other through the phase 
mass densities $ \rho_1 $ and $ \rho_2 $ as
\begin{equation}\label{eqn.examps.morphology}
\rho = \alpha \rho_1 + (1-\alpha)\rho_2, \qquad c = \frac{\alpha \rho_1}{\rho}.
\end{equation}
Obviously that the mass and volume fraction of the second phase are $ 1-c $ and 
$ 1-\alpha $ accordingly.

The kinematics of the mixture is characterized by two velocity fields
\begin{equation}
\vv = c \vv^1 + (1-c)\vv^2, \qquad \ww = \vv^1 - \vv^2,
\end{equation} 
which are the mixture velocity and the relative velocity of the phases. 

Thus, if one identifies the scalar $ \sigma $ in \eqref{eqn.heat} with $ \rho c 
$ and introduces the specific total energy $ \tes = \ted/\rho $ then system 
\eqref{eqn.heat} can be written as
\begin{subequations}\label{eqn.examps.2fluid}
	\begin{align}
&\frac{\pd  \rho  v_i}{\pd  t} + \pd_k  (\rho  v_i v_k + \rho^2 
\tes_\rho 
\delta _{ik}+\rho  w_i\tes_{w_k}) = 0,\\[1mm]
&\frac{\pd w_k}{\pd  t} + \pd_k  
(w_j v_j+\tes_c) + v_j(\pd_j w_k - \pd_k w_j)=-\frac{1}{\tau 
^w}\tes_{w_k},\\[1mm]
 &\frac{\pd \rho  c}{\pd  t} + \pd_k  (\rho  c v_k + \rho\tes_{w_k}) = 0,\\[1mm]
 &\frac{\pd \rho  }{\pd  t} + \pd_k (\rho v_k ) = 0,\label{eqn.mass.rho}\\[1mm]
&\frac{\pd \rho  \alpha }{\pd  t} + \pd_k (\rho \alpha v_k) =-\frac{1}{\tau 
^{\alpha}}\rho \tes_{\alpha },\label{eqn.mass.alpha}\\[1mm]
&\frac{\pd \rho  s }{\pd  t} + \pd_k (\rho s v_k) = \frac{1}{\tes_s}\left ( 
\frac{1}{\tau^A}\tes_{A_{ij}}\tes_{A_{ij}} + \frac{\rho^2}{\tau ^{\alpha
}} \tes_{\alpha } \tes_\alpha + \frac{1}{\tau ^w}\tes_{w_i} \tes_{w_i}\right ) 
\geq 0,\label{eqn.mass.s}
	\end{align}
\end{subequations}
where we also add the time evolution of $ \rho\alpha $, while the entropy 
time evolution 
now has the structure of fluxes not as in the heat conduction case in the 
previous section, i.e. as in 
the time evolution for the scalar $ \sigma $ in~\eqref{eqn.heat}, but as the 
mass conservation. Recall that PDEs with such a simple structure of fluxes as  
in \eqref{eqn.mass.rho}, \eqref{eqn.mass.alpha} and \eqref{eqn.mass.s} can be 
added in an arbitrary amount, see 
Section~\ref{sec.compl.parity}.

To close the system~\eqref{eqn.examps.2fluid}, one has to provide the specific 
total energy potential $ \tes $. For example, it can be taken as
\begin{equation}
E= c\, e^1\left(\rho _1,s\right)+(1-c)e^2\left(\rho 
_2,s\right)+\frac{1}{2}c(1-c) \ww^2 + E^{\text{mix}}(\rho 
,s,\alpha ,c) + \frac{1}{2}\vv^2,
\end{equation}
where $ e^1 $ and $ e^2 $ are the internal energies of the phases, $ 
E^{\text{mix}} $ defines the chemical interaction of the phase 
molecules and can be left unspecified for our purposes. Then, the mixture 
chemical potential is defined as the derivative of the energy with 
respect to the mass fraction
\begin{equation}
\tes_c=e^1+\rho _1 e_{\rho _1}^1-e^2-\rho _2 e_{\rho 
_2}^2+\frac{1}{2}(1-2c)\ww^2+\tes_c^{\text{mix}}=\mu _1-\mu 
_2+\frac{1}{2}(1-2c)\ww^2+\tes_c^{\text{mix}},
\end{equation}
where $\mu _i=e^i+\frac{p_i}{\rho _i}$, $i=1,2$ are the chemical 
potentials of 
the phases.

Taking the derivatives of the total energy with respect to $ \alpha $ we obtain 
the pressure relaxation term in the volume fraction time evolution
\begin{equation}\label{eqn.E_alpha}
\tes_{\alpha }=\frac{1}{\rho }(p_2-p_1 ),
\end{equation}
where $ p_i = \rho_i^2 \frac{\pd e^i}{\pd \rho_i}$ are the phase pressures.

Now, we are in position to show that the Fick law of diffusion is an 
inherent property of the model~\eqref{eqn.examps.2fluid}. As in the case of the 
Fourier law of heat conduction, it ca be 
shown that the Fick law is recovered in our hyperbolic model in the 
leading terms for small relaxation time $ \tau^w $ when the mass fraction time 
evolution reduces to 
\begin{equation}
\rho \frac{\rmd c}{\rmd t} = \nabla\cdot[\rho\tau^w\nabla(\mu_1 - \mu_2 + 
\tes_c^\text{mix})]
\end{equation}
from which one can recognize the Fick law
\begin{equation}
\rho \frac{\rmd c}{\rmd t} = \nabla\cdot\JJ, \qquad \JJ = \eta \nabla \tes_c
\end{equation}
or 
\begin{equation}
\rho \frac{\rmd c}{\rmd t} = \nabla\cdot(D\nabla c), \qquad D = \eta 
\tes_{cc}
\end{equation}
if to require 
\begin{equation}
\eta = \rho \tau^w,
\end{equation}
and hence the effective diffusion coefficient is $ D = \rho \tau^w \tes_{cc} $.

Eventually, we recall that the viscous momentum transfer is modeled by the 
distortion field in the SHTC formulation~\cite{HPR2016,DPRZ2016,DPRZ2017}. 
Thus, 
if one wants to generalize the above equations to model the viscous 
properties of mixtures then it is sufficient to add only one distortion field 
representing the distortion of the mixture elements. Then, the strain 
relaxation time, see~\cite{DPRZ2016}, should be composed of the phase 
relaxation times.  We hope to discuss this in a subsequent paper dedicated to 
modeling of multi-phase flows.

\subsection{Electrodynamics of slowly moving medium}
 The electrodynamics of slowly moving medium ($ |\vv| \ll c $, 
where $ c $ is the speed of light)
can be modeled by equations 
~\eqref{eqn.SHTC.momentum}, \eqref{eqn.SHTC.Efield}, \eqref{eqn.SHTC.Hfield} 
and \eqref{eqn.SHTC.rho}, see~\cite{DPRZ2017,elmag}. By means of adding the 
dissipative source term 
$-\frac{1}{\eta} \ted_{e_i} $ to the right hand side of the 
equation~\eqref{eqn.SHTC.Efield} which represents the dynamics of the electric 
field, one can describe dielectrics ($ \eta \rightarrow \infty $), ideal 
conductors ($ \eta \rightarrow 0 $), and 
resistive conductors ($0 < \eta < \infty $) as particular cases, where $ \eta $ 
is the resistivity, see details in~\cite{DPRZ2017,elmag}.

\section{Concluding remarks}

Continuum mechanics of fluids and solids with dislocations, with Cattaneo type 
heat conduction, with mass transfer, and with electromagnetic fields is 
formulated in the 
Hamiltonian and the Godunov type 
form. In order to emphasize that the Godunov type structure that we consider is 
an extension of the original Godunov structure of local conservation laws we 
also call it SHTC structure.
The Hamiltonian structure guarantees the physical well-posedness and the 
Godunov like structure the mathematical and the numerical well-posedness.

From the microscopic  point of view, the physical systems under consideration 
are  mechanical systems, their microscopic governing equations thus possess the 
Hamiltonian structure. This structure then passes also to the time reversible 
part of the dynamics in the continuum formulation. However,
due to the ignorance of microscopic details in the continuum  formulation,  a 
new dynamics, that is time irreversible and of gradient type, emerges. The 
governing equations of continuum mechanics are thus an appropriate combination 
of the Hamiltonian and the gradient dynamics (called GENERIC).

The governing equations in the SHTC form  are first order symmetric hyperbolic 
partial differential equations admitting a companion local conservation law. We 
have seen that the Hamiltonian continuum equations can be cast into the SHTC  
form. The companion local conservation law in the SHTC formulation appears  in 
the Hamiltonian formulation as a statement about the degeneracy of the 
Hamiltonian structure. The time irreversible part of the continuum dynamics is 
in both the GENERIC and the SHTC formulations of gradient type. In the SHTC  
formulation the time irreversible part  is moreover required to be free of 
spatial 
derivatives. The advantage of the SHTC formulation is the mathematical 
well-posedness in the sense that the partial differential equations can be 
symmetrized (due to the existence of the companion local conservation law) and 
as such the Cauchy initial value problem for them is well posed. The numerical 
well-posedness means that the Godunov finite volume discretization method, that 
keeps the physical content of the governing equations also in their discrete 
form, can be applied.

%

\section*{Acknowledgments}
I.P. acknowledges a financial support from ANR-11-LABX-0040-CIMI within the 
program ANR-11-IDEX-0002-02.
M.P. and M.G. were supported by Czech Science Foundation, project no. 
17-15498Y, by Natural Sciences and Engineering Research Council of Canada 
(NSERC). E.R. acknowledges a partial support by the Program N15 of the 
Presidium of RAS (project 121) and the Russian Foundation for Basic Research 
(grant No 16-29-15131).

\appendix

\section{Lagrange-to-Euler transformation}\label{sec.SHTC.LagToEu}

In this section we recover all the technical details omitted 
in~\cite{GodRom1996a}  of the transformation of 
the Lagrangian master system~\eqref{eqn.lagr.masterU} into its Eulerian 
counterpart~\eqref{eqn.SHTC}.


In what follows, we use the notations $ \AAA=\FF^{-1} $, $ w = \det(\FF) $, and 
we use primes to denote the Lagrangian fields, e.g. $ \mm' $ is the field 
appearing in equation~\eqref{eqn.lagr.masterU.Momentum}, etc. Yet the reference 
(Lagrangian) density will be denoted not as $ \rho' $ but $ \rho_0 $.

The Lagrangian and Eulerian fields relate to each other by the following 
formulas
\begin{subequations}
\begin{equation}
\label{eqn.m}
\mm'=w \,\mm, \qquad \FF=\AAA^{-1}, \qquad \rho_0 = w\,\rho,
\end{equation}
\begin{equation}\label{eqn.eh}
\ee' = w \AAA \ee, \qquad \hh' = w \AAA \hh,
\end{equation}
\begin{equation}\label{eqn.w}
\ww' = \FF^\transpose \ww, \qquad \sigma' = w \,\sigma,
\end{equation}
\noindent while the Lagrangian total energy density $ U $ is related to 
Eulerian 
total energy density $ \ted $ as
\begin{equation}\label{eqn.gray.potential}
U= w \,\ted.
\end{equation}
\end{subequations}

For the Lagrange-Euler transformation, we also need the following relations 
between the derivatives of $ \ted $ and $ U $ with respect to the state 
variables
\begin{subequations}
\begin{align}\label{eqn.derivatives.eh}
&\ted_{h_i} = A_{ji}U_{h'_j}, \qquad  \ted_{e_i} = A_{ji}U_{e'_j}, \\[1mm]
\label{eqn.derivatives.sw}
&\ted_{\sigma} = U_{\sigma'}, \qquad \ \ \ \  \ted_{w_k} = \frac{1}{\rho_0} 
F_{kj} 
U_{w'_j}.
\end{align}
\end{subequations}

\subsection{Auxiliary relations}
Here, we summarize the definitions and formulas used in the Lagrange-to-Euler 
transformation. The total 
deformation gradient $\FF= [F_{ij}] $, the distortion matrix $ \AAA=[A_{ij}] $ 
and the velocity are defined as
\begin{equation}\label{eq.gradient}
F_{ij}=\dfrac{\pd x_i}{\pd y_j},\qquad \AAA=\FF^{-1},\qquad w=\det 
(\FF)=\dfrac{\rho_0}{\rho},\qquad v_i=\frac{\pd x_i}{ \pd t}, \qquad
\frac{\rm d}{{\rm d} t} = \frac{\pd}{\pd t} + v_k\dfrac{\pd}{x_k}
\end{equation}
where  $ y_j $ are the Lagrangian coordinates and $ x_i $ are 
the Eulerian ones, $ \rho $ and $ \rho_0 $ are the actual
and the reference mass densities, respectively. The time evolution equation for 
$ F_{ij} $ in the Lagrangian coordinates
\begin{equation}\label{eq.gradient.time}
\frac{\rmd F_{ij}}{\rmd t} - \frac{\pd v_i}{\pd y_j}=0
\end{equation}
is a trivial consequence of definitions \eqref{eq.gradient}$ _1 $ and 
\eqref{eq.gradient}$ _4 $.

The following standard definitions and formulas  are also introduced

\begin{equation}\label{eq.CofDefinition}
\boldsymbol{C}={\rm cof}(\FF)=w \AAA^\transpose=\left[C_{{ij}}\right], \text{ \ 
or \ }A_{{ij}}=w^{-1}C_{{ji}},
\end{equation}

\begin{equation}\label{eq.DivCof}
\frac{\pd C_{i j}}{\pd y_j}=0,\qquad \varepsilon 
_{{ijk}}\frac{\pd  F_{{ik}}}{\pd y_j}=0,
\end{equation}

\begin{equation}\label{eq.DetTimeEvol}
\frac{{\rm d} w}{{\rm d} t}-\frac{\pd w A_{j k}v_k}{\pd y_j}=0,{\rm 
or\ using\ (\ref{eq.CofDefinition})\ and\ (\ref{eq.DivCof}):\ \ \ }\frac{{\rm 
d} w}{{\rm d} t}-w A_{j k}\frac{\pd v_k}{\pd y_j}=0,\ \ {\rm or}\ \ 
\frac{{\rm 
d} \rho}{{\rm d} t} + \rho A_{j k}\frac{\pd v_k}{\pd y_j}=0,
\end{equation}

\begin{equation}\label{eqn.constF.euler}
\frac{\pd w^{-1}F_{kj}}{\pd x_k}=0,\ \ \ \ \frac{\pd }{\pd t}\left ( 
\frac{1}{w}\right ) + \frac{\pd }{\pd x_k}\left( 
\frac{v_k}{w}\right) =0,
\end{equation}

\begin{equation}\label{eq.DetLeviCivita}
\varepsilon _{{ikl}}A_{{mi}}A_{{jk}}A_{{al}}=\varepsilon_{mja}w^{-1},
\end{equation}

\begin{equation}\label{eq.CofFormulaLeviCivita}
C_{{km}}=\frac{1}{2}\varepsilon _{{lnk}}\varepsilon _{{pqm}}F_{{lp}} F_{{nq}},
\end{equation}

\begin{equation}\label{eq.LeviCivitaKronecker}
\varepsilon _{{imn}}\varepsilon _{{jmn}}=2\delta _{{ij}}.
\end{equation}
Here, $ \varepsilon_{ijk} $ is the Levi-Civita symbol and $ \delta_{ik} $ is 
the 
Kronecker delta.

\subsection{$ (w_i,\sigma) $-pair}
Here, we work with the complimentary pair
\begin{subequations}
	\begin{align*}
	& \frac{\rmd w'_j}{{\rm d} t} + \frac{\pd U_\sigma'}{\pd y_j} = 0,\\[1mm]
	& \frac{\rmd \sigma'}{{\rm d} t} + \frac{\pd U_{w'_j}}{\pd y_j} = 0.\\[1mm]
	\end{align*}
\end{subequations}
This pair of equation can be used (if proper energy potential and dissipation 
source terms are 
introduced) to model heat transfer ($ \sigma $ 
then should be treated as the entropy while $ w_j $ as some heat vector) or 
multi-phase flows ($ \sigma $ should be treated as the mass fraction while $ 
w_j $ 
as the relative velocity vector).


Moreover, we also assume that $ \sigma = 
\rho s $ and $ \ted = \rho \tes $. Note that the specific energy $ \tes $ is
related to the Lagrangian energy density $ U $ as $ \tes = \rho_0 U $.

\subsubsection{Transformation \eqref{eqn.SHTC.entropy}$ \rightarrow 
$\eqref{eqn.lagr.masterU.scalar}}

\[\frac{\pd \sigma}{\pd t}+\frac{\pd \left( 
\sigma v_k+\ted_{w_k}\right)}{\pd x_k}=\frac{\pd \rho   s }{\pd t}+\frac{\pd 
\left(\rho  
 s v_k+\rho  \tes_{w_k}\right)}{\pd x_k}=
s \frac{\pd \rho }{\pd t}+\rho \frac{\pd  s }{\pd 
t}+\rho   s \frac{\pd v_k}{\pd x_k}+v_k\rho \frac{\pd  s }{\pd
x_k}+v_k s \frac{\pd \rho }{\pd x_k}+\frac{\pd \rho  
\tes_{w_k}}{\pd x_k}=\]

\[{=s \frac{{\rm d} \rho }{{\rm d} t}+\rho \frac{{\rm d}  s }{{\rm 
d} t}+\rho   s 
\frac{\pd
v_k}{\pd x_k}+\frac{\pd \rho  \tes_{w_k}}{\pd x_k}=s \frac{{\rm d} \rho 
}{{\rm d} t}+\rho \frac{{\rm d}  s }{{\rm d} t}+\rho   s A_{j k}\frac{\pd 
v_k}{\pd
y_j}+A_{j k}\frac{\pd \rho  \tes_{w_k}}{\pd y_j}}\]

We now use \eqref{eq.DetTimeEvol}$ _3 $ to substitute time derivative of the 
density:
\[
{- s \rho  A_{j k}\frac{\pd v_k}{\pd y_j}+\rho \frac{{\rm d}  s }{{\rm d} 
t}+\rho   s A_{j k}\frac{\pd v_k}{\pd y_j}+A_{j k}\frac{\pd
\rho  \tes_{w_k}}{\pd y_j}=\rho \frac{{\rm d}  s }{{\rm d} t}+A_{j k}\frac{\pd 
\rho  
\tes_{w_k}}{\pd y_j}=\rho _0w^{-1}\frac{{\rm d}  s }{{\rm d} t}+w^{-1}C_{k 
j}\frac{\pd
\rho  \tes_{w_k}}{\pd y_j}=0}
\]
Using constraint~\eqref{eq.DivCof}$ _1 $ for the cofactor matrix, the last 
equality can be rewritten as
\[\frac{{\rm d} \rho_0  s }{{\rm d} t}+\frac{\pd \rho C_{kj} \tes_{w_k}}{\pd 
y_j}=0,\]
and after using~\eqref{eqn.w} and $ w^{-1} C_{kj} = A_{jk} $, as
\[\frac{{\rm d} \sigma' }{{\rm d} t}+\frac{\pd A_{jk} \rho_0 \ted_{w_k}}{\pd 
y_j}=0.\]
Eventually, after noting that $ \ted_{w_k}= \left 
(\frac{1}{\rho_0}U(F_{ba}w_{a})\right )_{w_k} 
= \frac{1}{\rho_0} F_{kj} U_{w'_j}$, 
the latter equation becomes
\begin{equation}\label{eqn.scalar.final}
\frac{{\rm d} \sigma'}{{\rm d} t}+\frac{\pd U_{w'_j}}{\pd y_j}=0,
\end{equation}
which is the sought Lagrangian equation~\eqref{eqn.lagr.masterU.scalar}.

\subsubsection{Transformation \eqref{eqn.SHTC.w}$ \rightarrow 
$\eqref{eqn.lagr.masterU.vector}:}

Denoting $ \omega_i = \varepsilon_{ijk}\pd_j w_k $, equation \eqref{eqn.SHTC.w} 
reads as
\[\frac{\pd  w_k}{\pd t}+\frac{\pd  \left(v_{\alpha }w_{\alpha 
}+\ted_\sigma\right)}{\pd x_k}+\varepsilon _{k l j}\omega_l v_j=\frac{\pd
 w_k}{\pd t}+v_{\alpha }\frac{\pd  w_{\alpha }}{\pd 
 x_k}+w_{\alpha }\frac{\pd  v_{\alpha }}{\pd x_k}+\frac{\pd  
 \ted_\sigma}{\pd
x_k}+\varepsilon _{k l j}\omega_l v_j=\frac{\pd  w_k}{\pd t}+v_{\alpha 
}\frac{\pd  w_{\alpha }}{\pd x_k}+w_{\alpha }\frac{\pd
 v_{\alpha }}{\pd x_k}+\frac{\pd  \ted_\sigma}{\pd x_k}+\] 
 \[+v_{\alpha 
 }\left(\frac{\pd w_k}{\pd x_{\alpha }}-\frac{\pd w_{\alpha
}}{\pd x_k}\right)=\frac{\pd  w_k}{\pd t}+v_{\alpha 
}\left(\frac{\pd  w_k}{\pd x_{\alpha }}\right)+w_{\alpha 
}\frac{\pd
 v_{\alpha }}{\pd x_k}+\frac{\pd  \ted_\sigma}{\pd x_k}=\frac{{\rm d}w_k}{d 
 t}+w_{\alpha }\frac{\pd  v_{\alpha }}{\pd x_k}+\frac{\pd
\ted_\sigma}{\pd x_k}=\frac{{\rm d}w_k}{{\rm d} t}+A_{j k}\left(w_{\alpha 
}\frac{\pd  
v_{\alpha }}{\pd y_j}+\frac{\pd  \ted_\sigma}{\pd y_j}\right)\] 
The intermediate result is
\[
\frac{{\rm d}w_k}{{\rm d} t}+A_{j k}\left(w_{\alpha }\frac{\pd v_{\alpha }}{\pd 
y_j}+\frac{\pd \ted_\sigma}{\pd y_j}\right)=0.\]

By multiplying the last equation  by $ \FF^\transpose $, we obtain
\[
F_{k i}\frac{{\rm d}w_k}{{\rm d} t}+F_{k i}A_{j k}\left(w_{\alpha }\frac{\pd 
v_{\alpha }}{\pd y_j}+\frac{\pd \ted_\sigma}{\pd y_j}\right)=0.\]
Further, using the identity $A_{j k} F_{k i}=\delta_{ji}$ we can rewrite the 
last 
equations
as 
\[F_{k i}\frac{{\rm d}w_k}{{\rm d} t}+F_{k i}A_{j k}\left(w_{\alpha }\frac{\pd 
v_{\alpha }}{\pd y_j}+\frac{\pd \ted_\sigma}{\pd y_j}\right)=F_{k
i}\frac{{\rm d}w_k}{{\rm d} t}+\left(w_{\alpha }\frac{\pd v_{\alpha }}{\pd 
y_i}+\frac{\pd \ted_\sigma}{\pd y_i}\right)=0.\] 
Then using~\eqref{eq.gradient.time}, we get
\[F_{k i}\frac{{\rm d}
w_k}{{\rm d} t}+\left(w_{\alpha }\frac{\pd v_{\alpha }}{\pd 
y_i}+\frac{\pd
\ted_\sigma}{\pd y_i}\right)=F_{k i}\frac{{\rm d}w_k}{{\rm d} 
t}+\left(w_{\alpha 
}\frac{{\rm d}
F_{\alpha  i}}{{\rm d} t}+\frac{\pd \ted_\sigma}{\pd y_i}\right)=\frac{{\rm d}
(F_{k i} w_k)}{{\rm d} t}+\frac{\pd \ted_\sigma}{\pd y_i}=0\]

Then using that $ U_{\sigma'} = (w \ted)_{w \sigma'} =\ted_\sigma  $, we get 
the 
final 
result, i.e. sought Lagrangian equation~\eqref{eqn.lagr.masterU.vector},
\begin{equation}\label{eqn.w.final}
\frac{{\rm d} w_k'}{{\rm d} t}+\frac{\pd U_{\sigma'}}{\pd y_k}=0.
\end{equation}
Therefore, equations \eqref{eqn.scalar.final} and \eqref{eqn.w.final} suggest 
that the Lagrangian fields $ \ww' $, $ \sigma' $ and Eulerian $ \ww $, $ \sigma 
$ are 
related by 
\eqref{eqn.w}.

\subsection{$ (e,h) $-pair, transformation of the nonlinear Maxwell equations} 
\label{sec.app.elmag}

We shall transform \eqref{eqn.SHTC.Efield} into \eqref{eqn.lagr.masterU.Electr} 
while \eqref{eqn.SHTC.Hfield} transforms into \eqref{eqn.lagr.masterU.Magn} 
analogously. Thus, \eqref{eqn.SHTC.Efield} is equivalent to
\begin{equation}
\frac{\pd  e_i}{\pd  t}+v_k\frac{\pd e_i}{\pd  
x_k}+\frac{\pd v_k}{\pd  x_k} e_i-\frac{\pd
v_i}{\pd  x_k}e_k-\varepsilon _{{ikl}}\frac{\pd \ted_{h_l}}{\pd  
x_k}=0.
\end{equation}
Using \(\rmd/{\rmd t}=\pd /\pd t+v_k /\pd x_k\) we have

\begin{equation}
\frac{\rmd e_i}{{\rmd t}}+\frac{\pd v_k}{\pd  x_k} e_i-\frac{\pd 
v_i}{\pd  x_k}e_k-\varepsilon
_{{ikl}}\frac{\pd \ted_{h_l}}{\pd  x_k}=0.
\end{equation}
From \eqref{eq.gradient} it follows that $\frac{\pd}{\pd 
x_k}=A_{{jk}}\frac{\pd}{\pd y_j}$ and thus we change the variables $ x_k $ on $ 
y_j $

\begin{equation}
\frac{\rmd e_i}{{\rmd t}}+A_{{jk}}\frac{\pd v_k}{\pd  y_j} 
e_i-A_{{jk}}\frac{\pd v_i}{\pd  
y_j}e_k-\varepsilon_{{ikl}}A_{{jk}}\frac{\pd \ted_{h_l}}{\pd  y_j}=0.
\end{equation}
Applying (\ref{eq.DetTimeEvol})$_2$ to the second term and 
\eqref{eq.gradient.time} to the third term of the last equation, we have
\begin{equation}
\frac{\rmd e_i}{{\rmd t}}+\frac{1}{w}\frac{{\rmd w}}{{\rmd t}} 
e_i-A_{{jk}}\frac{\rmd F_{ij}}{\rmd 
t}e_k-\varepsilon_{{ikl}}A_{{jk}}\frac{\pd \ted_{h_l}}{\pd  y_j}=0,
\end{equation}
and then
\begin{equation}
\frac{\rmd  w e_i}{{\rmd t}}-w e_k A_{{jk}}\frac{\rmd F_{ij}}{\rmd t}-w 
\varepsilon _{{ikl}}A_{{jk}}\frac{\pd \ted_{h_l}}{\pd  y_j}=0.
\end{equation}

Now, we add \(0\equiv w e_k \dfrac{\rmd\delta_{ik}}{\rmd t}\) to the left hand 
side and then using \(\delta _{ik}=F_{ij}A_{jk}\) one can obtain that
\begin{equation}
\frac{\rmd w e_i}{{\rmd t}}+w e_k\frac{\rmd \delta_{ik}}{ \rmd t}-w e_k 
A_{{jk}}\frac{{\rmd F}_{{ij}}}{{\rmd t}}-w
\varepsilon _{{ikl}}A_{{jk}}\frac{\pd \ted_{h_l}}{\pd  y_j}=0,
\end{equation}
\begin{equation}
\frac{\rmd w e_i}{{\rmd t}}+w e_k\frac{\rmd F_{{ij}}A_{{jk}}}{{\rmd t}}-w e_k 
A_{{jk}}\frac{{\rmd F}_{{ij}}}{{\rmd t}}-w \varepsilon 
_{{ikl}}A_{{jk}}\frac{\pd \ted_{h_l}}{\pd  y_j}=0,
\end{equation}
\begin{equation}
\frac{\rmd w e_i}{{\rmd t}}+w e_kF_{{ij}}\frac{\rmd A_{{jk}}}{{\rmd t}}+w 
e_k\frac{\rmd F_{{ij}}}{{\rmd t}}A_{{jk}}-w e_k A_{{jk}}\frac{{\rmd 
F}_{{ij}}}{{\rmd t}}-w \varepsilon _{{ikl}}A_{{jk}}\frac{\pd 
\ted_{h_l}}{\pd  y_j}=0.
\end{equation}
After multiplying the last equation by $ A_{mi} $
\begin{equation}
A_{{mi}}\left(\frac{\rmd w e_i}{{\rmd t}}+w e_kF_{{ij}}\frac{\rmd 
A_{{jk}}}{{\rmd t}}-w \varepsilon _{{ikl}}A_{{jk}}\frac{\pd 
\ted_{h_l}}{\pd  y_j}\right)=0,
\end{equation}
\begin{equation}
A_{{mi}}\frac{\rmd w e_i}{{\rmd t}}+w e_k\frac{\rmd A_{{mk}}}{{\rmd t}}-w 
\varepsilon _{{ikl}}A_{{mi}}A_{{jk}}\frac{\pd \ted_{h_l}}{\pd  y_j}=0,
\end{equation}
we have an intermediate result:
\begin{equation}\label{eq.inter.result}
\frac{\rmd w A_{{mk}} e_k}{{\rmd t}}-w \varepsilon 
_{{ikl}}A_{{mi}}A_{{jk}}\frac{\pd \ted_{h_l}}{\pd
 y_j}=0.
\end{equation}

Now, we introduce the change of unknowns~\eqref{eqn.eh}: $w 
A_{{mk}}e_k=e'_m$, $w A_{{mk}}h_k=h'_m$, and we also change the energy 
potential $\ted(e_i,h_i)= 
\ted(w^{-1}F_{{ij}}e'_j,w^{-1}F_{{ij}}h'_j)=w^{-1}U(e'_j,h'_j)$. Hence, 
$ \ted_{h_i} = A_{ji}U_{h'_j}$. After this, the intermediate 
result~\eqref{eq.inter.result} reads as
\begin{equation}
\frac{\rmd e'_m}{{\rmd t}}-w \varepsilon 
_{{ikl}}A_{{mi}}A_{{jk}}\frac{\pd   A_{{al}} U_{h'_a}}{\pd  y_j}=0,
\end{equation}
\begin{equation}
\frac{\rmd e'_m}{{\rmd t}}-w \varepsilon 
_{{ikl}}A_{{mi}}A_{{jk}}A_{{al}}\frac{\pd   U_{h'_a}}{\pd  y_j}-w 
\varepsilon
_{{ikl}}A_{{mi}}A_{{jk}}\frac{\pd  A_{{al}}}{\pd  y_j}   U_{h'_a}=0.
\end{equation}
Applying (\ref{eq.DetLeviCivita}) and \eqref{eq.CofDefinition}$_3 $ to the 
second term, we get
\begin{equation}
\frac{\rmd e'_m}{{\rmd t}} - \varepsilon _{{mja}}\frac{\pd   
U_{h'_a}}{\pd  y_j}-w \varepsilon _{{ikl}}A_{{mi}}A_{{jk}}\frac{\pd
A_{{al}}}{\pd  y_j}   U_{h'_a}=0,
\end{equation}
Now using the cofactor definition (\ref{eq.CofDefinition}) and then applying 
(\ref{eq.CofFormulaLeviCivita}) to the third term, we have
\begin{equation}
\frac{\rmd e'_m}{{\rmd t}} - \varepsilon _{{mja}}\frac{\pd   
U_{h'_a}}{\pd  y_j}-\varepsilon _{{ikl}}A_{{mi}}C_{{kj}}\frac{\pd
A_{{al}}}{\pd  y_j}   U_{h'_a}=0,
\end{equation}
\begin{equation}
\frac{\rmd e'_m}{{\rmd t}} - \varepsilon _{{mja}}\frac{\pd   
U_{h'_a}}{\pd  y_j}-\frac{1}{2}\varepsilon _{{ikl}}\varepsilon 
_{{lnk}}\varepsilon _{{pqj}}F_{{lp}} F_{{nq}}A_{{mi}}\frac{\pd 
A_{{al}}}{\pd  y_j}   U_{h'_a}=0.
\end{equation}
Using (\ref{eq.LeviCivitaKronecker}) in the third term gives us
\begin{equation}
\frac{\rmd e'_m}{{\rmd t}} - \varepsilon _{{mja}}\frac{\pd   
U_{h'_a}}{\pd  y_j}-\delta _{{in}}\varepsilon _{{pqj}}F_{{lp}} 
F_{{nq}}A_{{mi}}\frac{\pd A_{{al}}}{\pd  y_j}   U_{h'_a}=0,
\end{equation}
and subsequently,
\begin{equation}
\frac{\rmd e'_m}{{\rmd t}} - \varepsilon _{{mja}}\frac{\pd   
U_{h'_a}}{\pd  y_j}-\varepsilon _{{pqj}}F_{{lp}} 
F_{{iq}}A_{{mi}}\frac{\pd A_{{al}}}{\pd  y_j}   U_{h'_a}=0,
\end{equation}
\begin{equation}
\frac{\rmd e'_m}{{\rmd t}} - \varepsilon _{{mja}}\frac{\pd   
U_{h'_a}}{\pd  y_j}-\varepsilon _{{mjp}}F_{{lp}}\frac{\pd 
A_{{al}}}{\pd  y_j}   U_{h'_a}=0.
\end{equation}

Now, adding $0\equiv \varepsilon _{{mjp}}\frac{\pd  F_{{lp}}}{\pd 
y_j} A_{{al}}   U_{h'_a}$ (see \eqref{eq.DivCof}$_2$), we get
\begin{equation}
\frac{\rmd e'_m}{{\rmd t}} - \varepsilon _{{mja}}\frac{\pd   
U_{h'_a}}{\pd  y_j}-\varepsilon _{{mjp}}F_{{lp}}\frac{\pd 
A_{{al}}}{\pd  y_j} U_{h'_a}-\varepsilon _{{mjp}}\frac{\pd  
F_{{lp}}}{\pd y_j}A_{{al}}  U_{h'_a}=0,
\end{equation}
\begin{equation}
\frac{\rmd e'_m}{{\rmd t}} - \varepsilon _{{mja}}\frac{\pd   
U_{h'_a}}{\pd  y_j} -   U_{h'_a}\left(\varepsilon _{{mjp}}\frac{\pd
A_{{al}}F_{{lp}}}{\pd  y_j}\right)=0,
\end{equation}
\begin{equation}
\frac{\rmd e'_m}{{\rmd t}} - \varepsilon _{{mja}}\frac{\pd   
U_{h'_a}}{\pd  y_j} -   U_{h'_a}\left(\varepsilon _{{mjp}}\frac{\pd 
\delta _{{ap}}}{\pd  y_j}\right)=0.
\end{equation}
Eventually, we have 
\begin{equation}
\frac{\rmd e'_m}{{\rmd t}} - \varepsilon _{{mja}}\frac{\pd    
U_{h'_a}}{\pd  y_j}=0,
\end{equation}
which is identical to \eqref{eqn.lagr.masterU.Electr}.

\subsection{$ (m,F) $-pair}

\subsubsection{Distortion matrix}
Here, we derive Eulerian equation~\eqref{eqn.SHTC.A} from its Lagrangian 
form~\eqref{eqn.lagr.masterU.Momentum}.

Recall that the velocity field $ v_i $ is conjugate to the momentum field, 
i.e. $ v_i=U_{m_i} $. Thus, \eqref{eqn.lagr.masterU.F} reads as
\begin{equation}
\frac{\rmd F_{ij}}{\rmd t} - \frac{\pd v_i}{\pd y_j}=0.
\end{equation}
Then, applying \eqref{eq.gradient}$ _1 $ and \eqref{eq.gradient}$ _5 $ we 
arrive 
at
\begin{equation}\label{eqn.F2}
\frac{\pd F_{ij}}{\pd t} + v_k\frac{\pd F_{ij}}{\pd x_k}-F_{kj}\frac{\pd 
v_i}{\pd x_k}=0.
\end{equation}
Using the identity
\begin{equation}\label{eqn.F3}
\bm{0}\equiv{\rm d}\II\equiv{\rm d}(\FF\AAA)\equiv\FF{\rm d}\AAA + ({\rm 
d}\FF)\AAA\,,
\end{equation}
\eqref{eqn.F2} can be rewritten as
\begin{equation}\label{eqn.F4}
\dfrac{\pd A_{ij}}{\pd t} + v_k\dfrac{\pd A_{ij}}{\pd x_k} +A_{ik}\dfrac{\pd 
v_k}{\pd x_j}=0,
\end{equation}
and eventually as
\begin{equation}\label{eqn.F5}
\frac{\pd A_{i k}}{\pd t}+\frac{\pd (A_{im} v_m)}{\pd 
x_k}+v_j\left(\frac{\pd A_{ik}}{\pd x_j}-\frac{\pd 
A_{ij}}{\pd x_k}\right) = 0,
\end{equation}
which is the sought equation~\eqref{eqn.SHTC.A}.

\subsubsection{Transformation of the momentum conservation} 
\label{app.EulertoLagr.momentum}

In this section, we demonstrate that Eulerian momentum conservation 
law~\eqref{eqn.SHTC.momentum}  can be derived from its Lagrangian 
counterpart~\eqref{eqn.lagr.masterU.Momentum}. As earlier, we use prime to 
denote Lagrangian fields. Now, 
\eqref{eqn.lagr.masterU.Momentum} reads as
\begin{equation}\label{app.momentum}
\frac{{\rm d}m_i'}{{\rm d}t}-\frac{\pd U_{F_{ij}}}{\pd y_j}=0.
\end{equation}
Using that $ {\rm d}/{\rm d}t = \pd /\pd t + v_k\pd/\pd x_k $ and $ 
F_{ij}=\pd x_i/\pd y_j $, equation~\eqref{app.momentum} can be rewritten as
\begin{equation}\label{app.momentum2}
\frac{\pd m_i'}{\pd t} + v_k\frac{\pd m_i'}{\pd x_k} - F_{kj}\frac{\pd 
U_{F_{ij}}}{\pd x_k}=0.
\end{equation}
Subsequently, using the Eulerian stationary constraint for $ F_{ij} $ and time 
evolution of $ w=\det(\FF) $~\eqref{eqn.constF.euler}:
\begin{equation}
\frac{\pd w^{-1}F_{kj}}{\pd x_k}=0,\ \ \ \ \frac{\pd }{\pd t}\left ( 
\frac{1}{w}\right ) + \frac{\pd }{\pd x_k}\left( 
\frac{v_k}{w}\right) =0,
\end{equation}
\eqref{app.momentum2} can be rewritten as
\begin{equation}\label{app.momentum3}
\frac{\pd }{\pd t}\left(\frac{m_i'}{w}\right)  + \frac{\pd }{\pd 
x_k}\left(\frac{v_k m_i' - F_{kj}U_{F_{ij}}}{w}\right) =0.
\end{equation}

We now introduce change of the potential 
\[ U = w \, \scU ,\]
and change of the momentum~\eqref{eqn.m},  $ m_i=w^{-1}m_i' $. Note that $ \scU 
$ equals to $ \ted 
$ but depends on the Lagrangian fields, i.e.
\begin{equation}\label{app.scU.mcE}
\scU(F_{ij},m_i',e_i',h_i',\sigma',w_i') = 
\ted(\rho,F_{ij},m_i,e_i,h_i,\sigma,w_i).
\end{equation}
After that, we arrive at
\begin{equation}\label{app.momentum4}
\frac{\pd m_i}{\pd t}  + \frac{\pd }{\pd 
x_k}\left(v_k m_i - \scU \delta_{ki} - F_{kj}\,\scU_{F_{ij}}\right) =0.
\end{equation}

Finally, it remains to transform the term $  
F_{kj}\,\scU_{F_{ij}} $ in the flux of 
equation~\eqref{app.momentum4}. Thus, according to~\eqref{app.scU.mcE} and 
change of the variables \eqref{eqn.m}, \eqref{eqn.eh} and \eqref{eqn.w}: $ m'_a 
= w m_a $, $e'_b = w A_{{ba}}e_a $, $h'_b = w A_{{ba}}h_a$, $\sigma' = w \sigma 
$ and $ w_b' = F_{ab}w_a $ , we have
\begin{equation}
\scU_{F_{ij}} = \frac{\pd}{\pd F_{ij}}\left( \ted(\rho,F_{ab},w^{-1} m'_a, 
w^{-1}F_{ab}e'_b, 
w^{-1}F_{ab}h'_b,
w^{-1}\sigma',A_{ba} w'_b) \right)\ \ ,
\end{equation}
and hence
\begin{equation}
F_{kj}\scU_{F_{ij}}=F_{kj}\left(-\rho  
A_{ji}\ted_{\rho 
}+\ted_{F_{ij}}+\ted_{m_a}(w^{-1}m'_a)_{F_{ij}} + 
\ted_{e_a}(w^{-1}F_{ab}e'_b 
)_{F_{ij}} + \ted_{\sigma }(w^{-1}\sigma' )_{F_{ij}} + 
\ted_{w_a}\left(A_{ba}w_b'\right)_{F_{ij}}\right).
\end{equation}
After using the formulas 
\begin{equation}
\frac{\pd w}{\pd F_{ij}} = w A_{ji},\qquad \frac{\pd A_{ab}}{\pd F_{cd}} = 
-A_{ac}A_{db}, 
\end{equation}
and doing some trivial algebra, we arrive at
\begin{equation}\label{eqn.momentum.stress}
 F_{kj}\,\scU_{F_{ij}} = -(\rho\ted_\rho + m_l\ted_{m_l} + \sigma\ted_\sigma + 
 e_l\ted_{e_l} + 
 h_l\ted_{h_l})\delta_{ki} + e_k\ted_{e_i} + h_k\ted_{h_i} - w_i\ted_{w_k} + 
 F_{kj}\ted_{F_{ij}}.
\end{equation}
Eventually, using that $ F_{kj}\ted_{F_{ij}} =-A_{jk}\ted_{A_{ji}} $, the 
momentum equation becomes
\begin{equation}\label{app.momentum5}
\frac{\pd m_i}{\pd t}  + \frac{\pd }{\pd 
x_k}\left(m_i \ted_{m_k} + (\rho\ted_\rho + m_l\ted_{m_l} + 
e_l\ted_{e_l} + 
 h_l\ted_{h_l} + \sigma\ted_\sigma - \ted)\delta_{ki} - e_k\ted_{e_i} - 
 h_k\ted_{h_i} + 
 w_i\ted_{w_k} + 
 A_{jk}\ted_{A_{ji}}\right) =0.
\end{equation}

\section{Derivation of the energy conservation law}\label{sec.energycons}
In this section, we prove that the energy conservation law~\eqref{eqn.energyL}
\begin{multline*}
\frac{\pd }{\pd t}(r L_r + v_i L_{v_i} + \alpha_{ij} L_{\alpha _{ij}} + d_i 
L_{d_i} + b_i L_{b_i} + \theta L_\theta + \eta_k L_{\eta_k} - L) + \\
\frac{\pd }{\pd x_k}\left(v_k\left(r L_r + v_i L_{v_i} + \alpha_{ij} 
L_{\alpha _{ij}} + d_i L_{d_i} + b_i L_{b_i} + \theta L_\theta + \eta_k 
L_{\eta_k} - L\right) + \right.\\
\left. v_i\left[\left(L - \alpha_{ab} L_{\alpha_{ab}} - \eta_a L_{\eta_a} 
\right) \delta_{ik} + \alpha_{nk} L_{\alpha_{ni}} - d_i L_{d_k} - b_i 
L_{b_k} + \eta_k L_{\eta_i}\right] + \varepsilon_{kij} 
d_i b_j +  \theta \eta_k \right)=0.
\end{multline*}
is compatible with the governing equations~\eqref{eqn.MasterEulerL}, that 
is it can be obtained as the consequence of all the 
equations~\eqref{eqn.MasterEulerL} multiplied by the corresponding conjugate 
state variables~\eqref{eqn.p}. 
From this prove, it will be clear that the 
energy flux, in fact, is not the consequence of only the fluxes in 
\eqref{eqn.MasterEulerL}, but the non-conservative terms in 
equations~\eqref{eqn.MasterEulerL.F}, \eqref{eqn.MasterEulerL.Electr},  
\eqref{eqn.MasterEulerL.Magn} and \eqref{eqn.MasterEulerL.Heat} contribute into 
the energy flux as well.

It is obvious that the time derivative of the energy conservation can be 
obtained as the sum of the time derivatives of the 
equations~\eqref{eqn.MasterEulerL} multiplied by the corresponding conjugate 
fields, i.e.
\begin{equation}
\frac{\pd (p_i L_{p_i} - L)}{\pd t} = p_i \frac{\pd L_{p_i}}{\pd t},
\end{equation}
while the main challenge is to show that the energy flux is 
\begin{multline}\label{eqn.appB.eq1}
\text{flux}\eqref{eqn.energyL} = r\cdot 
\text{flux}\eqref{eqn.MasterEulerL.Mass} + 
v_i\cdot\text{flux}\eqref{eqn.MasterEulerL.Momentum} + 
\alpha_{ik}\cdot\left (\text{flux}\eqref{eqn.MasterEulerL.F} + v_j(\pd_j 
L_{\alpha_{ik}} - \pd_k L_{\alpha_{ij}})\right ) + \\
d_i\cdot\left (\text{flux}\eqref{eqn.MasterEulerL.Electr} + v_i\pd_k 
L_{d_k} \right ) + 
b_i\cdot\left (\text{flux}\eqref{eqn.MasterEulerL.Magn} + v_i\pd_k 
L_{b_k} \right ) +\\ 
\theta\cdot \text{flux}\eqref{eqn.MasterEulerL.Entropy} +
\eta_k\cdot\left (\text{flux}\eqref{eqn.MasterEulerL.Heat} + v_j(\pd_j 
L_{\eta_k} - \pd_k L_{\eta_j})\right ).
\end{multline}
The right-hand side of \eqref{eqn.appB.eq1} is 
\begin{gather}\label{eqn.appB.eq2}
r\pd_k(v_k L_r) + v_i\pd_k \left(\delta_{ik} L + v_k L_{v_i}  + 
	\alpha_{mk}L_{\alpha_{mi}} - \delta_{ik}\alpha_{mn}L_{\alpha_{mn}} 
	- d_i L_{d_k} - b_i L_{b_k} + 
	\eta_{k}L_{\eta_{i}} - \delta_{ik}\eta_n L_{\eta_n}\right) + 
	\nonumber\\[1mm]
\alpha_{ik}\left (\pd_k  (v_m L_{\alpha_{im}}) + 
	v_j(\pd_j L_{\alpha_{ik}} - \pd_k L_{\alpha_{ij}})\right ) + 
	\nonumber\\[1mm]
d_i\left (\pd_k( v_k L_{d_i} - v_i L_{d_k} - \varepsilon_{ikl}b_l) + v_i \pd_k 
L_{d_k} \right ) +
b_i\left( \pd_k( v_k L_{b_i} - v_i L_{b_k} + \varepsilon_{ikl}d_l) + v_i \pd_k 
L_{b_k} \right) + \nonumber\\[1mm]
\theta \pd_k(v_k L_\theta + \eta_k) + 
\eta_k\left(\pd_k (v_i L_{\eta_i} + \theta) + v_j(\pd_j 
L_{\eta_k} - \pd_k L_{\eta_j})\right )  = 
\end{gather}
\begin{gather}\label{eqn.appB.eq3}
r\pd_k(v_k L_r) + v_i\pd_k ( v_k L_{v_i}) + \alpha_{mi} \pd_k(v_k 
L_{\alpha_{mi}}) + d_i \pd_k(v_k L_{d_i}) + b_i \pd_k(v_k L_{b_i}) + 
\theta\pd_k(v_k L_\theta) + \eta_i\pd_k(v_k L_{\eta_i}) - \nonumber \\
v_i \pd_k \left(\delta_{ik}\alpha_{mn}L_{\alpha_{mn}} - 
\alpha_{mk}L_{\alpha_{mi}} + d_i L_{d_k} + b_i L_{b_k} + \delta_{ik} \eta_n 
L_{\eta_n} - \eta_k L_{\eta_i} \right)  - \nonumber\\[1mm]
\left(\delta_{ik}\alpha_{mn}L_{\alpha_{mn}} - 
\alpha_{mk}L_{\alpha_{mi}} + d_i L_{d_k} + b_i L_{b_k} + \delta_{ik} \eta_n 
L_{\eta_n} - \eta_k L_{\eta_i} \right)\pd_k v_i + \nonumber\\[1mm]
v_k \pd_k L - \eps_{ikl}d_i\pd_k b_l + \eps_{ikl}b_i\pd_k d_l +
\theta\pd_k\eta_k + \eta_k\pd_k\theta = 
\end{gather}
\begin{gather}\label{eqn.appB.eq4}
\pd_k\left(v_k(r L_r + v_i L_{v_i} + \alpha_{mi}L_{\alpha_{mi}} + d_i L_{d_i} + 
b_i L_{b_i} + \theta L_\theta + \eta_k L_{\eta_k})\right) - \nonumber\\[1mm]
v_k\left(L_r\pd_k r + L_{v_i}\pd_k v_i + L_{\alpha_{mi}}\pd_k \alpha_{mi} +  
L_{d_i} \pd_k d_i + L_{b_i} \pd_k b_i  + L_\theta \pd_k \theta + L_{\eta_k}
\pd_k \eta_k \right) + v_k\pd_k L - \nonumber\\[1mm]
\pd_k \left(v_i (\delta_{ik}\alpha_{mn}L_{\alpha_{mn}} - 
\alpha_{mk}L_{\alpha_{mi}} + d_i L_{d_k} + b_i L_{b_k} + \delta_{ik} \eta_n 
L_{\eta_n} - \eta_k L_{\eta_i}) \right) + \pd_k (\eps_{kij}d_i b_j + \theta 
\eta_k),
\end{gather}
which eventually transforms exactly into the energy flux~\eqref{eqn.energyL}.

\section{Derivation of the time evolution for $ \nabla\times w $ and $ 
\nabla\times A $}\label{app.evol.rotw}

Here, we derive the time evolutions \eqref{eqn.invol.constr.rotA} 
and 
\eqref{eqn.invol.constr.rotw} for $ \nabla\times\AAA $ and $ \nabla\times\ww $ 
accordingly. In fact, because the time evolution for each column $ A_{ik} $, $ 
k=1,2,3 $ has the same structure as the time evolution for $ w_i $ (apart of 
the constitutive flux $ \ted_\sigma $) we can perform the derivation only for $ 
w_i $.

We start by rewriting \eqref{eqn.SHTC.w} (we also assume the presence of an 
algebraic source term $ -f_k $ in the right-hand side)
\begin{equation}\label{eqn.inv.eq1}
	\frac{\pd w_k}{\pd t} + \pd_k \left(v_a w_a 
	+ \ted_{\sigma}\right) + v_j(\pd_j w_k - \pd_k w_j) = -f_k
\end{equation}
in an equivalent form
\begin{equation*}
	\frac{\pd w_k}{\pd t} + w_a \pd_k v_a + \pd_k \ted_{\sigma} + v_a \pd_a w_k 
	= -f_k.
\end{equation*}
Then, applying $ \pd_m $ to the $ k $-th equation and $ \pd_k $ to the $ m $-th 
equation
and subtracting the later from the former we get
\begin{equation}\label{eqn.inv.eq2}
\frac{\pd }{\pd t}(\pd _mw_k-\pd _kw_m)+v_a\pd _a(\pd _mw_k-\pd _kw_m)+\pd 
_mw_a\pd
_kv_a-\pd _kw_a\pd _mv_a-\pd _aw_m\pd _kv_a+\pd _aw_k \pd _mv_a 
=-(\pd_m f_k - \pd_k f_m).
\end{equation}
This equation can be written in an elegant form if all the derivatives of the 
components of $ w_i $ are arranged in such a way that the resulting equation 
contains only the components of the vector $ \bm{\Omega} = \nabla\times\ww $. 
Indeed, it suffices to explain how to handle with last four terms in the 
left-hand side of 
\eqref{eqn.inv.eq2} for some  $ m $ and $ k $. We consider only the case $ m = 
1 $, $ k = 2 $ and 
write explicitly the summation with respect to $ a $:
\begin{gather}
\pd _1w_a\pd _2v_a+\pd _aw_2 \pd _1v_a-\pd _2w_a\pd _1v_a-\pd _aw_1\pd 
_2v_a=\nonumber\\
\pd _1w_1\pd _2v_1+\pd _1w_2 \pd _1v_1-\pd _2w_1\pd _1v_1-\pd _1w_1\pd 
_2v_1+\pd _1w_2\pd _2v_2+\pd _2w_2 \pd _1v_2-\nonumber\\
\pd _2w_2\pd _1v_2-\pd _2w_1\pd _2v_2+\pd _1w_3\pd _2v_3+\pd _3w_2 \pd 
_1v_3-\pd _2w_3\pd _1v_3-\pd_3w_1\pd _2v_3=\nonumber\\
\left(\pd _1w_2-\pd _2w_1\right)\pd _1v_1+\left(\pd _3w_2-\pd _2w_3\right)\pd 
_1v_3+\left(\pd _1w_3-\pd _3w_1\right)\pd _2v_3+\left(\pd _1w_2-\pd 
_2w_1\right)\pd _2v_2=\\
\left(\pd _1w_2-\pd _2w_1\right)\left(\pd _1v_1+\pd _2v_2\right)+\left(\pd 
_3w_2-\pd _2w_3\right)\pd _1v_3+\left(\pd_1w_3-\pd _3w_1\right)\pd 
_2v_3=\nonumber\\
\Omega _3\left(\pd _1v_1+\pd _2v_2\right)-\Omega _1\pd 
_1v_3-\Omega _2\pd _2v_3=\Omega _3\left(\pd _1v_1+\pd 
_2v_2+\pd
_3v_3\right)-\Omega _1\pd _1v_3-\Omega _2\pd _2v_3-\Omega _3\pd 
_3v_3=\nonumber\\
\Omega _3\pd _av_a-\Omega _a\pd _av_3\nonumber.
\end{gather}

Finally, we conclude that 
\begin{equation}\label{eqn.inv.eq23}
\frac{\pd \Omega _i}{\pd t} + v_k\pd _k\Omega _i+\Omega _i\pd 
_k v_k - \Omega_k\pd_k v_i + \eps_{i k l} \pd_kf_l = 0 
\end{equation}
or equivalently
\begin{equation}\label{eqn.inv.eq3}
\frac{\pd \Omega _i}{\pd t} + \pd_k\left(\Omega_i v_k - \Omega_kv_i + \eps_{i k 
l} f_l \right) + v_i\pd_k\Omega_k = 0,
\end{equation}
which apparently has the same structure as equation~\eqref{eqn.SHTC.Hfield} of 
the Eulerian SHTC system and the sought equation~\eqref{eqn.invol.constr.O} but 
with an additional term in the flux conditioned by the presence of the source 
term $ f_k $ in \eqref{eqn.inv.eq1}.

Eventually we make an important remark. By construction, $ \pd_k\omega_k = 0 $ 
because $ \bm{\Omega} = \nabla\times\ww $ and hence, one may want to write 
\eqref{eqn.inv.eq3} in a conservative form by eliminating the 
non-conservative term $ v_i\pd_k\Omega_k = 0 $. This however should be avoided 
because the obtained conservative equation 
\begin{equation}\label{eqn.inv.eq4}
\frac{\pd \Omega _i}{\pd t} + \pd_k\left(v_k\Omega _i-\Omega _kv_i + \eps_{i k 
l} f_l \right) = 0
\end{equation}
has different characteristic speeds, e.g. see~\cite{Powell1999}, and thus is 
not equivalent to the original equations~\eqref{eqn.inv.eq23} and 
\eqref{eqn.inv.eq3}. Moreover, in contrast to~\eqref{eqn.inv.eq23} and 
\eqref{eqn.inv.eq3}, conservative equation \eqref{eqn.inv.eq4} is not Galilean 
invariant.

\bibliographystyle{spmpsci}       
\bibliography{library}

\begin{thebibliography}{10}
\providecommand{\url}[1]{{#1}}
\providecommand{\urlprefix}{URL }
\expandafter\ifx\csname urlstyle\endcsname\relax
  \providecommand{\doi}[1]{DOI~\discretionary{}{}{}#1}\else
  \providecommand{\doi}{DOI~\discretionary{}{}{}\begingroup
  \urlstyle{rm}\Url}\fi

\bibitem{Abarbanel}
Abarbanel, H.D.I., Brown, R., Yang, Y.M.: {Hamiltonian formulation of inviscid
  flows with free boundaries}.
\newblock Physics of Fluids \textbf{31}, 2802 (1988)

\bibitem{Arnold1}
Arnold, V.I.: {Sur la g{\'{e}}ometrie diff{\'{e}}rentielle des groupes de Lie
  de dimension infini et ses applications dans l'hydrodynamique des fluides
  parfaits}.
\newblock Ann. Inst. Fourier \textbf{16}, 319 (1966)

\bibitem{Barton2013}
Barton, P.T., Deiterding, R., Meiron, D., Pullin, D.: {Eulerian adaptive
  finite-difference method for high-velocity impact and penetration problems}.
\newblock Journal of Computational Physics \textbf{240}, 76--99 (2013)

\bibitem{BartonRom2010}
Barton, P.T., Drikakis, D., Romenski, E.I.: {An Eulerian finite-volume scheme
  for large elastoplastic deformations in solids}.
\newblock International journal for numerical methods in engineering
  \textbf{81(4)}, 453--484 (2010)

\bibitem{Serre2007}
Benzoni-Gavage, S., Serre, D.: {Multidimensional Hyperbolic Partial
  Differential Equations}.
\newblock Oxford University Press, Oxford (2007)

\bibitem{Van-book}
Berezovski, A., V{\'{a}}n, P.: {Internal Variables in Thermoelasticity}.
\newblock Solid Mechanics and Its Applications. Springer International
  Publishing (2017).
\newblock \urlprefix\url{https://books.google.de/books?id=vv3MDgAAQBAJ}

\bibitem{Beris1994}
Beris, A.N., Edwards, B.J.: {Thermodynamics of Flowing Systems: With Internal
  Microstructure}.
\newblock Oxford University Press, USA (1994)

\bibitem{Bobylev1982}
Bobylev, A.: {The Chapman-Enskog and Grad methods for solving the Boltzmann
  equation}.
\newblock Akademiia Nauk SSSR Doklady \textbf{262}, 71--75 (1982)

\bibitem{Boillat1974}
Boillat, G.: {Sur l'existence et la recherche d'{\'{e}}quations de conservation
  suppl{\'{e}}ment aires pour les syst{\'{e}}mes hyperboliques}.
\newblock C. R. Acad. Sc. Paris, S{\'{e}}r A \textbf{278} (1974)

\bibitem{bolmatov2013thermodynamic}
Bolmatov, D., Brazhkin, V.V., Trachenko, K.: {Thermodynamic behaviour of
  supercritical matter}.
\newblock Nature communications \textbf{4} (2013)

\bibitem{Bolmatov2015a}
Bolmatov, D., Zav'yalov, D., Zhernenkov, M., Musaev, E.T., Cai, Y.Q.: {Unified
  phonon-based approach to the thermodynamics of solid, liquid and gas states}.
\newblock Annals of Physics \textbf{363}, 221--242 (2015).
\newblock \doi{10.1016/j.aop.2015.09.018}.
\newblock \urlprefix\url{http://dx.doi.org/10.1016/j.aop.2015.09.018}

\bibitem{bolmatov2015revealing}
Bolmatov, D., Zhernenkov, M., Zav'yalov, D., Stoupin, S., Cai, Y.Q., Cunsolo,
  A.: {Revealing the Mechanism of the Viscous-to-Elastic Crossover in Liquids}.
\newblock The journal of physical chemistry letters \textbf{6}(15), 3048--3053
  (2015)

\bibitem{Bolmatov2016}
Bolmatov, D., Zhernenkov, M., Zav'yalov, D., Stoupin, S., Cunsolo, A., Cai,
  Y.Q.: {Thermally triggered phononic gaps in liquids at THz scale}.
\newblock Scientific Reports \textbf{6}(November 2015), 19,469 (2016).
\newblock \doi{10.1038/srep19469}.
\newblock \urlprefix\url{http://www.nature.com/articles/srep19469}

\bibitem{Boscheri2016}
Boscheri, W., Dumbser, M., Loub{\`{e}}re, R.: {Cell centered direct
  Arbitrary-Lagrangian-Eulerian ADER-WENO finite volume schemes for nonlinear
  hyperelasticity}.
\newblock Computers {\&} Fluids \textbf{134-135}, 111--129 (2016).
\newblock \doi{10.1016/j.compfluid.2016.05.004}.
\newblock
  \urlprefix\url{http://linkinghub.elsevier.com/retrieve/pii/S004579301630144X}

\bibitem{brazhkin2012two}
Brazhkin, V.V., Fomin, Y.D., Lyapin, A.G., Ryzhov, V.N., Trachenko, K.: {Two
  liquid states of matter: A dynamic line on a phase diagram}.
\newblock Physical Review E \textbf{85}(3), 31,203 (2012)

\bibitem{Clebsch}
Clebsch, A.: {{\"{U}}ber die Integration der hydrodynamische Gleichungen}.
\newblock J. Reine Angew. Math. \textbf{56}(1) (1859)

\bibitem{DPRZ2016}
Dumbser, M., Peshkov, I., Romenski, E., Zanotti, O.: {High order ADER schemes
  for a unified first order hyperbolic formulation of continuum mechanics:
  Viscous heat-conducting fluids and elastic solids}.
\newblock Journal of Computational Physics \textbf{314}, 824--862 (2016).
\newblock \doi{10.1016/j.jcp.2016.02.015}.
\newblock
  \urlprefix\url{http://www.sciencedirect.com/science/article/pii/S0021999116000693}

\bibitem{DPRZ2017}
Dumbser, M., Peshkov, I., Romenski, E., Zanotti, O.: {High order ADER schemes
  for a unified first order hyperbolic formulation of Newtonian continuum
  mechanics coupled with electro-dynamics}.
\newblock Journal of Computational Physics \textbf{348}, 298--342 (2017).
\newblock \doi{10.1016/j.jcp.2017.07.020}.
\newblock \urlprefix\url{http://dx.doi.org/10.1016/j.jcp.2017.07.020
  http://www.sciencedirect.com/science/article/pii/S0021999117305284}

\bibitem{DupretMarchal1986}
Dupret, F., Marchal, J.: {Loss of evolution in the flow of viscoelastic
  fluids}.
\newblock Journal of Non-Newtonian Fluid Mechanics \textbf{20}, 143--171 (1986)

\bibitem{Dzyaloshinskii1980}
Dzyaloshinskii, I.E., Volovick, G.E.: {Poisson brackets in condensed matter
  physics}.
\newblock Annals of Physics \textbf{125}(1), 67--97 (1980).
\newblock \doi{10.1016/0003-4916(80)90119-0}

\bibitem{Romenski2002}
{E. I. Romenski}: {Thermodynamics and Balance Laws for Processes of Inelastic
  Deformations}.
\newblock In: Proceedings "WASCOM 2001" 11th Conference on Waves and Stability
  in Continuous Media, pp. 484--495. World Scientific (2002)

\bibitem{elmag}
Esen, O., Pavelka, M., Grmela, M.: {Hamiltonian coupling of electromagnetic
  field and matter}.
\newblock International Journal of Advances in Engineering Sciences and Applied
  Mathematics  (2017).
\newblock \doi{10.1007/s12572-017-0179-4}.
\newblock \urlprefix\url{http://link.springer.com/10.1007/s12572-017-0179-4}

\bibitem{Favrie2011}
Favrie, N., Gavrilyuk, S.: {Dynamics of shock waves in elastic-plastic solids}.
\newblock In: ESAIM proceedings, vol.~30, pp. 50--67 (2011).
\newblock \doi{10.1051/proc/201133005}.
\newblock
  \urlprefix\url{https://www.esaim-proc.org/component/makeref/?task=show{\&}type=html{\&}doi=10.1051/proc/201133005}

\bibitem{Favrie2017}
Favrie, N., Gavrilyuk, S.: {A rapid numerical method for solving
  Serre–Green–Naghdi equations describing long free surface gravity waves}.
\newblock Nonlinearity \textbf{30}(7), 2718--2736 (2017).
\newblock \doi{10.1088/1361-6544/aa712d}.
\newblock
  \urlprefix\url{http://stacks.iop.org/0951-7715/30/i=7/a=2718?key=crossref.139c98587b84970534e28823dcd579eb}

\bibitem{Fecko}
Fecko, M.: {Differential Geometry and Lie Groups for Physicists}.
\newblock Cambridge University Press (2006).
\newblock \urlprefix\url{https://books.google.de/books?id=vQR0mN1dgUEC}

\bibitem{Frenkel1955}
Frenkel, J.: {Kinetic theory of liquids}.
\newblock Dover (1955)

\bibitem{Friedrichs1958}
Friedrichs, K.O.: {Symmetric positive linear differential equations}.
\newblock Communications on Pure and Applied Mathematics1 \textbf{11}(3),
  333--418 (1958)

\bibitem{FriedLax1971}
Friedrichs, K.O., Lax, P.D.: {Systems of conservation equations with a convex
  extension}.
\newblock Proceedings of the National Academy of Sciences \textbf{68}(8),
  1686--1688 (1971)

\bibitem{Gavrilyuk}
Gavrilyuk, S.L., Makarenko, N.I., Sukhinin, S.V.: {Waves in Continuous Media}.
\newblock Lecture Notes in Geosystems Mathematics and Computing. Springer
  International Publishing (2017)

\bibitem{GodRom1996a}
Godunov, S., Mikhailova, T., Romenskii, E.: {Systems of thermodynamically
  coordinated laws of conservation invariant under rotations}.
\newblock Siberian Mathematical Journal \textbf{37}(4), 690--705 (1996)

\bibitem{GodPesh2010}
Godunov, S., Peshkov, I.: {Thermodynamically Consistent Nonlinear Model of
  Elastoplastic Maxwell Medium}.
\newblock Computational Mathematics and Mathematical Physics \textbf{50}(8),
  1409--1426 (2010).
\newblock \doi{10.1134/S0965542510080117}

\bibitem{GodRom1995}
Godunov, S., Romensky, E.: {Thermodynamics, conservation laws and symmetric
  forms of differential equations in mechanics of continuous media}.
\newblock In: Computational Fluid Dynamics Review 1995, vol.~95, pp. 19--31.
  John Wiley, NY (1995).
\newblock \doi{10.1142/7799}

\bibitem{Godunov1996}
Godunov, S., {Yu Mikhailova}, T., Romenskii, E.: {Systems of thermodynamically
  coordinated laws of conservation invariant under rotations}.
\newblock Siberian Mathematical Journal \textbf{37}(4), 790--806 (1996)

\bibitem{God1959}
Godunov, S.K.: {A difference method for numerical calculation of discontinuous
  solutions of the equations of hydrodynamics}.
\newblock Matematicheskii Sbornik \textbf{89}(3), 271--306 (1959)

\bibitem{God1961}
Godunov, S.K.: {An interesting class of quasilinear systems}.
\newblock Dokl. Akad. Nauk SSSR \textbf{139(3)}, 521--523 (1961)

\bibitem{God1962}
Godunov, S.K.: {The problem of a generalized solution in the theory of
  quasilinear equations and in gas dynamics}.
\newblock Russian Mathematical Surveys \textbf{17}(3), 145--156 (1962)

\bibitem{God1972MHD}
Godunov, S.K.: {Symmetric form of the magnetohydrodynamic equation}.
\newblock Numerical Methods for Mechanics of Continuum Medium \textbf{3}(1),
  26--34 (1972).
\newblock
  \urlprefix\url{http://citeseerx.ist.psu.edu/viewdoc/download?doi=10.1.1.55.9645{\&}rep=rep1{\&}type=pdf}

\bibitem{God1972}
Godunov, S.K.: {Symmetric form of the magnetohydrodynamic equation}.
\newblock Numerical Methods for Mechanics of Continuum Medium \textbf{3}(1),
  26--34 (1972).
\newblock
  \urlprefix\url{http://citeseerx.ist.psu.edu/viewdoc/download?doi=10.1.1.55.9645{\&}rep=rep1{\&}type=pdf}

\bibitem{God1978}
Godunov, S.K.: {Elements of mechanics of continuous media}, 1st russia edn.
\newblock Nauka (1978)

\bibitem{GodRom1972}
Godunov, S.K., Romenskii, E.I.: {Nonstationary equations of nonlinear
  elasticity theory in Eulerian coordinates}.
\newblock Journal of Applied Mechanics and Technical Physics \textbf{13(6)},
  868--884 (1972)

\bibitem{GodRom2003}
Godunov, S.K., Romenskii, E.I.: {Elements of continuum mechanics and
  conservation laws}.
\newblock Kluwer Academic/Plenum Publishers (2003)

\bibitem{GodRom1998}
Godunov, S.K., Romenskii, E.I.: {Elements of mechanics of continuous media}.
\newblock Kluwer Academic/Plenum Publishers, New York (2003)

\bibitem{GodRom1996}
Godunov, S.K., Romensky, E.I.: {Symmetric forms of thermodynamically compatible
  systems of conservation laws in continuum mechanics}.
\newblock In: ECCOMAS Conference on numerical methods in engineering, pp.
  54--57 (1996)

\bibitem{GC}
Grmela, M.: {Particle and Bracket Formulations of Kinetic Equations}.
\newblock Contemporary Mathematics \textbf{28}, 125--132 (1984)

\bibitem{GPhD}
Grmela, M.: {Bracket formulation of diffusion-convection equations}.
\newblock Phys. D \textbf{21}, 179--212 (1986)

\bibitem{Miroslav-PLA}
Grmela, M.: {A framework for elasto-plastic hydrodynamics}.
\newblock Physics letters A \textbf{312}, 134--146 (2003)

\bibitem{Grmela2012-PhysD}
Grmela, M.: {Fluctuations in extended mass-action-law dynamics}.
\newblock Physica D: Nonlinear Phenomena \textbf{241}(10), 976--986 (2012)

\bibitem{Grmela2014a}
Grmela, M.: {Contact geometry of mesoscopic thermodynamics and dynamics}.
\newblock Entropy \textbf{16}(3), 1652--1686 (2014).
\newblock \doi{10.3390/e16031652}

\bibitem{Grmela2011a}
Grmela, M., Lebon, G., Dubois, C.: {Multiscale thermodynamics and mechanics of
  heat}.
\newblock Physical Review E - Statistical, Nonlinear, and Soft Matter Physics
  \textbf{83}(6), 1--15 (2011).
\newblock \doi{10.1103/PhysRevE.83.061134}

\bibitem{GrmelaOttingerI}
Grmela, M., {\"{O}}ttinger, H.C.: {Dynamics and thermodynamics of complex
  fluids. I. Development of a general formalism}.
\newblock Physical Review E \textbf{56}(6), 6620--6632 (1997).
\newblock \doi{10.1103/PhysRevE.56.6620}

\bibitem{dGM}
de~Groot, S.R., Mazur, P.: {Non-equilibrium Thermodynamics}.
\newblock Dover Books on Physics. Dover Publications (1984)

\bibitem{Gurtin}
Gurtin, M.E.: {An Introduction to Continuum Mechanics}.
\newblock Mathematics in Science and Engineering. Elsevier Science (1982)

\bibitem{Holm-EMHD}
Holm, D.D.: {Hamiltonian dynamics of a charged fluid, including electro-and
  magnetohydrodynamics}.
\newblock Phys. Lett. A \textbf{114}(3), 137--141 (1986)

\bibitem{Hron-B}
Hron, J., Milo{\v{s}}, V., Pr$\backslash$ru{\v{s}}a, V., Sou{\v{c}}ek, O.,
  T$\backslash$ruma, K.: {On thermodynamics of viscoelastic rate type fluids
  with temperature dependent material coefficients}.
\newblock Int. J. Non-Linear Mech. \textbf{95}, 193--208 (2017).
\newblock \doi{10.1016/j.ijnonlinmec.2017.06.011}

\bibitem{hutter-plastic}
H{\"{u}}tter, M., Svendsen, B.: {Thermodynamic model formulation for
  viscoplastic solids as general equations for non-equilibrium
  reversible–irreversible coupling}.
\newblock Continuum Mechanics and Thermodynamics \textbf{24}(3), 211--227
  (2012).
\newblock \doi{10.1007/s00161-011-0232-7}.
\newblock \urlprefix\url{http://link.springer.com/10.1007/s00161-011-0232-7}

\bibitem{hutter2013}
H{\"{u}}tter, M., Svendsen, B.: {Quasi-linear versus potential-based
  formulations of force–flux relations and the GENERIC for irreversible
  processes: comparisons and examples}.
\newblock Continuum Mechanics and Thermodynamics \textbf{25}(6), 803--816
  (2013).
\newblock \doi{10.1007/s00161-012-0289-y}

\bibitem{JNET-EntProdMax}
Jane{\v{c}}ka, A., Pavelka, M.: {Gradient dynamics and entropy production
  maximization}.
\newblock Journal of Non-Equilibrium Thermodynamics  (2017)

\bibitem{Kato1975}
Kato, T.: {The Cauchy problem for quasi-linear symmetric hyperbolic systems}.
\newblock Archive for Rational Mechanics and Analysis \textbf{58}(3), 181--205
  (1975).
\newblock \doi{10.1007/BF00280740}.
\newblock \urlprefix\url{http://link.springer.com/10.1007/BF00280740}

\bibitem{kauf}
Kaufman, A.N.: {Dissipative Hamiltonian systems: A unifying principle}.
\newblock Physics Letters A \textbf{100}(8), 419--422 (1984)

\bibitem{kroeger2010}
Kroeger, M., Huetter, M.: {Automated symbolic calculations in nonequilibrium
  thermodynamics}.
\newblock Comput. Phys. Commun. \textbf{181}, 2149--2157 (2010)

\bibitem{Landau-Lifshitz6}
Landau, L.D., Lifshitz, E.M.: {Fluid Mechanics, Course of Theoretical Physics,
  Volume 6}.
\newblock Elsevier Butterworth-Heinemann, Oxford (2004)

\bibitem{Landau1984electrodynamics}
{Landau L.D. Lifshitz}, E.M.: {Electrodynamics of continuous media}, vol.~8, 2
  edn.
\newblock elsevier (1984)

\bibitem{Malek-B}
M{\'{a}}lek, J., Pr$\backslash$ru{\v{s}}a, V.: {Derivation of equations for
  continuum mechanics and thermodynamics of fluids}.
\newblock In: Y.~Giga, A.~Novotn{\'{y}} (eds.) Handbook of Mathematical
  Analysis in Mechanics of Viscous Fluids, pp. 1--70. Springer (2017).
\newblock \doi{10.1007/978-3-319-10151-4_1-1}

\bibitem{MaWe}
Marsden, J.E., Weinstein, A.: {Coadjoint orbits, vortices and Clebsch variables
  for incompressible fluids}.
\newblock Physica D \textbf{7}, 305--323 (1983)

\bibitem{Mazaheri2016}
Mazaheri, A., Ricchiuto, M., Nishikawa, H.: {A first-order hyperbolic system
  approach for dispersion}.
\newblock Journal of Computational Physics \textbf{321}, 593--605 (2016).
\newblock \doi{10.1016/j.jcp.2016.06.001}.
\newblock
  \urlprefix\url{http://www.sciencedirect.com/science/article/pii/S0021999116302261}

\bibitem{Mielke2014}
Mielke, A., Peletier, M.A., Renger, D.R.M.: {On the Relation between Gradient
  Flows and the Large-Deviation Principle, with Applications to Markov Chains
  and Diffusion}.
\newblock Potential Analysis \textbf{41}(4), 1293--1327 (2014)

\bibitem{mor}
Morrison, P.J.: {Bracket formulation for irreversible classical fields}.
\newblock Physics Letters A \textbf{100}(8), 423--427 (1984)

\bibitem{Morrison1998}
Morrison, P.J.: {Hamiltonian description of the ideal fluid}.
\newblock Reviews of Modern Physics \textbf{70}(2), 467--521 (1998).
\newblock \doi{10.1103/RevModPhys.70.467}.
\newblock \urlprefix\url{http://link.aps.org/doi/10.1103/RevModPhys.70.467}

\bibitem{MullerRuggeri1998}
Muller, I., Ruggeri, T.: {Rational Extended Thermodynamics}, vol.~16.
\newblock Springer (1998)

\bibitem{Ottinger1998}
{\"{O}}ttinger, H.C.: {On the structural compatibility of a general formalism
  for nonequilibrium dynamics with special relativity}.
\newblock Physica A: Statistical Mechanics and its Applications
  \textbf{259}(1-2), 24--42 (1998).
\newblock \doi{10.1016/S0378-4371(98)00298-2}.
\newblock
  \urlprefix\url{http://linkinghub.elsevier.com/retrieve/pii/S0378437198002982}

\bibitem{Ottinger-book}
{\"{O}}ttinger, H.C.: {Beyond Equilibrium Thermodynamics}.
\newblock Wiley (2005)

\bibitem{GrmelaOttingerII}
{\"{O}}ttinger, H.C., Grmela, M.: {Dynamics and thermodynamics of complex
  fluids. II. Illustrations of a general formalism}.
\newblock Physical Review E \textbf{56}(6), 6633--6655 (1997).
\newblock \doi{10.1103/PhysRevE.56.6633}.
\newblock \urlprefix\url{http://link.aps.org/doi/10.1103/PhysRevE.56.6633}

\bibitem{Pavelka2016}
Pavelka, M., Klika, V., Esen, O., Grmela, M.: {A hierarchy of Poisson brackets
  in non-equilibrium thermodynamics}.
\newblock Physica D: Nonlinear Phenomena \textbf{335}, 54--69 (2016).
\newblock \doi{10.1016/j.physd.2016.06.011}.
\newblock
  \urlprefix\url{http://linkinghub.elsevier.com/retrieve/pii/S0167278915301019}

\bibitem{Pavelka2014a}
Pavelka, M., Klika, V., Grmela, M.: {Time reversal in nonequilibrium
  thermodynamics}.
\newblock Physical Review E \textbf{90}(6), 1--19 (2014).
\newblock \doi{10.1103/PhysRevE.90.062131}.
\newblock \urlprefix\url{http://link.aps.org/doi/10.1103/PhysRevE.90.062131}

\bibitem{PRE15}
Pavelka, M., Klika, V., Grmela, M.: {Time reversal in nonequilibrium
  thermodynamics}.
\newblock Phys. Rev. E \textbf{90}(062131) (2014)

\bibitem{PeshGrmRom2015}
Peshkov, I., Grmela, M., Romenski, E.: {Irreversible mechanics and
  thermodynamics of two-phase continua experiencing stress-induced solid-fluid
  transitions}.
\newblock Continuum Mechanics and Thermodynamics \textbf{27}(6), 905--940
  (2015).
\newblock \doi{10.1007/s00161-014-0386-1}

\bibitem{HPR2016}
Peshkov, I., Romenski, E.: {A hyperbolic model for viscous Newtonian flows}.
\newblock Continuum Mechanics and Thermodynamics \textbf{28}(1-2), 85--104
  (2016).
\newblock \doi{10.1007/s00161-014-0401-6}

\bibitem{HYP2016}
Peshkov, I., Romenski, E., Dumbser, M.: {A unified hyperbolic formulation for
  viscous fluids and elastoplastic solids}.
\newblock ArXiv e-prints (Accepted for Springer Proceedings in Mathematics and
  Statistics, XVI International Conference on Hyperbolic Problems)  (2017).
\newblock \urlprefix\url{http://arxiv.org/abs/1705.02151}

\bibitem{Powell1999}
Powell, K.G., Roe, P.L., Linde, T.J., Gombosi, T.I., {De Zeeuw}, D.L., Keck,
  W.M.: {A Solution-Adaptive Upwind Scheme for Ideal Magnetohydrodynamics}.
\newblock Journal of Computational Physics \textbf{154}, 284--309 (1999).
\newblock \urlprefix\url{http://www.idealibrary.com}

\bibitem{Rajagopal-B}
Rajagopal, K.R., Srinivasa, A.R.: {A thermodynamic frame work for rate type
  fluid models}.
\newblock J. Non-Newton. Fluid Mech. \textbf{88}(3), 207--227 (2000).
\newblock \doi{10.1016/S0377-0257(99)00023-3}

\bibitem{Romenski2016}
Romenski, E., Belozerov, A., Peshkov, I.: {Conservative formulation for
  compressible multiphase flows}.
\newblock Quarterly of Applied Mathematics \textbf{74}(1), 1--24 (2016).
\newblock \doi{10.1090/qam/1409}

\bibitem{RomDrikToro2010}
Romenski, E., Drikakis, D., Toro, E.: {Conservative Models and Numerical
  Methods for Compressible Two-Phase Flow}.
\newblock Journal of Scientific Computing \textbf{42(1)}, 68--95 (2010)

\bibitem{RomToro2007}
Romenski, E., Resnyansky, A.D., Toro, E.F.: {Conservative hyperbolic model for
  compressible two-phase flow with different phase pressures and temperatures}.
\newblock Quarterly of applied mathematics \textbf{65(2)}, 259--279 (2007)

\bibitem{Romenski2011}
Romenski, E.I., Sadykov, A.D.: {On Modeling the Frequency Transformation Effect
  in Elastic Waves}.
\newblock Journal of Applied and Industrial Mathematics \textbf{5}(2), 282--289
  (2011).
\newblock \doi{10.1134/S1990478911020153}

\bibitem{Rom1989}
Romenskii, E.I.: {Hyperbolic equations of Maxwell's nonlinear model of
  elastoplastic heat-conducting media}.
\newblock Siberian Mathematical Journal \textbf{30}(4), 606--625 (1989)

\bibitem{Rom1998}
Romensky, E.I.: {Hyperbolic systems of thermodynamically compatible
  conservation laws in continuum mechanics}.
\newblock Mathematical and computer modelling \textbf{28(10)}, 115--130 (1998)

\bibitem{Rom2001}
Romensky, E.I.: {Thermodynamics and hyperbolic systems of balance laws in
  continuum mechanics}.
\newblock In: Godunov Methods: Theory and Applications, pp. 745--761 (2001)

\bibitem{Ruggeri2005}
Ruggeri, T.: {Global existence of smooth solutions and stability of the
  constant state for dissipative hyperbolic systems with applications to
  extended thermodynamics}.
\newblock In: S.~Rionero, G.~Romano (eds.) Trends and Applications of
  Mathematics to Mechanics, pp. 215--224. Springer, Milano (2005).
\newblock \doi{10.1007/88-470-0354-7_17}

\bibitem{Ruggeri1981Euler}
Ruggeri, T., Strumia, A.: {Main field and convex covariant density for
  quasi-linear hyperbolic systems. Relativistic Fluid Dynamics}.
\newblock Ann. Inst. Henri Poincar{\'{e}} \textbf{34}(1), 65--84 (1981)

\bibitem{StruchtrupTorrilhonR13}
Struchtrup, H., Torrilhon, M.: {Regularization of Grad's 13 moment equations:
  Derivation and linear analysis}.
\newblock Physics of Fluids \textbf{15}(9), 2668--2680 (2003).
\newblock \doi{10.1063/1.1597472}

\bibitem{Torrilhon2016a}
Torrilhon, M.: {Modeling Nonequilibrium Gas Flow Based on Moment Equations}.
\newblock Annual Review of Fluid Mechanics \textbf{48}(1), 429--458 (2016).
\newblock \doi{10.1146/annurev-fluid-122414-034259}.
\newblock \urlprefix\url{http://dx.doi.org/10.1146/annurev-fluid-122414-034259}

\bibitem{Van2017a}
V{\'{a}}n, P., Berezovski, A., F{\"{u}}l{\"{o}}p, T., Gr{\'{o}}f, G.,
  Kov{\'{a}}cs, R., Lovas, {\'{A}}., Verh{\'{a}}s, J.: {Guyer-Krumhansl–type
  heat conduction at room temperature}.
\newblock EPL (Europhysics Letters) \textbf{118}(5), 50,005 (2017).
\newblock \doi{10.1209/0295-5075/118/50005}.
\newblock \urlprefix\url{http://stacks.iop.org/0295-5075/118/i=5/a=50005}

\end{thebibliography}

\end{document}